\documentclass[sigconf]{acmart}

\AtBeginDocument{%
  \providecommand\BibTeX{{%
    \normalfont B\kern-0.5em{\scshape i\kern-0.25em b}\kern-0.8em\TeX}}}

\copyrightyear{2021}
\acmYear{2021}
\setcopyright{acmlicensed}\acmConference[SIGMOD '21]{Proceedings of the 2021 International Conference on Management of Data}{June 20--25, 2021}{Virtual Event, China}
\acmBooktitle{Proceedings of the 2021 International Conference on Management of Data (SIGMOD '21), June 20--25, 2021, Virtual Event, China}
\acmPrice{15.00}
\acmDOI{10.1145/3448016.3457300}
\acmISBN{978-1-4503-8343-1/21/06}

\usepackage{tikz}
\usepackage{mathtools}
\usepackage{amsfonts, amsthm} 
\usepackage{wasysym}
\usepackage{algorithm}
\usepackage[noend]{algpseudocode}
\usepackage[caption=false,font=scriptsize]{subfig}
\usepackage{balance}
\usepackage[percent]{overpic}

\usepackage{filecontents}

\usepackage{enumitem}
\usepackage{etoolbox}
\usepackage{tikz}
\usetikzlibrary{tikzmark}
\usetikzlibrary{calc}
\errorcontextlines\maxdimen

\newcommand{\ALGtikzmarkcolor}{black}
\newcommand{\ALGtikzmarkextraindent}{4pt}
\newcommand{\ALGtikzmarkverticaloffsetstart}{-.5ex}
\newcommand{\ALGtikzmarkverticaloffsetend}{-.5ex}
\makeatletter
\newcounter{ALG@tikzmark@tempcnta}

\newcommand\ALG@tikzmark@start{%
    \global\let\ALG@tikzmark@last\ALG@tikzmark@starttext%
    \expandafter\edef\csname ALG@tikzmark@\theALG@nested\endcsname{\theALG@tikzmark@tempcnta}%
    \tikzmark{ALG@tikzmark@start@\csname ALG@tikzmark@\theALG@nested\endcsname}%
    \addtocounter{ALG@tikzmark@tempcnta}{1}%
}

\def\ALG@tikzmark@starttext{start}
\newcommand\ALG@tikzmark@end{%
    \ifx\ALG@tikzmark@last\ALG@tikzmark@starttext
    \else
        \tikzmark{ALG@tikzmark@end@\csname ALG@tikzmark@\theALG@nested\endcsname}%
        \tikz[overlay,remember picture] \draw[\ALGtikzmarkcolor] let \p{S}=($(pic cs:ALG@tikzmark@start@\csname ALG@tikzmark@\theALG@nested\endcsname)+(\ALGtikzmarkextraindent,\ALGtikzmarkverticaloffsetstart)$), \p{E}=($(pic cs:ALG@tikzmark@end@\csname ALG@tikzmark@\theALG@nested\endcsname)+(\ALGtikzmarkextraindent,\ALGtikzmarkverticaloffsetend)$) in (\x{S},\y{S})--(\x{S},\y{E});%
    \fi
    \gdef\ALG@tikzmark@last{end}%
}

\apptocmd{\ALG@beginblock}{\ALG@tikzmark@start}{}{\errmessage{failed to patch}}
\pretocmd{\ALG@endblock}{\ALG@tikzmark@end}{}{\errmessage{failed to patch}}
\makeatother

\begin{document}
\title{Hybrid Edge Partitioner: Partitioning Large Power-Law Graphs under Memory Constraints}

\author{Ruben Mayer}
\affiliation{%
  \institution{Technical University of Munich}
}
\email{ruben.mayer@tum.de}

\author{Hans-Arno Jacobsen}
\affiliation{%
  \institution{University of Toronto}
}
\email{jacobsen@eecg.toronto.edu}

\begin{abstract}
Distributed systems that manage and process graph-structured data internally solve a \emph{graph partitioning} problem to minimize their communication overhead and query run-time. Besides computational complexity---optimal graph partitioning is NP-hard---another important consideration is the memory overhead. Real-world graphs often have an immense size, such that loading the complete graph into memory for partitioning is not economical or feasible. Currently, the common approach to reduce memory overhead is to rely on streaming partitioning algorithms. While the latest streaming algorithms lead to reasonable partitioning quality on some graphs, they are still not completely competitive to in-memory partitioners. In this paper, we propose a new system, Hybrid Edge Partitioner (HEP), that can partition graphs that fit \emph{partly} into memory while yielding a high partitioning quality. HEP can flexibly adapt its memory overhead by separating the edge set of the graph into two sub-sets. One sub-set is partitioned by NE++, a novel, efficient in-memory algorithm, while the other sub-set is partitioned by a streaming approach. Our evaluations on large real-world graphs show that in many cases, HEP outperforms both in-memory partitioning and streaming partitioning at the same time. Hence, HEP is an attractive alternative to existing solutions that cannot fine-tune their memory overheads. Finally, we show that using HEP, we achieve a significant speedup of distributed graph processing jobs on Spark/GraphX compared to state-of-the-art partitioning algorithms.
\end{abstract}

\ccsdesc[300]{Information systems~Graph-based database models}
\ccsdesc[500]{Theory of computation~Graph algorithms analysis}

\keywords{Graph partitioning; distributed graph processing}

\maketitle

\begin{tikzpicture}
\begin{scope}[overlay]
\node[text width=17.5cm] at ([yshift=-20cm,xshift=-11cm]current page.south) {(c) Owner 2021. This is the authors' version of the work. It is posted here for your personal use. Not for redistribution. \newline The definitive version is published in Proceedings of the 2021 International Conference on Management of Data (SIGMOD ’21), June 20–25, 2021, Virtual Event, China. ACM, New York, NY, USA. https://doi.org/10.1145/3448016.3457300.};
\end{scope}
\end{tikzpicture}

\fancyhead{}

\section{Introduction}
\label{sec:introduction}
In the past decade, many specialized frameworks for managing and processing graph-structured data have emerged, such as Pregel~\cite{pregel}, Giraph~\cite{giraph}, GraphLab~\cite{10.14778/2212351.2212354}, PowerGraph~\cite{powergraph}, Spark/GraphX~\cite{graphx}, Neo4j~\cite{10.1145/2384716.2384777}, and Trinity~\cite{10.1145/2463676.2467799}. All of these system share a fundamental property: They keep a large graph distributed onto multiple machines and compute global queries that can potentially span the entire graph, such that communication between the machines is unavoidable in order to answer the queries. Thereby, the induced amount of communication in query processing depends on how the graph is placed on the machines: The lower the \emph{cut size}\footnote{Informally, ``cut size'' means how many times an edge or vertex spans two different graph partitions. A formal definition is provided in Section~\ref{sec:problem}.} through the graph, the lower is the communication volume and the faster runs the query processing. This led to a renaissance of a classical problem in mathematics and computer science, being now researched in the context of optimizing specialized graph data processing systems: \emph{graph partitioning}~\cite{kernighan70, Karypis:1998:FHQ:305219.305248, gp-survey}.

Fueled by the developments described above, the focus of graph partitioning research has shifted toward highly skewed graphs, i.e., graphs where the distribution of the vertex degrees roughly follows a power-law~\cite{powergraph, Petroni:2015:HSP:2806416.2806424, Zhang:2017:GEP:3097983.3098033}. Such graphs are naturally emerging in many real-world situations, such as web graphs and online social networks. While the classical formulation of graph partitioning problems is centered around vertex partitioning, i.e., separating the vertices of the graph by cutting through the edges, it has been shown that \emph{edge partitioning} is more effective in reducing the communication volume of distributed query processing on skewed power-law graphs~\cite{Bourse:2014:BGE:2623330.2623660, 10.1145/1060590.1060674}. In edge partitioning, the edges of the graph are separated by cutting through the vertices. Unfortunately, the edge partitioning problem is NP-hard~\cite{Zhang:2017:GEP:3097983.3098033}. Hence, a number of heuristic algorithms have been proposed to solve the edge partitioning problem for large graphs \cite{Zhang:2017:GEP:3097983.3098033, 8416335, verma-vldb, Pacaci:2019:EAS:3299869.3300076,Stanton:2012:SGP:2339530.2339722}.

Existing partitioning algorithms can be divided into two categories: In-memory algorithms~\cite{schlag2019scalable, Margo:2015:SDG:2824032.2824046, Zhang:2017:GEP:3097983.3098033, dne} and streaming algorithms~\cite{powergraph, dbh, grid, Petroni:2015:HSP:2806416.2806424, 8416335}. In-memory algorithms load the complete graph into memory, and, hence, have full flexibility to assign any edge to any partition at any time. On the other hand, streaming algorithms pass through the edge stream, only looking at one (or a small number of) edge(s) at a time. None of these two ways of graph partitioning is entirely satisfactory. In-memory algorithms may yield very good partitioning quality even on challenging graphs, but they consume a lot of memory. Streaming algorithms consume little memory, but even though they have been improved by sophisticated techniques such as window-based streaming~\cite{8416335} and multi-pass streaming~\cite{Nishimura:2013:RGP:2487575.2487696}, they do not yield the same partitioning quality on all graphs as the best in-memory algorithms. In current graph partitioning systems, the user has to decide for one of the two options, and then either provide a very large machine (or a cluster of machines) and get good partitioning quality~\cite{schlag2019scalable, Margo:2015:SDG:2824032.2824046, Zhang:2017:GEP:3097983.3098033, dne} or a small machine and get worse partitioning quality~\cite{powergraph, dbh, grid, Petroni:2015:HSP:2806416.2806424, 8416335}. 

A further shortcoming in existing work is that many of the graph partitioning algorithms are described from a rather theoretical viewpoint without discussing their design and implementation~\cite{Petroni:2015:HSP:2806416.2806424, 8416335, Zhang:2017:GEP:3097983.3098033}. This leaves opportunities for optimizations unexplored. For instance, in their implementation of the neighborhood expansion (NE) algorithm~\cite{Zhang:2017:GEP:3097983.3098033}, the authors solve complex problems like the prevention of ``double assignments'' of edges (cf. Section~\ref{sec:lazyedge}) in a rather straight-forward and unoptimized manner.

In this paper, we propose a new hybrid approach of edge partitioning that fine-tunes the trade-off between memory consumption and partitioning quality. In doing so, we challenge the common pattern of partitioning algorithms and break out of the dichotomy of pure in-memory and pure streaming algorithms. Instead, we perform different partitioning strategies on different sub-graphs. Edges incident to at least one \emph{low-degree} vertex are partitioned with a novel efficient in-memory partitioning algorithm called NE++. Edges incident to two \emph{high-degree} vertices, however, are partitioned with stateful streaming partitioning. In doing so, partitioning state from in-memory partitioning is exploited in the streaming phase to increase the partitioning quality. 

In detail, our contributions are the following:

\begin{itemize}[noitemsep]
	\item[(1)] \textbf{We propose Hybrid Edge Partitioner (HEP)}, a novel graph partitioning system that follows the hybrid partitioning model outlined above. In many cases, HEP outperforms both in-memory partitioning and streaming partitioning at the same time. The code is available online: \url{https://github.com/mayerrn/hybrid_edge_partitioner}.
	\item[(2)] \textbf{We introduce NE++}, a new, highly \emph{memory-efficient} and \emph{fast} extension of the well-known NE algorithm~\cite{Zhang:2017:GEP:3097983.3098033}. In particular, NE++ shows significantly lower run-time and memory overhead than NE while yielding the same partitioning quality. We achieve this by introducing two novel optimizations, namely \emph{graph pruning} and \emph{lazy edge removal} which make the in-memory partitioning phase of HEP extremely resource-efficient.
	\item[(3)] We describe in detail the \textbf{design and implementation} we employ for graph partitioning in both phases, i.e., in-memory and streaming. This closes a gap in the literature, where papers mostly focus on the partitioning algorithms but fall short in discussing their efficient implementation.  
	\item[(4)] \textbf{Extensive evaluations} on seven real-world graphs of up to 64 billion edges show the efficiency of HEP compared to seven strong competitors. Furthermore, we show a \textbf{speed-up} in the state-of-the-art production-grade distributed \textbf{graph processing} system Spark/GraphX in many cases when using HEP instead of previously proposed partitioning algorithms. Finally, based on our evaluations, we make a recommendation when it is most beneficial to use HEP.
\end{itemize}

The rest of the paper is organized as follows. In Section~\ref{sec:problem}, we formalize and discuss the edge partitioning problem. In Section~\ref{sec:approach}, we introduce the HEP system, including the novel NE++ algorithm and further optimizations. We provide a theoretical analysis of HEP in Section~\ref{sec:analysis}. Finally, we evaluate HEP against seven strong baseline systems in Section~\ref{sec:evaluation}, discuss related work in Section~\ref{sec:related} and conclude the paper in Section~\ref{sec:conclusion}.

\section{Efficient Edge Partitioning}
\label{sec:problem}

The problem setting that HEP addressed relates to the type of partitioning and graphs targeted as well as to resource efficiency.


\begin{figure}
\centering
  \includegraphics[width=0.9\linewidth]{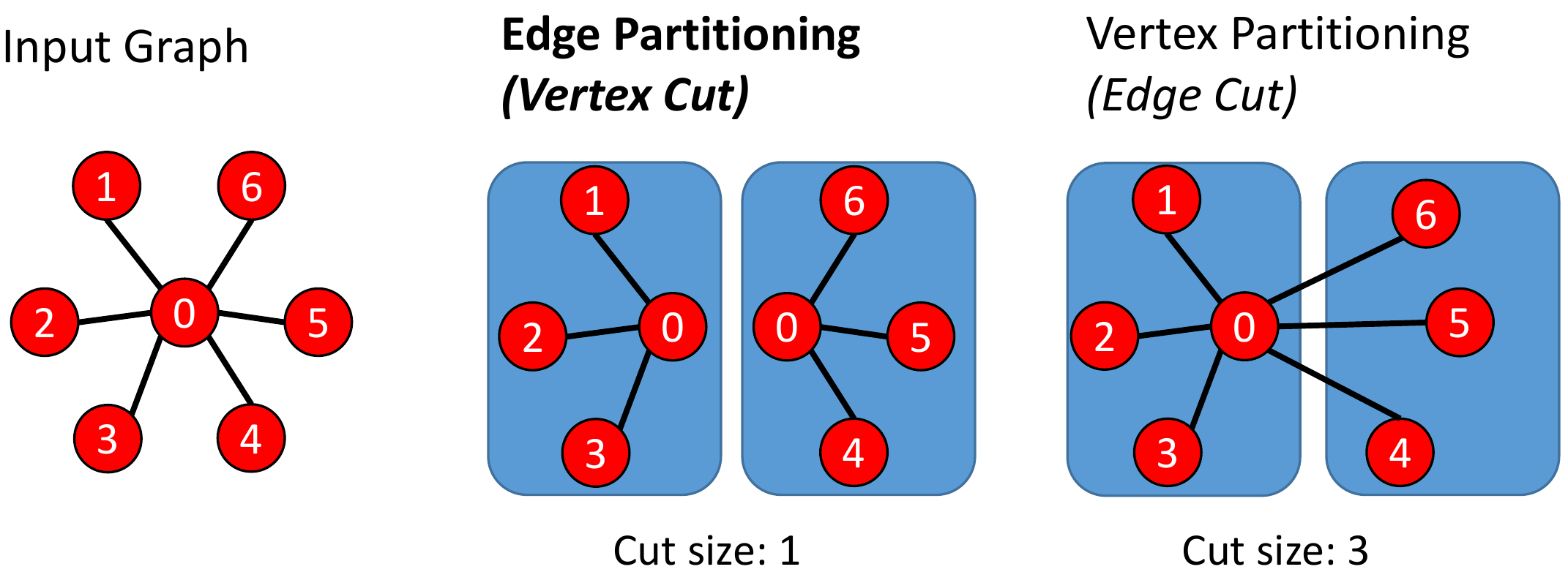}
  \vspace{-0.4cm}
  \caption{Edge partitioning vs. vertex partitioning.}
\vspace{-0.3cm}
  \label{fig:cut_examples}
\end{figure}

\emph{Cut Type}.
We tackle the \emph{edge partitioning problem}, as specified in the following (cf. ~\cite{Zhang:2017:GEP:3097983.3098033, Bourse:2014:BGE:2623330.2623660}). Let there be an undirected, unweighted graph $G = (V, E)$ with a set of vertices $V$ and a set of edges $E \subseteq V \times V$. The goal of edge partitioning is to divide $E$ into $k>1, k\in\mathbb{N}$ components that are called \emph{partitions} $P = \{p_1, ..., p_k\}$  such that $\bigcup_{i=1,...,k} p_i = E$ and  $p_i \cap p_j = \emptyset, i \neq j$. The size of the partitions is limited by a balancing constraint: $\forall p_i \in P : |p_i| \leq \alpha  * \frac{|E|}{k} $ for a given $\alpha \geq 1, \alpha \in \mathbb{R}$. Figure~\ref{fig:cut_examples} illustrates edge partitioning with $k=2$ partitions on an exemplary star-shaped graph. When an edge $(u,v)$ is assigned to a partition $p_i$, both $u$ and $v$ are \emph{covered} by $p_i$. For a partition $p_i$, denote $V(p_i)=\{x | (x,y) \in p_i\}$ as the set of vertices covered by $p_i$. When a vertex $v$ is covered by a partition, we also say that $v$ is \emph{replicated} on that partition. In the example in Figure~\ref{fig:cut_examples}, vertex 0 is replicated on both partitions, while the other vertices are replicated only on a single partition. Now, the optimization objective in graph partitioning is to minimize the \emph{replication factor} RF($p_1, \dots, p_k$)$ = \frac{1}{|V|} \sum_{i=1,...,k}{|V(p_i)|}$. In the given example, replicating vertex 0 induces communication between the two machines that execute a query that spans the complete graph. For instance, the state of vertex 0 would need to be synchronized in each iteration of a distributed graph processing algorithm~\cite{powergraph}. By minimizing the replication factor, the amount of synchronization between the distributed compute nodes that process a query is thus minimized.

Edge partitioning stands in contrast to vertex partitioning, where the vertices are assigned to partitions and the edges are cut (cf. right side in Figure~\ref{fig:cut_examples}). Here, the edge cut induces the communication in query processing; the machines exchange vertex states across the cut edges. As can be seen in the example of the star-shaped graph, such an edge cut might be much larger than the corresponding best vertex cut. Bourse et al.~\cite{Bourse:2014:BGE:2623330.2623660} have analyzed the difference between both cut types, proving that vertex cuts are smaller than edge cuts on power-law graphs. 


\emph{Graph Type.}
In many natural graphs, e.g., social networks and web graphs, the degree distribution of the vertices follows a \emph{power-law}~\cite{RevModPhys.74.47}. In power-law graphs, a large fraction of edges in the graph are incident to only a few vertices which have a very high degree. This property can be exploited by assigning edges to partitions in such a way that the replication of a high-degree vertex is preferred to the replication of a low-degree vertex. The rationale for this is that high-degree vertices are incident to so many edges that they are likely to be replicated anyway; by focusing on placing low-degree vertices with a low replication factor, the overall replication factor can be decreased~\cite{powerlyra, Petroni:2015:HSP:2806416.2806424, dbh}.

\emph{Efficiency.}
Resource efficiency is desirable, e.g., in order to reduce monetary costs. Given a compute node with constrained memory capacities (which may be less than the size of the graph data), the graph partitioning algorithm should be able to provide the best possible replication factor in the lowest possible run-time.

\section{Hybrid Edge Partitioner}
\label{sec:approach}

HEP is a hybrid edge partitioning system that combines in-memory partitioning with streaming partitioning. In-memory partitioning can achieve a low replication factor, because the problem is less constrained than streaming partitioning. On the other hand, streaming partitioning induces very low memory overhead, because there is no need to keep the complete graph in memory. Our objective is to combine both paradigms by partitioning one sub-part of the graph with in-memory partitioning and the other sub-part with streaming partitioning. To do so, we have to find a sub-graph that can be partitioned with streaming partitioning while not compromising too much in terms of the replication factor. 

\subsection{Approach Overview}
\label{sec:dividing_the_graph}

\begin{figure}
	\centering	
		\subfloat[LJ graph.]{\label{a}   \includegraphics[width=0.49\linewidth]{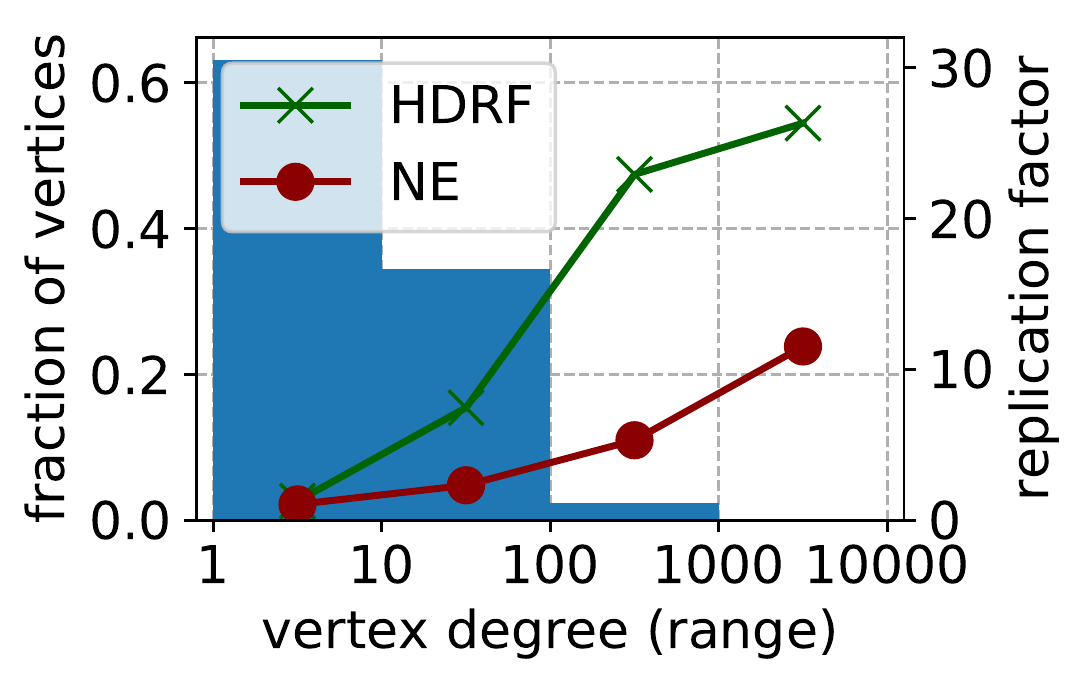}}
	\subfloat[WI graph.]{\label{b}   \includegraphics[width=0.49\linewidth]{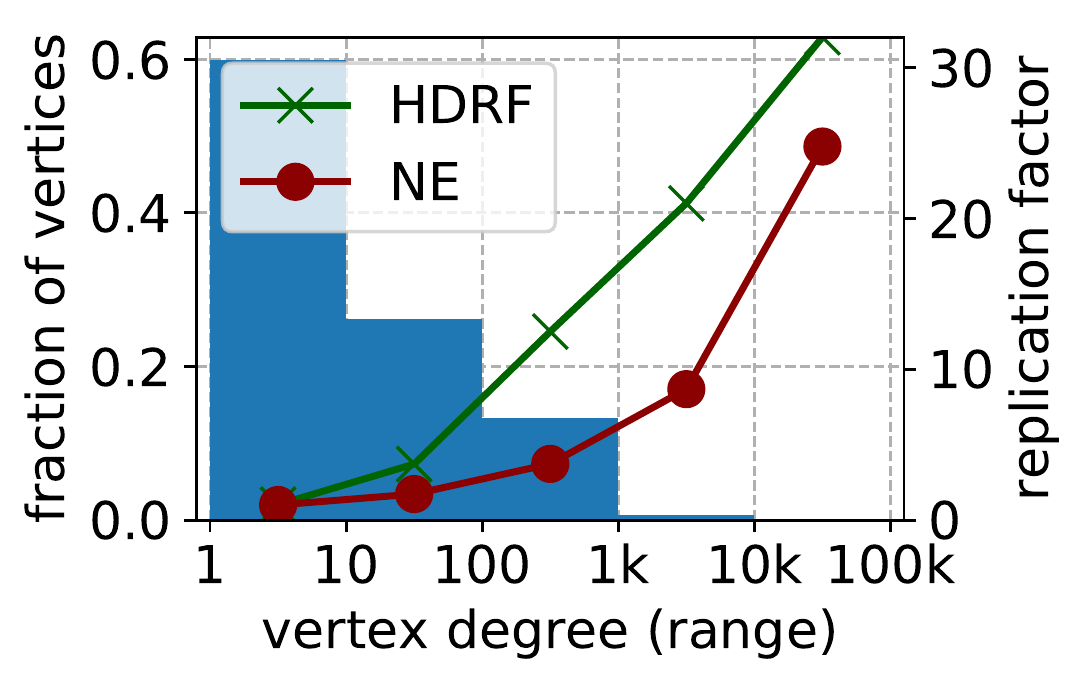}} 
	\vspace{-7pt}
	\caption{Degree vs. replication factor for LJ and WI graph (cf. Table~\ref{tab:graphs}) at $k=32$ partitions.}
	\label{eval:degree_vs_rf}
	\vspace{-5pt}
\end{figure}

We divide the graph into two sub-graphs by considering the vertex degrees. To motivate this, we perform experiments on two different real-world graphs (cf. Figure~\ref{eval:degree_vs_rf}). We measure the replication factor of vertices that have a degree in a specific range ($[1,10], [11,100],$ etc.) using both the streaming partitioner HDRF~\cite{Petroni:2015:HSP:2806416.2806424} as well as the in-memory partitioner NE~\cite{Zhang:2017:GEP:3097983.3098033} for building $k=32$ partitions. In the same figure, we also plot the degree distribution of the respective graphs. We can observe that the replication factor obtained by either partitioner heavily depends on the vertex degree: Vertices with a low degree are less often replicated than vertices with a high degree. Furthermore, the number of vertices of a low degree is much higher than the number of vertices of a high degree. Both HDRF and NE are \emph{degree-aware} partitioners: They focus on trying to reduce the replication factor of the many low-degree vertices rather than the few high-degree vertices.

\begin{algorithm}[t]
\caption{NE: Basic Algorithm.}
\begin{algorithmic}[1]
\footnotesize
\Procedure{Initialize}{$p_i$}
\State $v \gets $ any vertex in $V$, such that $v \not\in C$
\State \textsc{MoveToCore}($v$) 
\EndProcedure
\vspace{0.05cm}
\Procedure{PartitionGraph}{}
\For{\textbf{each} $p_i \in P$}
	\While{$|p_i| < \frac{|E|}{k}$}
		\If{$S_i \setminus C \neq \emptyset$}
			\State $v_{\mathit{min}} \gets   \underset{v \in S_i \setminus C}{\operatorname{arg\,min}}\, d_{\mathit{ext}}(v, S_i) $ 
			\State \textsc{MoveToCore}($v_{\mathit{min}}$)
		\Else
			\State \textsc{Initialize}($p_i$)
		\EndIf 
	\EndWhile
\EndFor
\EndProcedure
\vspace{0.05cm}

\Procedure{MoveToCore}{Vertex $v$}
	\State $C \gets C \cup v$
	\For{\textbf{each} $u \in N(v) \setminus (C \cup S_i) $} \Comment{$N(v)$ is the set of $v$'s neighbors.}
		\State \textsc{MoveToSecondary}($u$) 
	\EndFor
\EndProcedure

\vspace{0.05cm}

\Procedure{MoveToSecondary}{Vertex $v$}
	\State $S_i \gets S_i \cup v$
	\State $d_{\mathit{ext}}(v, S_i) \gets d(v)$
	\For{\textbf{each} $u \in N(v) \cap (C \cup S_i) $}
		\State $d_{\mathit{ext}}(u, S_i) \gets d_{\mathit{ext}}(u, S_i) - 1$
		\State $d_{\mathit{ext}}(v, S_i) \gets d_{\mathit{ext}}(v, S_i) - 1$
		\If{$|p_i| < \frac{|E|}{k}$}
			\State $p_i \gets p_i \cup (u, v)$ \Comment{assign edge to $p_i$}
			\State \textsc{RemoveEdge}($(u, v)$)
		\Else \Comment{spill over to next partition}
			\State $p_{i+1} \gets p_{i+1} \cup (u, v)$
			\State \textsc{RemoveEdge}($(u, v)$)
			\State $S_{i+1} \gets S_{i+1} \cup \{u, v\}$
		\EndIf
	\EndFor
\EndProcedure
\end{algorithmic}
\label{alg:ne}
\end{algorithm}

Our strategy in HEP is similar to existing partitioners in the sense that we try to achieve very good partitioning quality for low-degree vertices to keep the overall replication factor low. However, we add the new aspect of memory efficiency by allowing to compromise on the replication factor of high-degree vertices in order to reduce memory overhead:

(1) Edges that are incident to at least one \emph{low-degree} vertex are partitioned with in-memory partitioning. This significantly reduces the overall replication factor, because in power-law graphs, the number of low-degree vertices is much higher than the number of high-degree vertices.

(2) Edges  that are incident to two \emph{high-degree} vertices ($E_{h2h}$) are partitioned with streaming partitioning. Even if there are many edges between high-degree vertices, the number of vertices affected will still be low, because of the skewedness of power-law graphs. Hence, a large number of edges can be partitioned with streaming partitioning while only a small number of vertices is affected. 

But what is a ``high'' degree? We define a \emph{threshold factor} $\tau$ (tau) that separates high-degree vertices $V_h$ from low-degree vertices $V_l$, such that $V_h \cup V_l = V$ and $V_h \cap V_l = \emptyset$, as follows:

\vspace{-5pt}
\[ v \in
  \begin{cases}
   V_h       & \quad \text{if } d(v) > \tau * \diameter_d, \\
    V_l  & \quad \text{else,}
  \end{cases}
\]


where $\diameter_d$ is the mean degree of all vertices in the graph. By setting $\tau$, one can control the memory overhead of the partitioning algorithm: A lower $\tau$ means that more edges are incident to two high-degree vertices, so that the memory overhead is reduced because more edges are partitioned with streaming partitioning.

\subsection{In-Memory Partitioning via NE++}
 
The first phase of HEP is in-memory partitioning via NE++. NE++ is a new, highly \emph{memory-efficient} and \emph{fast} extension of the well-known NE (neighborhood expansion) algorithm~\cite{Zhang:2017:GEP:3097983.3098033}. In particular, NE++ introduces two novelties compared to NE. (1) NE++ employs a \emph{pruned graph represention}, which significantly reduces the memory overhead. (2) NE++ introduces \emph{lazy edge removal}, an efficient technique to avoid the assignment of the same edge to multiple partitions without eagerly keeping books on past edge assignments.

\begin{figure}
\centering
  \includegraphics[width=0.85\linewidth]{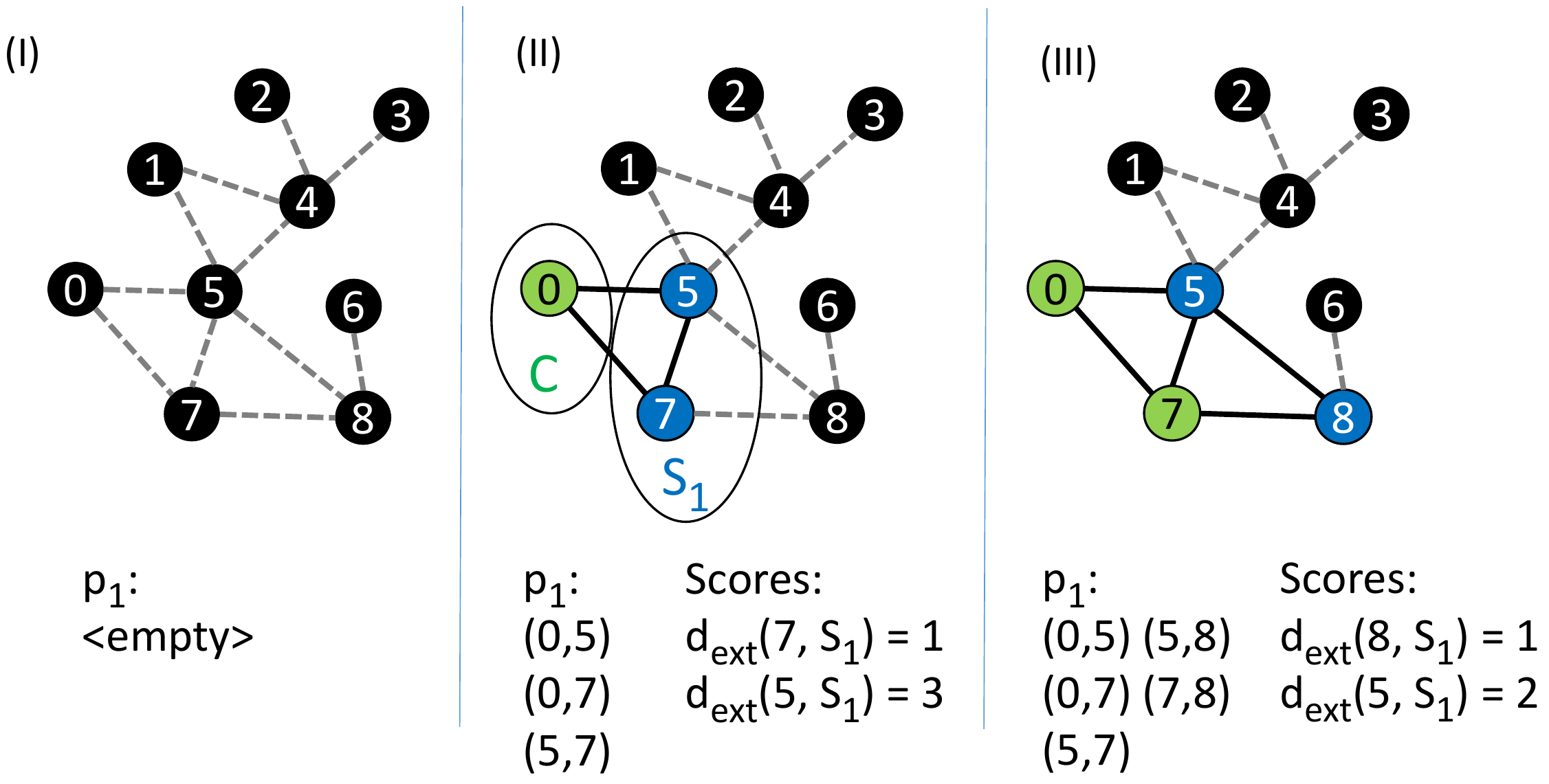}
\vspace{-0.15cm}
  \caption{Example of expansion step. (I) Input graph. (II) Initialization. (III) Partitioning state after first expansion step.}
\vspace{-0.25cm}
  \label{fig:ne_scheme}
\end{figure}

\paragraph{Basic NE Algorithm}
Before we detail the technical contributions of NE++, we quickly review the basic NE algorithm~\cite{Zhang:2017:GEP:3097983.3098033} and analyze its shortcomings. In the NE algorithm, partitions are built in sequence. To build partition $p_i$, a seed vertex is chosen (Algorithm~\ref{alg:ne}, lines 1--3). This seed vertex is placed in a vertex sub-set that is called \emph{core} set, denoted by $C$; hence, the seed vertex must not have been in $C$ before. All neighboring vertices of the seed vertex in $p_i$ are placed in another vertex sub-set that is called \emph{secondary} set, denoted by $S_i$. Edges that connect vertices between or within $C$ and $S_i$ are assigned to the current partition $p_i$. Now, in an \emph{expansion step} (lines 6--11), the vertex from $S_i$ is selected that has the lowest \emph{external degree} $d_{\mathit{ext}}(v,S_i)$, i.e., the lowest number of neighbors that are neither in $C$ nor in $S_i$ (line 8). This vertex is moved from $S_i$ to $C$ (line 9) and its external neighbors are added to $S_i$ (lines 12--15). Whenever a vertex $v$ is added to $S_i$, the algorithm decrements the external degree of neighboring vertices that are already in $S_i$ and calculates the external degree of $v$ (lines 19--25). Finally, edges between $v$ and vertices in $S_i$ or $C$ are assigned to the current partition $p_i$ (line 23); such edges are then removed from the graph to avoid double assignments (line 24). This ends the expansion step. Then, the next expansion step is performed by again determining the vertex in $S_i$ with the lowest degree and moving it to $C$ (lines 8--9). The expansion phase of a partition $p_i$ stops when the capacity bound of $p_i$ is reached. Then, the next partition $p_{i+1}$ is built. If the capacity bound of a partition is reached within an expansion step, further edges assigned in the expansion step are ``spilled over'' to the next partition $p_{i+1}$ (lines 26--28). The overall algorithm terminates when all partitions have been built.
In Figure~\ref{fig:ne_scheme}, we provide a detailed example of the first expansion step in a partition $p_1$. Vertices in $C$ are marked in green, vertices in $S_1$ are marked in blue. We further list the set of edges already assigned to $p_1$ and the external degrees of all vertices in $S_1$.

\emph{Limitations of NE.} We identify the following limitations of NE. (1) The complete graph must be loaded into memory. This prevents NE from being used to partition graphs that exceed the memory capacity of the available compute node. (2) NE eagerly keeps book on which edge has been assigned to which partition. This induces high memory and run-time overhead.



\begin{figure}
\centering
  \includegraphics[width=0.95\linewidth]{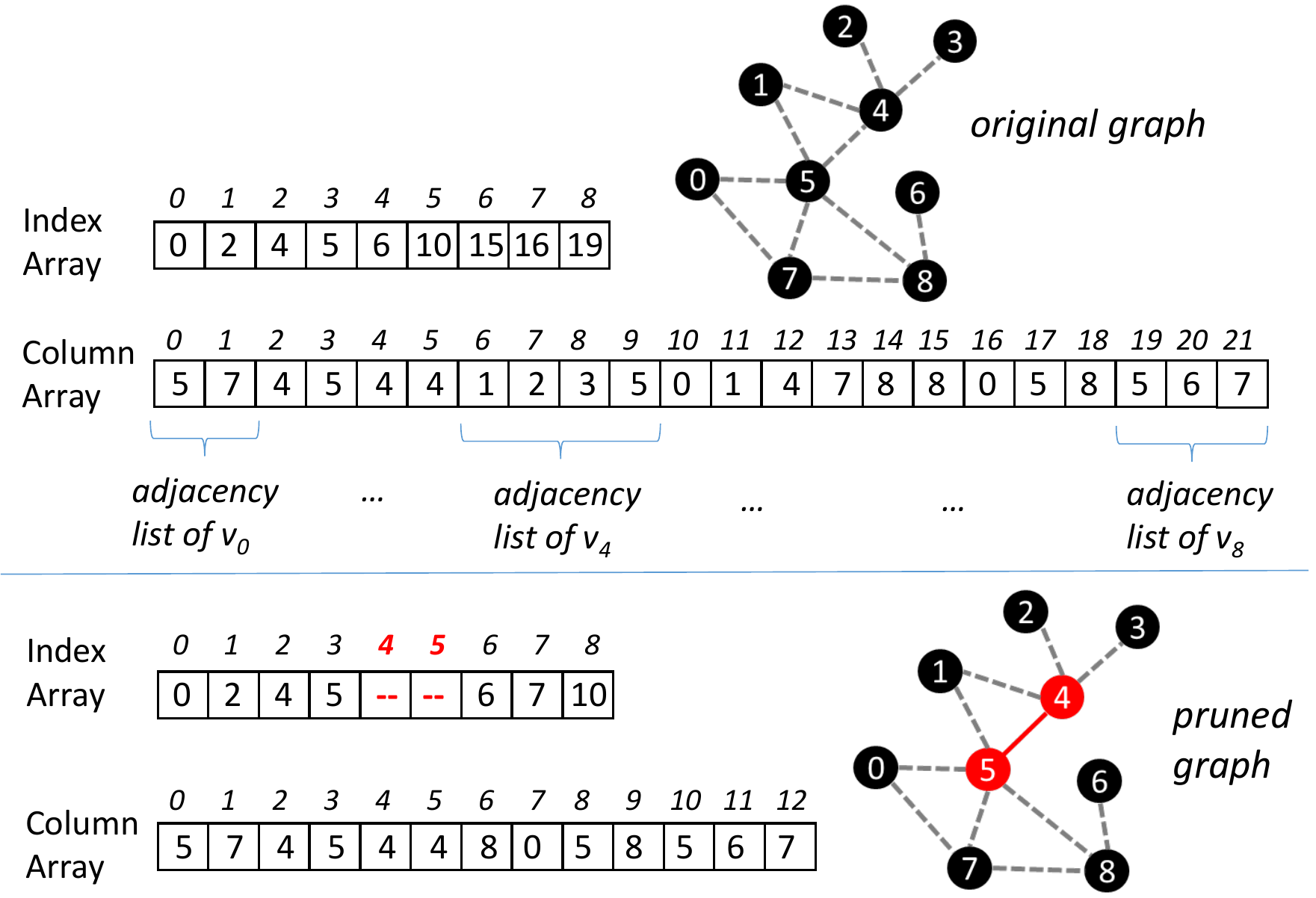}
  \caption{CSR representation of original and of pruned graph. Red-shaded vertices are high-degree.}
  \label{fig:csr}
\end{figure}

\subsubsection{Pruned Graph Representation}
\label{sec:graphconstruction}

NE++ employs an adaptation of the compressed sparse row (CSR) graph representation~\cite{10.1145/1583991.1584053, 1447941}\footnote{In our CSR implementation, we store edges in both directions (in and out) for each vertex in the column array. We denote the entries in the column array that represent the neighborhood of a vertex $v$ also as $v$'s \emph{adjacency list}.}. However, when we build the CSR from the input graph file---a binary edge list---, we leave out certain parts of the column array. In particular, \emph{we do not store the adjacency lists of high-degree vertices in the column array}. Edges between a low-degree and a high-degree vertex can still be reached via the adjacency list of the low-degree vertex. However, edges between two high-degree vertices are not represented in the column array at all. We, hence, write out edges between two high-degree vertices to an external file while building the CSR. 
Figure~\ref{fig:csr} depicts an example. The original graph contains 9 vertices and 11 undirected edges; the average degree $\diameter_d$ is 2.4. In the column array, this results in 22 entries. Now, with a degree threshold of $\tau = 1.5$, all vertices with a degree of 4 or more are considered high-degree (in the example, those are $v_4$ and $v_5$). In the column array of the pruned graph, adjacency lists of $v_4$ and $v_5$ are, hence, omitted. To not lose the edge $(v_4, v_5)$, we write it out into an external edge file that is later partitioned by a streaming algorithm. The column array of the pruned graph is much smaller (in the example, 13 entries instead of 22), i.e., we save memory.

\begin{figure}
\centering
  \includegraphics[width=0.8\linewidth]{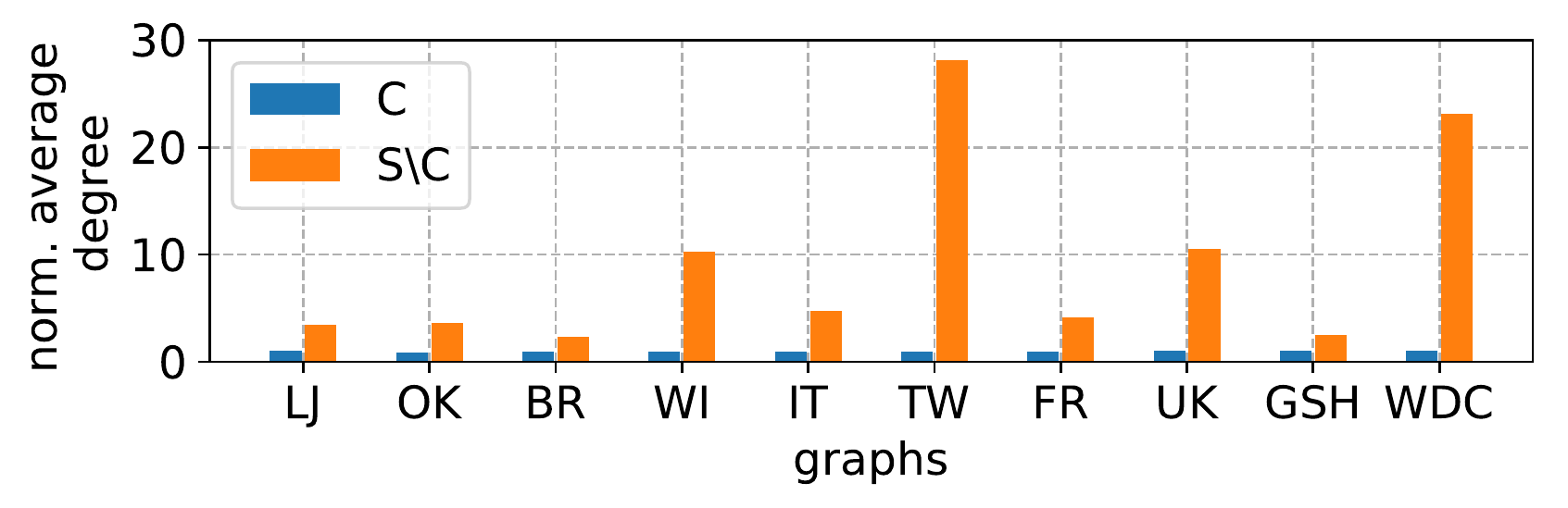}
 \vspace{-10pt}
  \caption{Average degree of vertices in $C$ and $S \setminus C$ at $k=32$ partitions, normalized to the total average degree (cf. Table~\ref{tab:graphs}).}
 \vspace{-10pt}
  \label{fig:eval:average_degrees}
\end{figure}

To adapt the NE algorithm to the pruned graph representation, we have to completely avoid access to the adjacency lists of high-degree vertices. We observe that in NE, high-degree vertices have a tendency to remain in the secondary set $S_i$ until the end of the expansion of a partition $p_i$, as they are less likely to have the lowest external degree compared to low-degree vertices. In Figure~\ref{fig:eval:average_degrees}, we compare the normalized average degree between vertices in $C$ and in $S \setminus C$ on different real-world graphs  when performing partitioning with NE at $k = 32$ partitions. The average degree of vertices that remain in the secondary set is very high, while for vertices that are moved to the core set, it is much lower. In other words, there are many high-degree vertices that remain in $S$ throughout the partition expansion and are not moved to $C$. 

In NE++, we exploit this property as follows: We assign high-degree vertices to the secondary set \emph{a priori} and also do not move them to the core set (\emph{``no expansion via a high-degree vertex''}), i.e., high-degree vertices are always in the secondary set. This implies that we do not iterate over the adjacency lists of high-degree vertices, so that the pruned graph representation in NE++ is sufficient. As moving a high-degree vertex to $C$ is unlikely in NE, NE++ stays close to the original behavior of the algorithm.

\subsubsection{Lazy Edge Removal}
\label{sec:lazyedge}

A central property of a valid edge partitioning is that an edge must be assigned \emph{exactly once}, i.e., to exactly one single partition. However, in an undirected graph, an edge $(u, v)$ is represented \emph{twice} in the CSR column array: if $v$ is a neighbor of $u$, then $u$ is also a neighbor of $v$. For instance, in Figure~\ref{fig:csr}, $v_7$ is in the adjacency list of $v_0$, and $v_0$ is in the adjacency list of $v_7$. To prevent an edge from being assigned to multiple partitions, the assignment of the edge to a partition must be taken into account in the further partitioning process, i.e., the edge must be removed or marked as invalid. There are two rather naive ways to implement this. First, one could remove the edge from two different locations in the column array. However, this induces computational overhead and leads to poor cache locality, as different (random) locations in the column array must be accessed subsequently. Second, one could introduce an auxiliary data structure that holds for each edge the information of whether it is still valid or not---the reference implementation of NE follows this approach. However, this induces additional memory overhead for the auxiliary data structure and it also leads to poor cache access locality, because such a data structure cannot be a contiguous representation of the adjacency lists of both vertices $u$ and $v$ at the same time.

In NE++, we introduce a new method, called \emph{lazy edge removal}, to solve this problem more efficiently. Lazy edge removal is both memory-efficient, as it foregoes introducing auxiliary data structures, and leads to a good cache locality, as it only accesses adjacency lists in a contiguous way. Before we introduce the method, we first analyze which vertices are visited in which phase of the expansion algorithm. If a vertex is not visited again by any other partition, removing edges in this vertex's adjacency list is not necessary. 

The following theorem (Theorem~\ref{theorem1}) proves that only vertices that remain in the secondary set of a partition at the end of the partition's expansion phase can be visited again in the expansion phase of a subsequent partition; vertices that are moved to the core set will not be visited again.

\vspace{-0.1cm}
\begin{theorem}
\label{theorem1}
\textbf{Access of Column Array.}
After a vertex $v$ has been moved to the secondary set $S_i$ of a partition $p_i$, the adjacency list of $v$ in the column array is accessed in a subsequent partition again only if $v$ is still in $S_i$ after $p_i$ has been completed.
\end{theorem}
\vspace{-0.2cm}
\begin{proof}
See Appendix~B.
\end{proof}
\vspace{-0.1cm}

As the adjacency lists of vertices moved to $C$ will not be accessed again, we can restrict edge removal for a partition $p_i$ to those vertices that remain in $S_i$. Hence, we introduce a \emph{clean-up phase} which is executed after the completion of each partition. It removes those entries from the column array that may be accessed again in a following partition $p_j, j>i$, but represent edges that have been assigned to $p_i$. This affects only adjacency lists of vertices in $S_i$. 

\begin{algorithm}[t]
\caption{Clean-up Algorithm.}
\begin{algorithmic}[1]
\footnotesize
\Procedure{CleanUp}{Partition $p_i$}
\For{\textbf{each} $v \in S_i$}
	\For{\textbf{each} $u \in N(v) \cap (C \cup S_i) $}
		\State $\mathit{v.adjacency\_list}.$\textsc{Remove}($u$)
	\EndFor	
\EndFor
\EndProcedure
\end{algorithmic}
\label{alg:cleanup}
\end{algorithm}

\begin{figure}
\centering
  \includegraphics[width=0.8\linewidth]{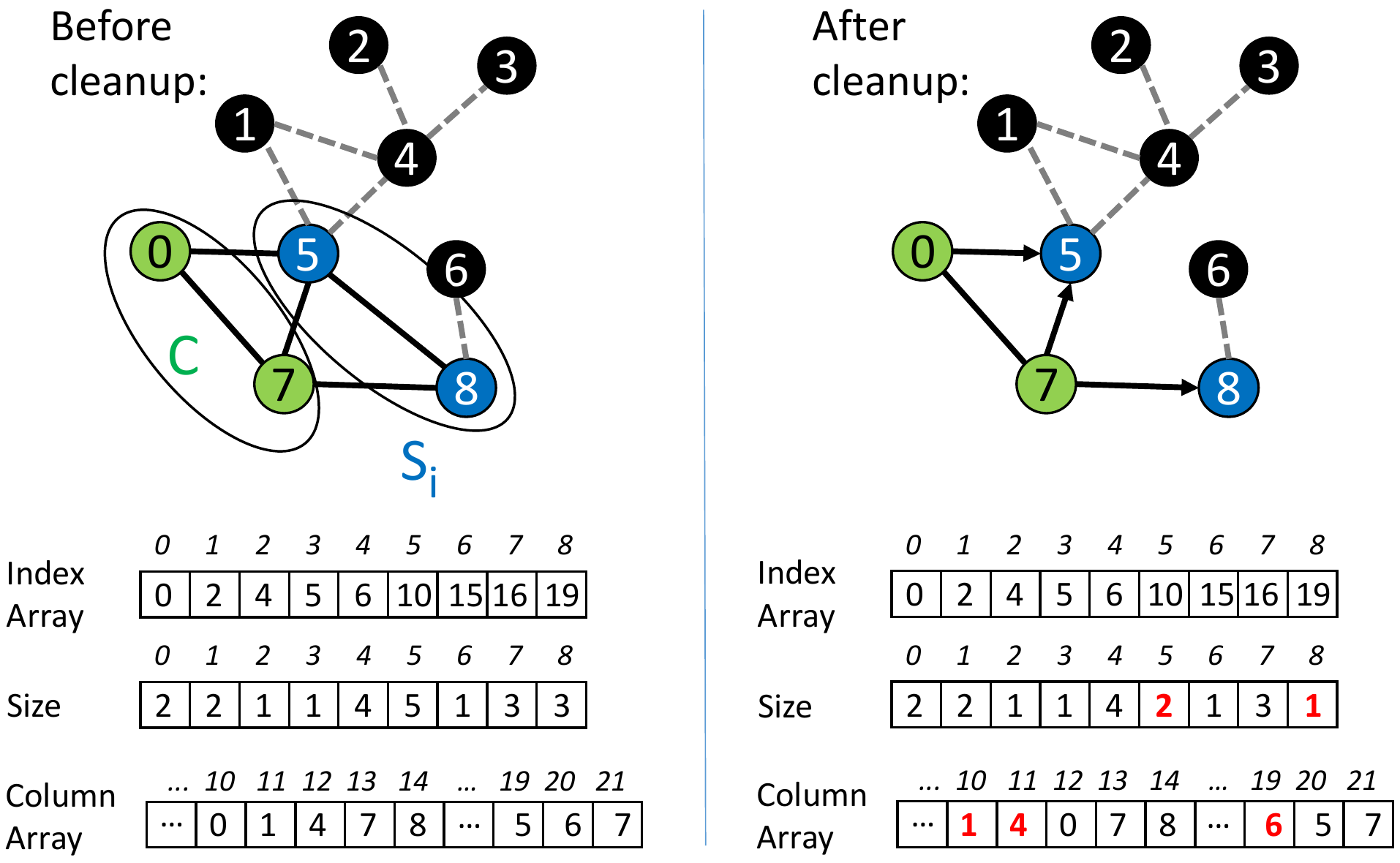}
 \vspace{-9pt}
  \caption{Clean-up phase: Example.}
 \vspace{-10pt}
  \label{fig:cleanup}
\end{figure}

The clean-up algorithm (Algorithm~\ref{alg:cleanup}) iterates through all vertices $v$ in $S_i$ and removes those edges from the column array that have been assigned to $p_i$ (i.e., neighbors of $v$ that are in $C$ or in $S_i$). Edge removal can be implemented efficiently by swapping the last valid entry of the adjacency list with the to-be-removed entry and then decrementing the ``size'' (a field indicating the number of valid entries) of the adjacency list. This is a constant-time operation.  Figure~\ref{fig:cleanup} depicts an example of a graph before and after the clean-up algorithm has been executed. The figure visualizes how the clean-up algorithm ``separates'' the core set from the rest of the graph; subsequent expansions cannot enter the core set of the graph any more, as the clean-up algorithm removes all links into it. Therefore, none of the edges that connect two vertices in the core set will be visited again and, hence, double-assignment of those edges is impossible. As the heuristic in NE++ tries to keep $S_i$ as small as possible, only a small number of vertices is visited in the clean-up phase.

\begin{figure}
\centering
  \includegraphics[width=0.8\linewidth]{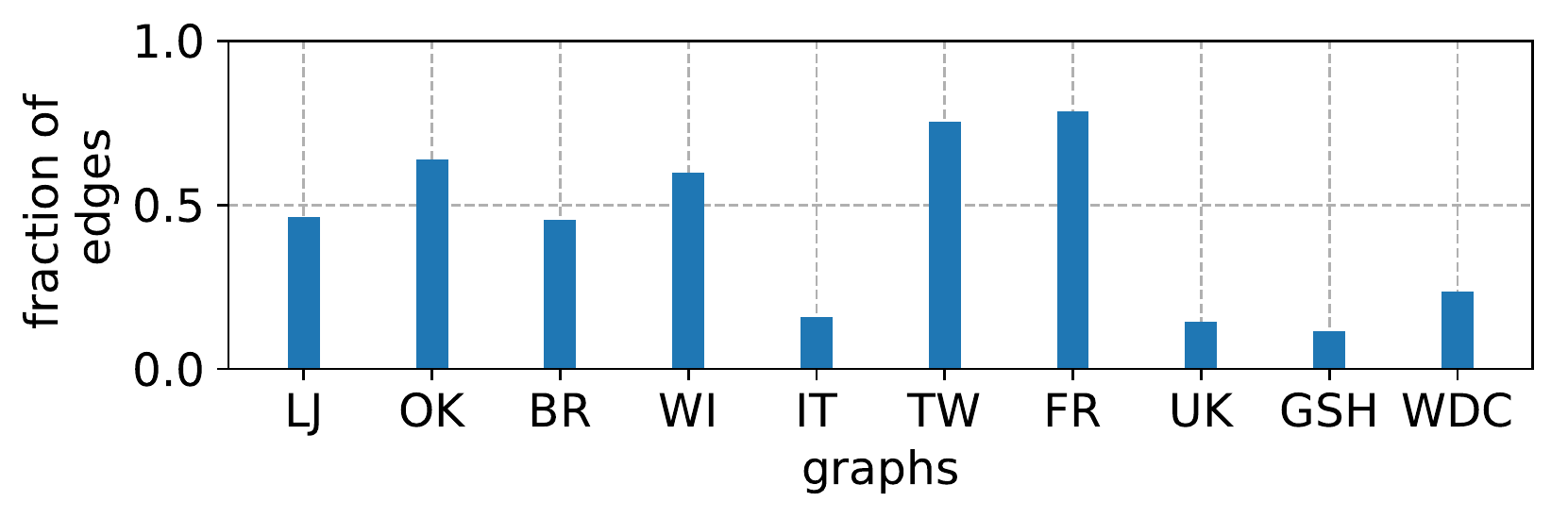}
 \vspace{-13pt}
  \caption{Total fraction of edges removed from column array during clean-up at $k=32$ partitions (graphs cf. Table~\ref{tab:graphs}).}
 \vspace{-7pt}
  \label{fig:eval:fraction_edges}
\end{figure}

To show the effectiveness of lazy edge invalidation, we evaluate on different graphs the overall fraction of edges in the column array which are removed in the clean-up process of NE++ (at $k = 32$ partitions). The results in Figure~\ref{fig:eval:fraction_edges} show that only a small fraction of edges are actually removed. The rest of the column array remains untouched. This is particularly pronounced on web graphs (IT, UK, GSH, WDC). Different from that, eager invalidation would remove all column array entries to prevent double assignment. 

\subsubsection{Further Optimizations for NE++}
\label{sec:optimizations}

\paragraph{Initialization.}
The initialization function is called in the expansion of a partition $p_i$ when there are no more vertices in $S_i$ that can be moved to $C$. The goal is to find a suitable vertex outside of $S_i$ that can be moved to $C$, i.e., the vertex must neither be a high-degree vertex nor a vertex in $C$. There are three scenarios when such an initialization is needed. First, when a partition's expansion is started. Second, when the graph is split into disconnected components and the expansion of a partition has assigned all edges of one of these components. Third, when there are only high-degree vertices in $S_i$ and, hence, no more vertices can be moved to $C$. The first case only happens once, but the other cases can occur frequently, depending on the graph structure and the number of high-degree vertices. Hence, it is important that the initialization procedure finds a suitable vertex quickly and efficiently. Using randomized vertex selection---as done in the reference implementation of NE---comes with the problem that the chance of hitting a suitable vertex becomes lower over time as the number of vertices that are not in $C$ decreases. Hence, the overhead grows as the partitioning process progresses. In NE++, we instead perform a sequential search through the vertices based on their ID. A vertex that was once found to be unsuitable for the initialization is thus never visited again in the initialization procedure. 

\emph{Adapted Partition Capacity Bound.}
As a consequence to graph pruning, edges incident to two high-degree vertices are not partitioned by NE++, but by streaming partitioning. Hence, in NE++, we adapt the capacity bound for the partitions, such that the in-memory edges are distributed to the partitions in a balanced fashion. Instead of a capacity bound of $\frac{|E|}{k}$, the new capacity bound is $\frac{|E \setminus E_{h2h}|}{k}$.

\begin{algorithm}[t]
\caption{Building the last partition in NE++.}
\begin{algorithmic}[1]
\footnotesize

\Procedure{AssignRemainingInMemoryEdges}{}
	\For{\textbf{each} $v \in V$}
		\If{$v \notin C$}
			\For{\textbf{each} $u \in N_{\textit{out}}(v)$}
				\State $p_i \gets p_i \cup (u,v)$
			\EndFor
			\If{$v$ has high-degree neighbors}
				\For{\textbf{each} $u \in N_{\textit{in}}(v)$}
					\If{$u \in V_h$}
						\State $p_i \gets p_i \cup (v,u)$
					\EndIf
				\EndFor
			\EndIf
		\EndIf
		\If{$|p_i| \geq \frac{|E|}{k}$}
			\State $i \gets i + 1$
		\EndIf
	\EndFor
\EndProcedure

\end{algorithmic}
\label{alg:remaining}
\end{algorithm}

\begin{algorithm}[t]
\caption{Streaming partitioning in HEP.}
\begin{algorithmic}[1]
\footnotesize

\Procedure{StatefulStreamingPartitioning}{}
	\For{\textbf{each} $e \in E_{\mathit{h2h}}$}
		\State $\mathit{bestScore} \gets 0$
		\State $\mathit{bestScore} \gets $ NULL
		\For{\textbf{each} $p_i \in P : |p_i| < \alpha  * \frac{|E|}{k} $} 
		\State \emph{score} $\gets$ $\mathit{scoring\_function}$($e$, $p_i$) 
		\If{\emph{score} $>$ \emph{bestScore}}
			\State \emph{bestScore} $\gets$ \emph{score}
			\State \emph{target\_p} $\gets p_i$ 
		\EndIf
	\EndFor
	\State \emph{target\_p} $\gets$ \emph{target\_p} $ \cup\ e$ 
	\EndFor
\EndProcedure

\end{algorithmic}
\label{alg:streaming}
\end{algorithm}

\emph{Building the Last Partition.}
For building the last partition, it is sufficient to simply assign all remaining in-memory edges to the last partition. To this end, NE++ iterates through the adjacency lists of all low-degree vertices that are not yet in the core set (cf. Algorithm~\ref{alg:remaining}). These adjacency lists contain all the remaining in-memory edges. There are two cases which must be treated differently: A remaining in-memory edge can be an edge between two low-degree vertices, or an edge between a low-degree and a high-degree vertex---edges between two high-degree vertices are external-memory edges by definition and are treated with streaming partitioning. 

When assigning the remaining edges between two low-degree vertices, we may approach an edge $(u,v)$ from two different directions (from $u$ or from $v$), but should only assign it once. To solve that issue, we assign edges between two low-degree vertices always from the perspective of the left-hand side vertex of the edge in the original edge list (i.e., from $u$ if the edge is $(u,v)$ in the input edge list). To do so, we need to track the ``direction'' of each edge in the CSR, i.e., whether an edge between vertex $u$ and vertex $v$ is $(u,v)$ (will be assigned from $u$'s side) or $(v,u)$ (will be assigned from $v$'s side). We can do that in the graph construction phase (cf. Section~\ref{sec:graphconstruction}) by building a second index array and dividing the adjacency list of each vertex into a set of out-edges and a subsequent set of in-edges. The first index array contains the reference to the beginning of the out-list of a vertex in the column array, while the second index array refers to the corresponding in-list. 
An alternative solution would be to always assign edges between two low-degree vertices from the perspective of the vertex with the smaller ID. This would save memory (no need for a second index array to separate out-list from in-list), but requires to pass through a vertex's complete adjacency list instead of only the out-list. 

The assignment of edges between a low-degree vertex and a high-degree vertex is straight-forward. As the adjacency list of the high-degree vertex is not present in the column array, the edge can safely be assigned from the low degree vertex's side.

\subsection{\emph{Informed} Stateful Streaming Partitioning}
\label{sec:streaming}

In HEP, we use stateful streaming partitioning and exploit the partitioning state from in-memory partitioning to improve the partitioning quality. Stateful streaming partitioning rates the placement of a given edge $e$ on each partition $p_i$ via a \emph{scoring function} $\sigma : (e,p_i) \to s$. Then, the edge is assigned to the partition that yields the highest score value $s$ (cf. Algorithm~\ref{alg:streaming}). 

In our implementation of HEP, we use the HDRF scoring function~\cite{Petroni:2015:HSP:2806416.2806424}, which takes into account the vertex degrees, the vertex replication on each partition, and the load (in number of edges) of each partition. The vertex degrees and the vertex replication of each partition are used as a result of NE++: A vertex is replicated in partition $p_i$ exactly if it is in $S_i$. This way, we overcome the ``uninformed assignment problem''~\cite{8416335} of stateful streaming partitioning where early edges in the stream are assigned to a partition without any knowledge of the rest of the graph. HDRF is the partitioning algorithm that fits best for the streaming phase of HEP and can be easily integrated without any need for adaptations. Having said this, the streaming phase of HEP could also employ other stateful streaming edge partitioning algorithms, such as Greedy~\cite{powergraph} or ADWISE~\cite{8416335}. However, the Greedy strategy is clearly outperformed by HDRF~\cite{Petroni:2015:HSP:2806416.2806424}. ADWISE invests additional run-time to overcome the uninformed assignment problem which is, however, already tackled in HEP by the in-memory phase via NE++.

\section{Theoretical Analysis}
\label{sec:analysis}

We perform a theoretical analysis of HEP's run-time and memory overhead as well as the obtained replication factor. Finally, we discuss strategies for setting $\tau$ so that a memory bound can be met.

\subsection{Time Complexity}

We analyze the time complexity of the following steps of HEP separately: building the CSR graph representation, in-memory partitioning with NE++, and streaming partitioning. 

\textbf{Graph Building.}
The run-time complexity of graph building is $\mathcal{O}(|E| + |V|)$. There are two passes over the edge list, each having constant complexity per edge. In the first pass, the degree of each vertex is determined; then, the index array is built by performing a pass over the vertices' degrees and computing a running sum. In the second pass, the edges are inserted into the column array or, if an edge is incident to two high-degree vertices, into the external memory edge file. 

\renewcommand{\arraystretch}{1.25}
\begin{table}
{\footnotesize
	\begin{center}
		\begin{tabular}{|l|l|p{3.4cm}|}
			\hline
			Name & Type & Time Complexity \\	
			\hline\hline
			HEP & Hybrid & $\mathcal{O}(|E|*(\log{}|V| + k) + |V|)$  \\ \hline\hline
			Greedy~\cite{powergraph} & Stateful Streaming & $\Theta(|E|*k)$  \\ \hline
			HDRF~\cite{Petroni:2015:HSP:2806416.2806424} & Stateful Streaming & $\Theta(|E|*k)$  \\ \hline
			ADWISE~\cite{8416335} & Stateful Streaming & $\Theta(|E|*k)$  \\ \hline \hline
			DBH~\cite{dbh} & Stateless Streaming & $\Theta(|E|)$  \\ \hline 
			Grid~\cite{grid} & Stateless Streaming & $\Theta(|E|)$  \\ \hline \hline
			NE/NE++ & In-Memory & $\mathcal{O}(|E|*(\log{}|V| + k) + |V|)$ \\ \hline
			DNE~\cite{dne} & In-Memory & $\mathcal{O}(\frac{d *|E| * (k+d)}{n * k})$ with $d = $ maximum vertex degree, $n = $ number of CPU cores \\ \hline
			METIS~\cite{Karypis:1998:FHQ:305219.305248} & In-Memory & $\mathcal{O}((|V|+|E|)*\log{}k)$ \\ \hline
		\end{tabular}
	\end{center}
}
\caption{Time complexity of different partitioners.}
\label{tab:time_complexity}
\vspace{-18pt}
\end{table}

\textbf{NE++.}
To determine the vertex $v\_{\mathit{min}}$ with the minimum external degree, we store vertices in $S_i$ in a binary min-heap, sorted by their external degree. In the worst case, $S_i$ contains all vertices of the entire graph. Every time the external degree of a vertex changes, the min-heap is updated to retain the heap property. Hence, each min-heap update operation has a complexity in $\mathcal{O}(\log{}|V|)$. In total, there will be up to $2*|E|$ updates of the min-heap, resulting in $\mathcal{O}(|E|*\log{}|V|)$ complexity to maintain the min-heap. The other operations of the main expansion algorithm have the following complexity. Assigning each edge to a partition is in $\mathcal{O}(|E|)$. Traversing the neighbors of a vertex in order to update the min-heap is in $\mathcal{O}(|E|*k)$, as each neighborhood is maximally traversed $k$ times (in case a vertex is in every secondary set $S_i$). Initialization has a  complexity of $\mathcal{O}(|V|)$, as each vertex is visited at most once. Finally, the clean-up procedure is executed $k-1$ times, each time iterating through all adjacency lists of all vertices in $S_i$ and performing a constant-time operation on them. Hence, clean-up has a time complexity of $\mathcal{O}(|E|*k)$. The overall time complexity of NE++ is $\mathcal{O}(|E|*(\log{}|V| + k) + |V|)$.

\textbf{Streaming.}
Stateful streaming partitioning computes a scoring function for each edge and each partition; the computation of each score is a constant-time operation. Hence, the streaming phase has a time complexity of $\mathcal{O}(|E|*k)$.

\textbf{Comparison and Discussion.} The overall time complexity of HEP is dominated by the NE++ phase, and, hence, is in $\mathcal{O}(|E|*(\log{}|V| + k) + |V|)$. This is higher than the stateful streaming partitioners Greedy, HDRF and ADWISE. However, the case that each vertex is in every secondary set is highly unlikely, as this would mean that each vertex is replicated on every partition. In contrast to this, the time complexity bounds of Greedy, HDRF and ADWISE are \emph{tight}, as they involve the computation of a scoring function on every combination of edges and partitions.

\subsection{Memory Overhead}
\label{sec:memory_analysis}

Let the ID of a vertex be stored in $b_{\mathit{id}}$ bytes. For instance, for a graph with up to $2^{32}$ (i.e., 4.29 B) vertices, we can store vertex IDs in 4 byte unsigned integers. The following data structures are necessary in order to implement HEP efficiently. 

(1) \emph{Two indices} for the index array to separate incoming and outgoing edges. Both contain one field for each vertex, which totals in $2 * |V| * b_{\mathit{id}}$ bytes.
(2) A \emph{column array} to store the edges. The size of the column array is the sum over the sizes of the adjacency lists of the low-degree vertices: $\sum_{v_i \in V_l} d(v_i) * b_{\mathit{id}}$ bytes.
(3) \emph{Size fields} for each in and out adjacency list to indicate the number of valid entries (this is used for efficient edge removal, as discussed above). Each size field occupies $b_{\mathit{id}}$ bytes, totalling in $|V| * 2 * b_{\mathit{id}}$ bytes.
(4) $k$ \emph{dense bitsets} to track the vertices in the secondary set $S_i$ for each partition $p_i$ and one dense bitset to track the vertices in the core set $C$. A dense bit set works as follows: For each vertex id, the bit with the corresponding index in the bit set signals whether the vertex is part of the set. Overall, these data structures occupy $|V| * \frac{k+1}{8}$ bytes.
(5) A \emph{min heap} to store the external degrees of vertices in $S_i$ and a \emph{lookup table} to directly access the entry of a vertex in the min heap by its ID. Together, they require $2 * |V| * b_{\mathit{id}}$ bytes.

The total size of all data structures is: \\
\centerline{$\sum_{v_i \in V_l} d(v_i) * b_{\mathit{id}} + 6 * |V| * b_{\mathit{id}} + |V| * \frac{k+1}{8}$ bytes.}


\subsection{Bounds on Replication Factor}
\label{sec:rep_factor}

The worst-case replication factor obtained with HEP is between the bounds of NE~\cite{Zhang:2017:GEP:3097983.3098033} and of HDRF~\cite{Petroni:2015:HSP:2806416.2806424}. NE has a better bound on replication factor than HDRF for power-law graphs~\cite{Zhang:2017:GEP:3097983.3098033}, i.e., a higher setting of $\tau$ leads to a better replication factor as more edges are partitioned with NE++.

\subsection{Setting of $\boldsymbol{\tau}$ to Keep Memory Bounds}
\label{sec:tau_setting}

A unique characteristic of hybrid partitioning with HEP is its ability to flexibly reduce memory overhead by setting $\tau$. The lower $\tau$ is set, the more vertices are considered high-degree and, thus, the column array---which is the dominant data structure---is pruned more aggressively. Hence, $\tau$ can be set with respect to the memory capacities of the compute node. To do so, one can perform a pre-computation step and build the cumulative sum of the size of the adjacency lists of the respective low-degree vertices for different values of $\tau$; this is a trivially parallelizable process. Then, one chooses the maximal value of $\tau$ that keeps the memory bound (e.g., memory capacity of the compute node). Once $\tau$ is determined, graph partitioning with HEP is performed.

We measured the run-time to determine the memory requirements for a specific setting of $\tau$ for different graphs (Table~\ref{tab:time_memory}). Compared to graph partitioning run-time (Section~\ref{ref:Graph Partitioning}), this is negligible, so that pre-computing an optimal value of $\tau$ is feasible and practical.

\renewcommand{\arraystretch}{1.25}
\begin{table}
{\footnotesize
	\begin{center}
		\begin{tabular}{|ll|ll|ll|ll|}
			\hline
			OK & 1 s & IT & 7 s & TW & 41 s  & FR & 45 s   \\ \hline\hline
			UK & 24 s & GSH & \multicolumn{2}{l|}{260 s}  & \multicolumn{3}{l|}{WDC \quad 868 s}     \\ \hline
		\end{tabular}
	\end{center}
}
\caption{Run-time to pre-compute memory footprint.}
\label{tab:time_memory}
\vspace{-15pt}
\end{table}

\section{Evaluation}
\label{sec:evaluation}
\subsection{Experimental Setup}

\textbf{Evaluation Platform.} 
We perform all experiments on a server with 4 x 16 Intel(R) Xeon(R) CPU E5-4650~@~2.70GHz, about 500 GiB of main memory and an HDD disk. Using a large server allows us to evaluate many baseline partitioners even for the larger graphs. We use Ubuntu 18.04.2 LTS as operating system.

\renewcommand{\arraystretch}{1.15}
\begin{table}
{\scriptsize
	\begin{center}
		\begin{tabular}{l|l|l|l|l}
			\hline
			Name & \textbf{$|V|$} & \textbf{$|E|$} & Size & Type \\	
			\hline
			com-livejournal (LJ) & 4.0 M & 35 M & 265 MiB & Social \\
			com-orkut (OK) & 3.1 M & 117 M & 895 MiB & Social \\
			brain (BR) & 784 k & 268 M & 2 GiB & Biological \\
			wiki-links (WI) & 12 M & 378 M & 3 GiB & Web \\
			it-2004 (IT) & 41 M & 1.2 B & 9 GiB & Web \\
			twitter-2010 (TW) & 42 M & 1.5 B & 11 GiB & Social \\
			com-friendster (FR) & 66 M & 1.8 B & 14 GiB & Social \\
			uk-2007-05 (UK) & 106 M & 3.7 B & 28 GiB & Web \\
			gsh-2015 (GSH) & 988 M & 33 B & 248 GiB & Web \\
			wdc-2014 (WDC) & 1.7 B & 64 B & 478 GiB & Web \\	
			\hline
		\end{tabular}
	\end{center}
}
\caption{Real-world graphs. \textit{Size} refers to binary edge lists with 32-bit vertex ids. }
\label{tab:graphs}
\vspace{-10pt}
\end{table}

\textbf{Real-World Graph Datasets.} We use 7 different graphs (cf. Table~\ref{tab:graphs}) that have different sizes and stem from different web repositories, having been crawled by different organizations. These are social network graphs (OK~\cite{orkut, snapnets, 6413740}, TW~\cite{twitter,Kwak:2010:TSN:1772690.1772751, snapnets} and FR~\cite{friendster, 6413740, snapnets}) as well as web graphs (IT~\cite{it, BMSB, BRSLLP, BoVWFI},  UK~\cite{uk, BMSB, BRSLLP, BoVWFI}, GSH~\cite{gsh, BMSB, BRSLLP, BoVWFI} and WDC~\cite{wdc}). The reasons that we chose these graphs are as follows: First, many of them are commonly used as baselines in related papers~\cite{8416335, Zhang:2017:GEP:3097983.3098033, dne}. Second, most of them are very large (billions of edges), and hence, are particularly challenging in terms of memory overheads even on a large machine. Indeed, we show that not all partitioners can handle graphs of that size on our evaluation platform. Third, the graphs have a large diversity in terms of how ``easy'' they are to partition; for some of them, even the best partitioners reach only relatively high replication factors.

\textbf{Baselines.} We compare HEP to 7 of the most recent and best streaming and in-memory partitioners. From the group of streaming partitioners, we compare to ADWISE~\cite{8416335}, DBH~\cite{dbh}, HDRF~\cite{Petroni:2015:HSP:2806416.2806424}, and SNE~\cite{Zhang:2017:GEP:3097983.3098033}. From the group of in-memory partitioners, we compare to NE~\cite{Zhang:2017:GEP:3097983.3098033}, DNE~\cite{dne} and METIS~\cite{Karypis:1998:FHQ:305219.305248}. There exist dozens of other partitioners, and we cannot compare to all of them. However, other partitioners are consistently outperformed by the chosen baselines either in terms of replication factor, run-time, scalability or memory efficiency. Hence, we argue that we made a reasonable and fair choice of baselines. See Section~\ref{sec:related} for a comprehensive discussion of other partitioners.

\begin{figure*}
	\centering	
	\captionsetup[subfloat]{captionskip=-2pt}
	\subfloat[OK: Replication factor.]{\includegraphics[width=0.31\textwidth]{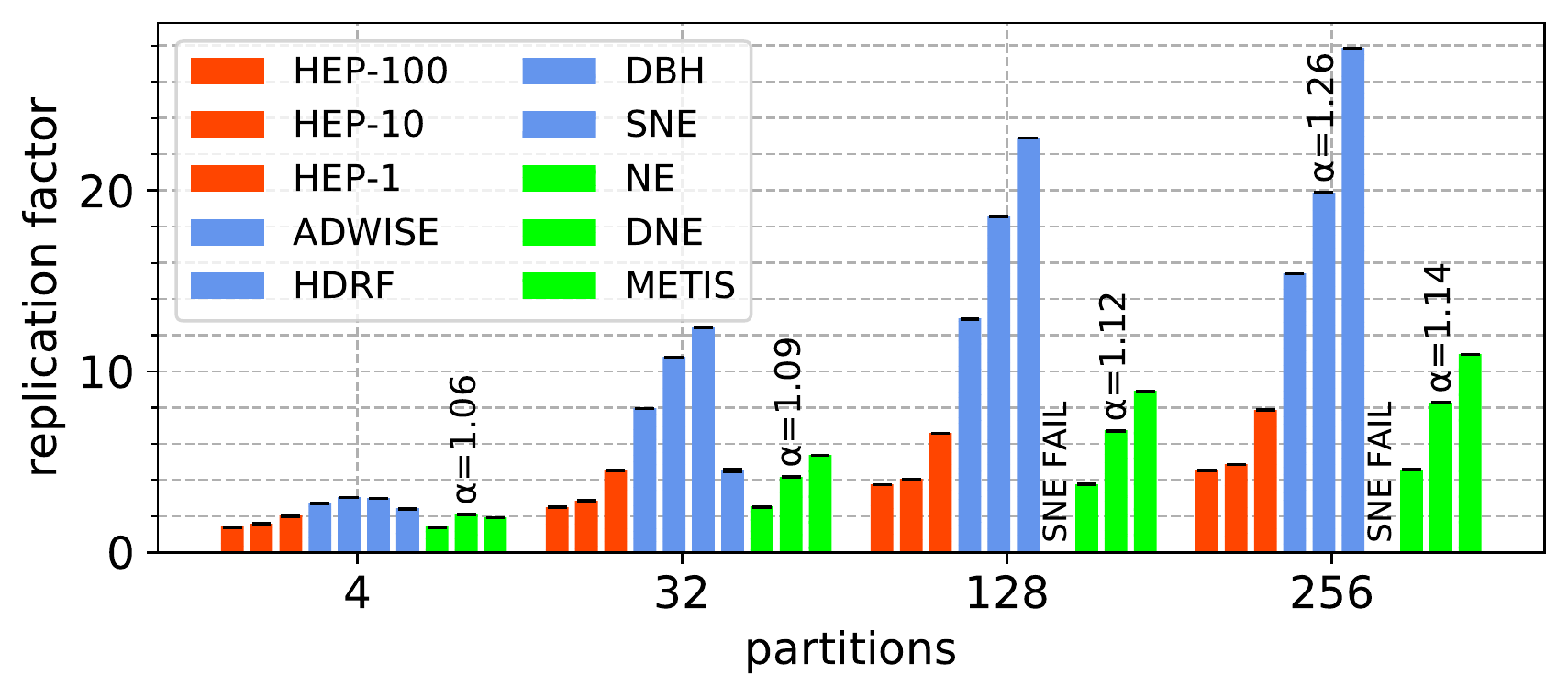}}
	\subfloat[OK: Run-time (logscale).]{\label{b}   \includegraphics[width=0.31\textwidth]{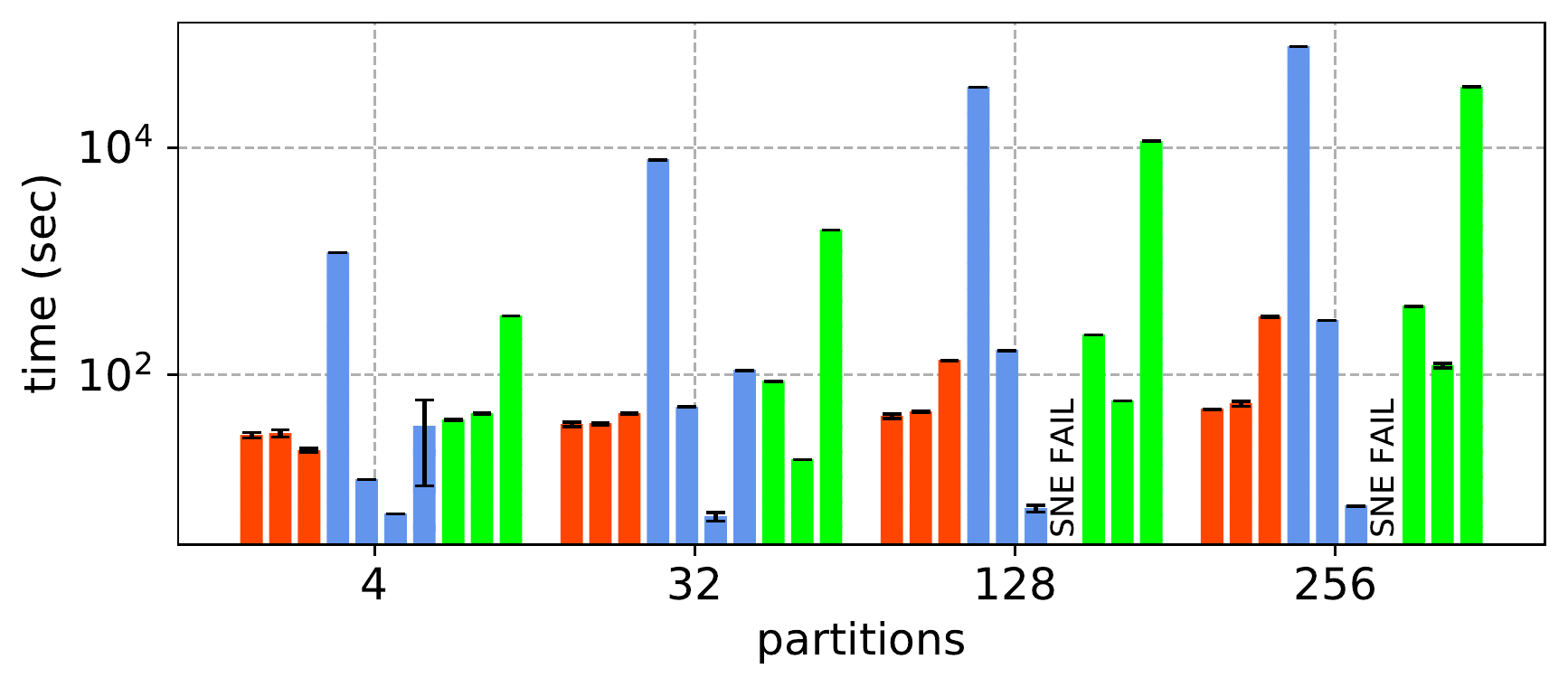}} 
	\subfloat[OK: Memory overhead (logscale).]{\label{c}   \includegraphics[width=0.31\textwidth]{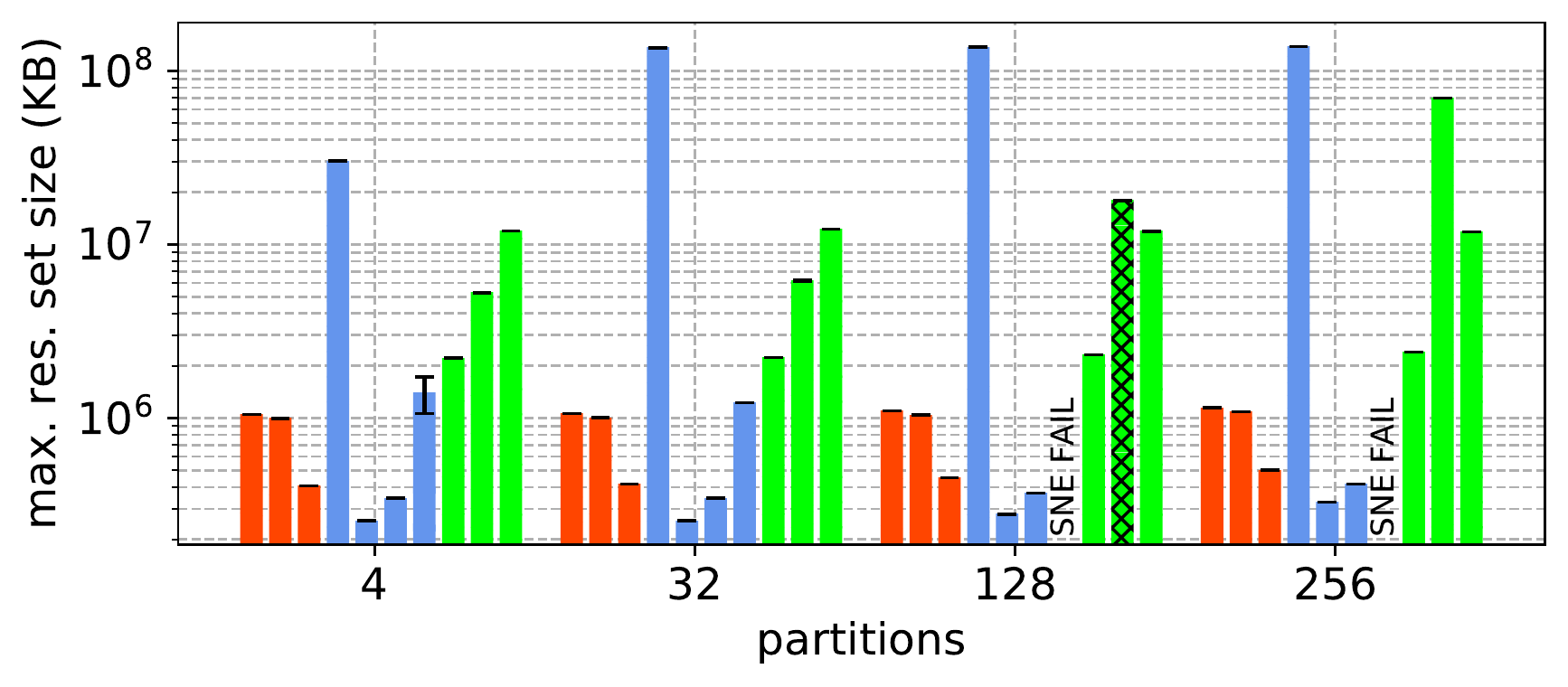}}\\
	\vspace{-0.35cm}
	\subfloat[IT: Replication factor.]{\includegraphics[width=0.31\textwidth]{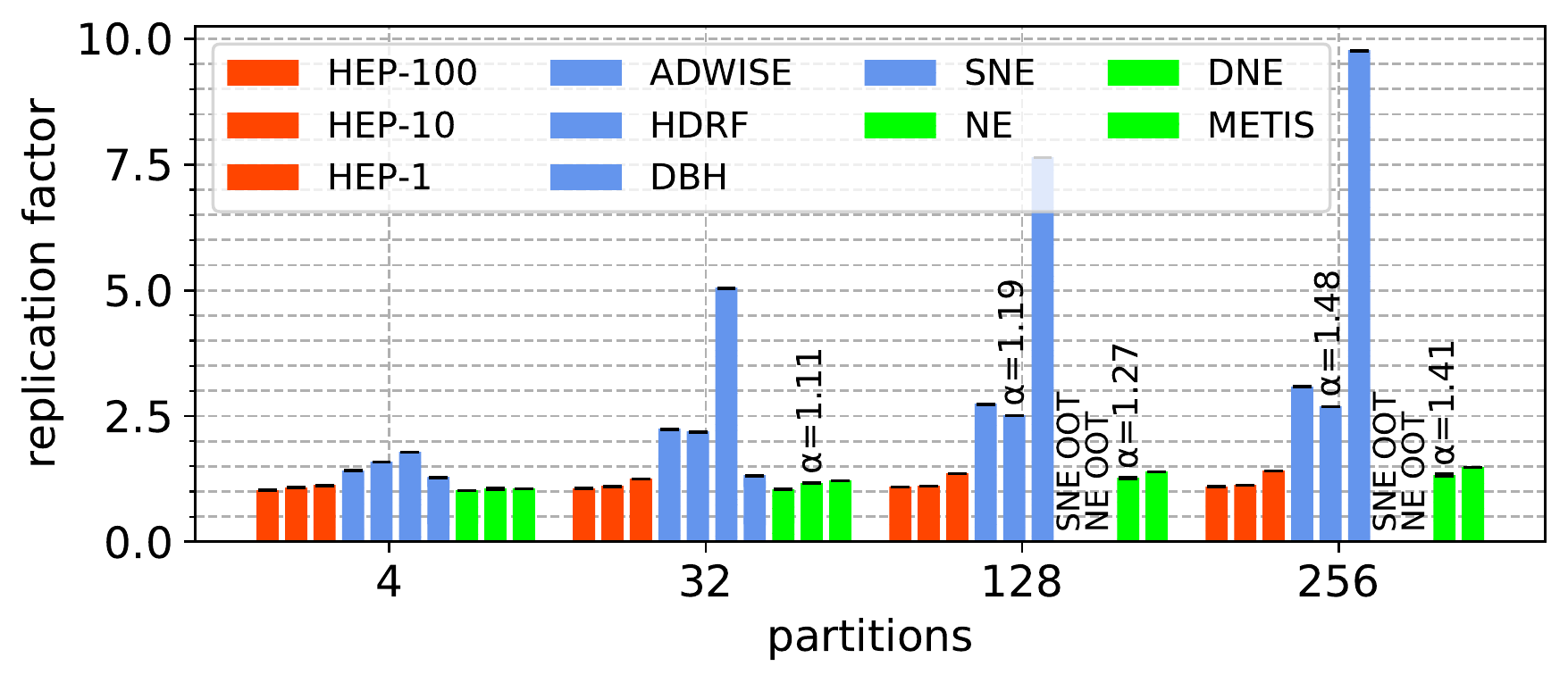}}
	\subfloat[IT: Run-time (logscale).]{\label{b}   \includegraphics[width=0.31\textwidth]{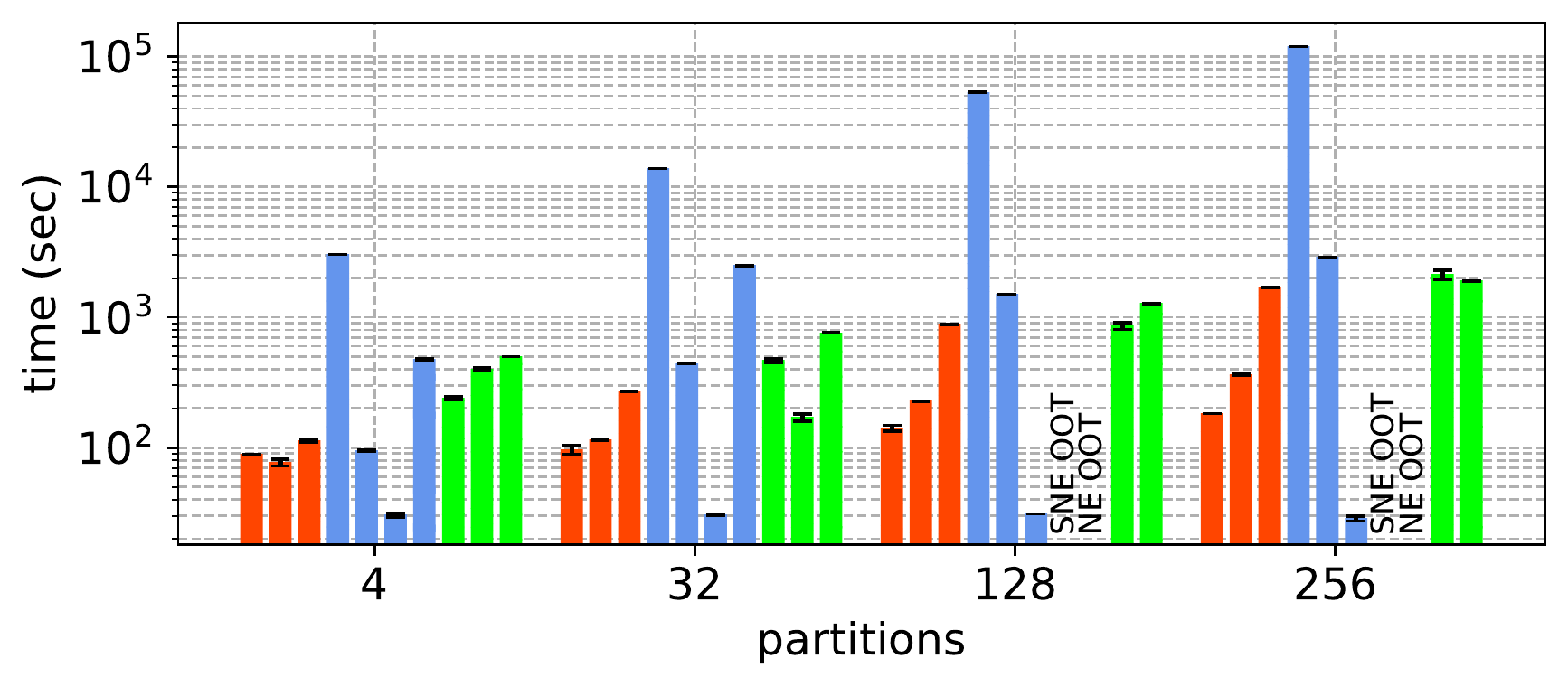}} 
	\subfloat[IT: Memory overhead (logscale).]{\label{c}   \includegraphics[width=0.31\textwidth]{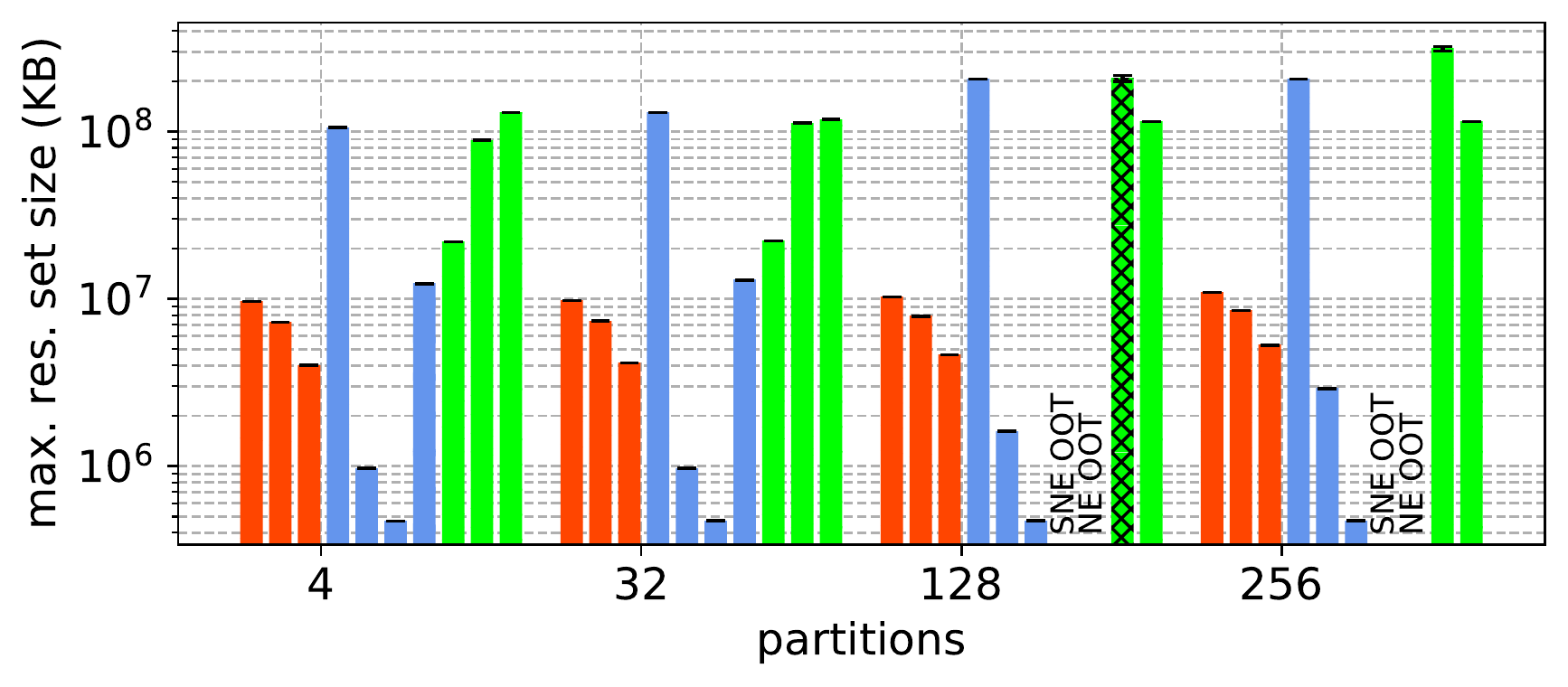}}\\
	\vspace{-0.35cm}
	\subfloat[TW: Replication factor.]{\label{a}   \includegraphics[width=0.31\textwidth]{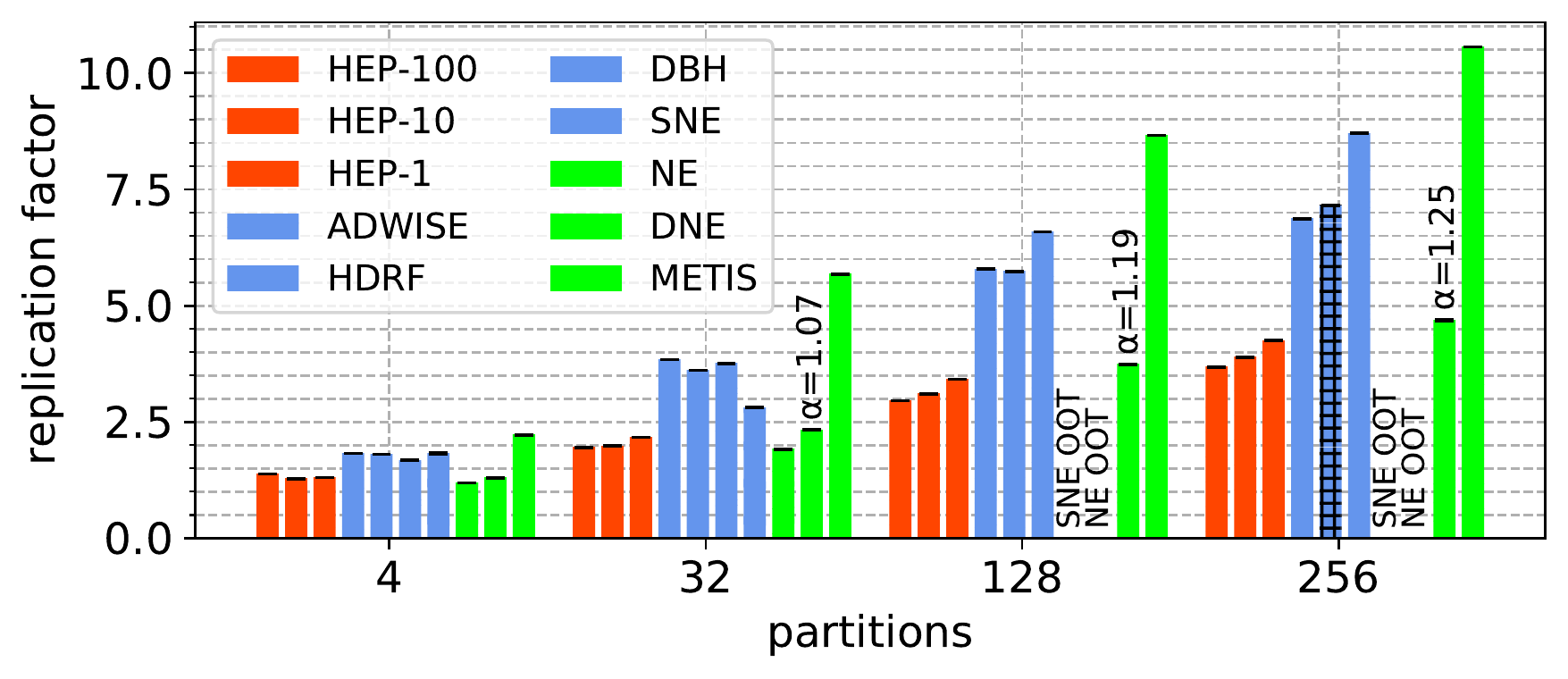}}
	\subfloat[TW: Run-time (logscale).]{\label{b}   \includegraphics[width=0.31\textwidth]{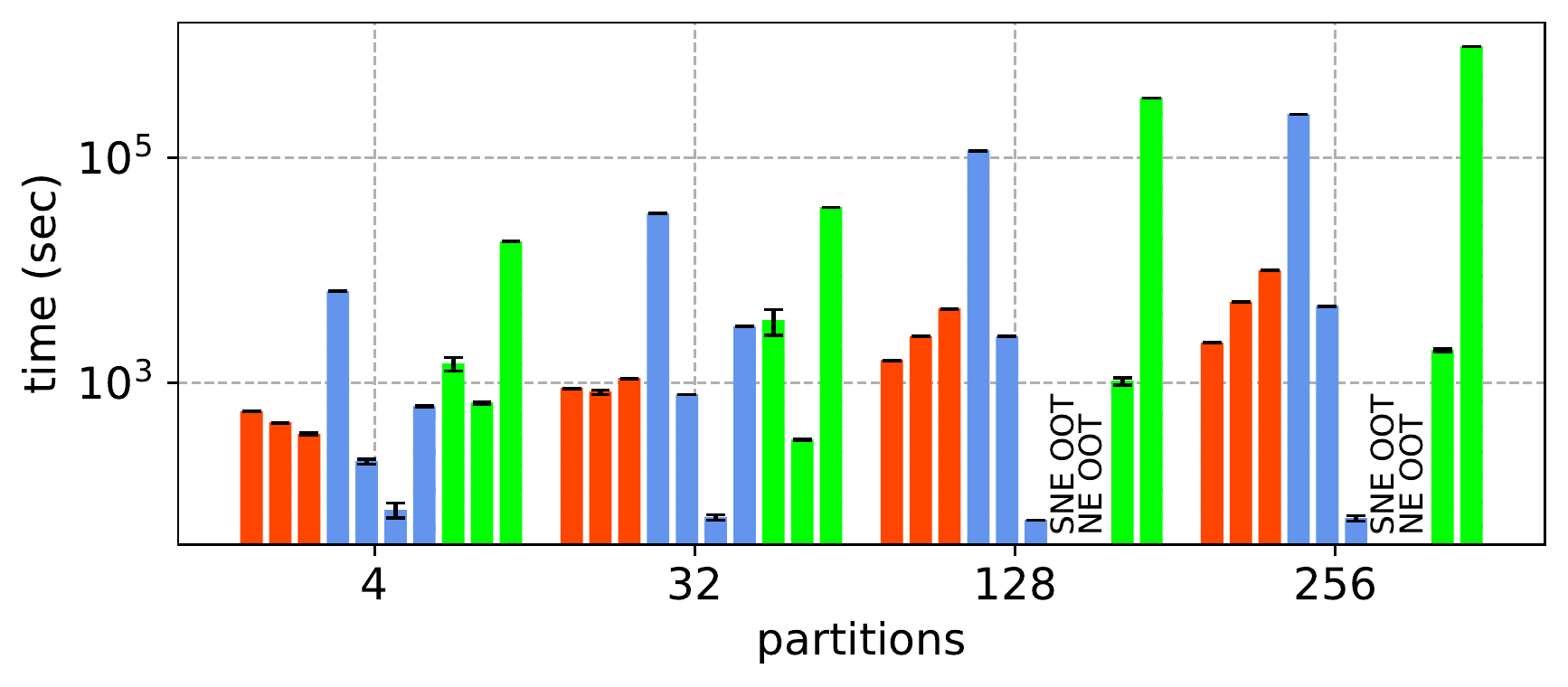}}
	\subfloat[TW: Memory overhead (logscale).]{\label{c}   \includegraphics[width=0.31\textwidth]{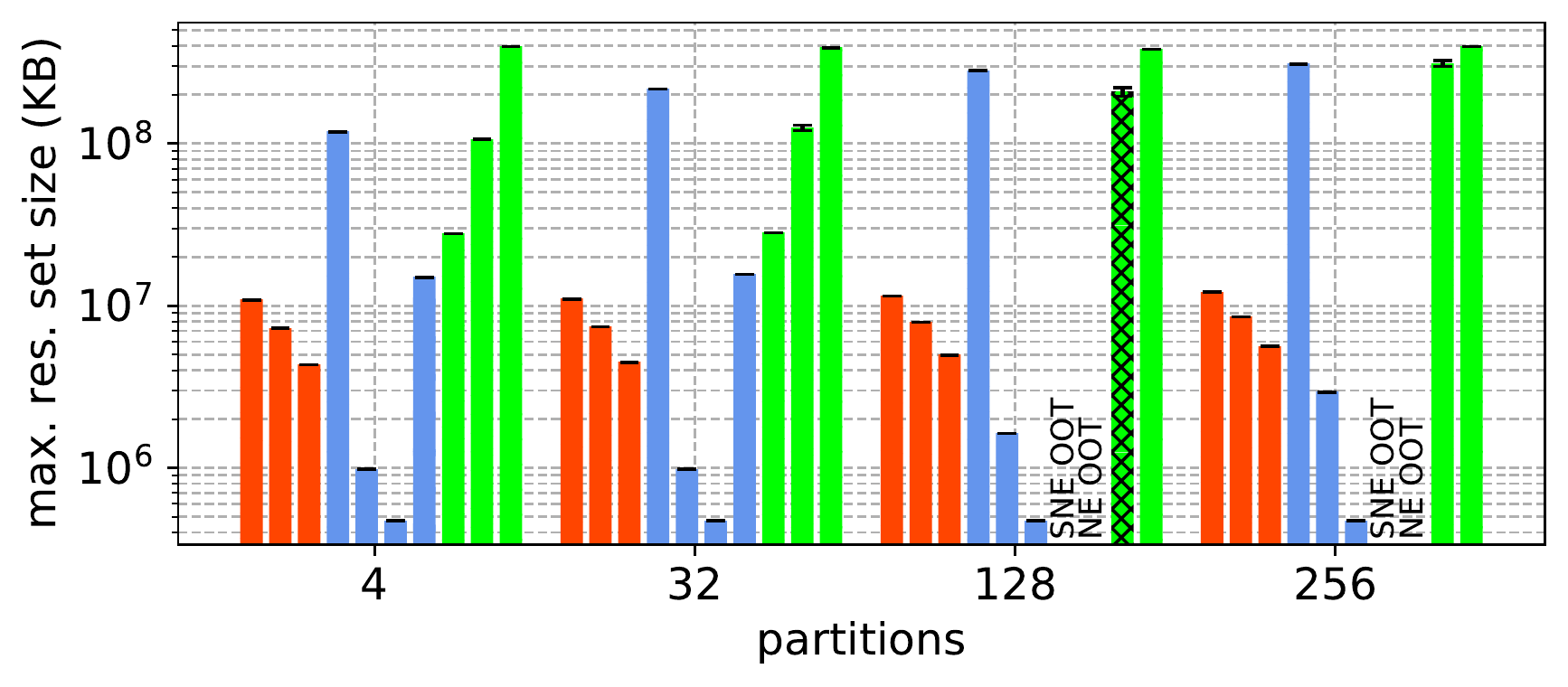}} \\
	\vspace{-0.35cm}
	\subfloat[FR: Replication factor.]{\label{a}   \includegraphics[width=0.31\textwidth]{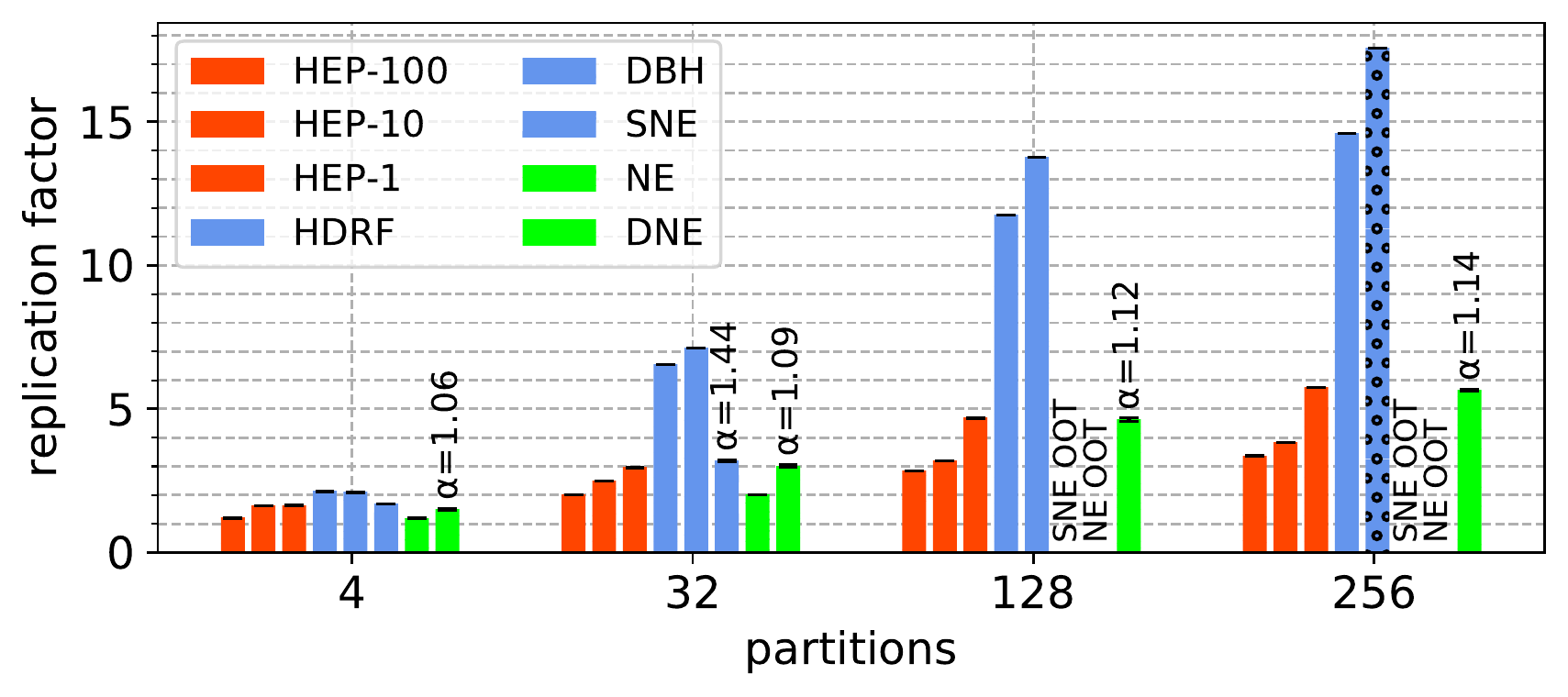}}
	\subfloat[FR: Run-time (logscale).]{\label{b}   \includegraphics[width=0.31\textwidth]{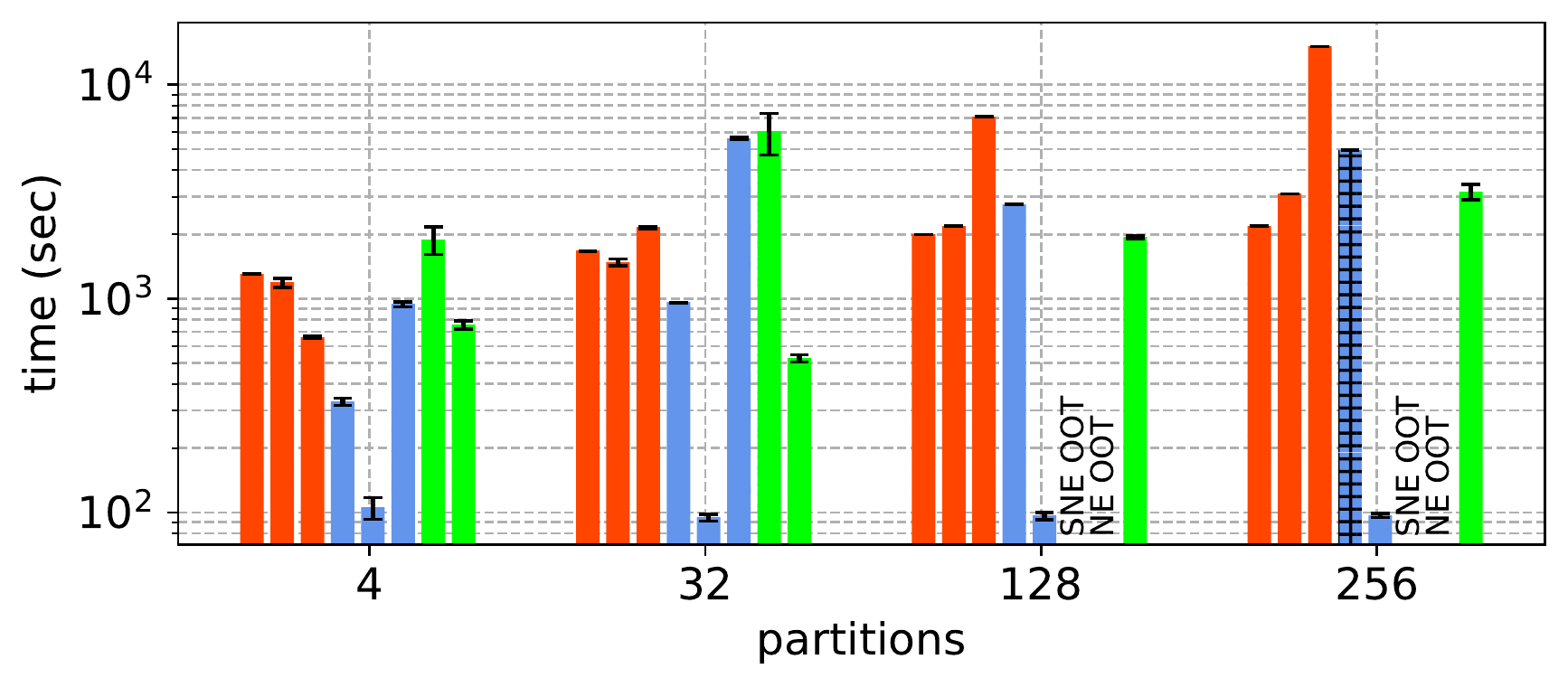}}
	\subfloat[FR: Memory overhead (logscale).]{\label{c}   \includegraphics[width=0.31\textwidth]{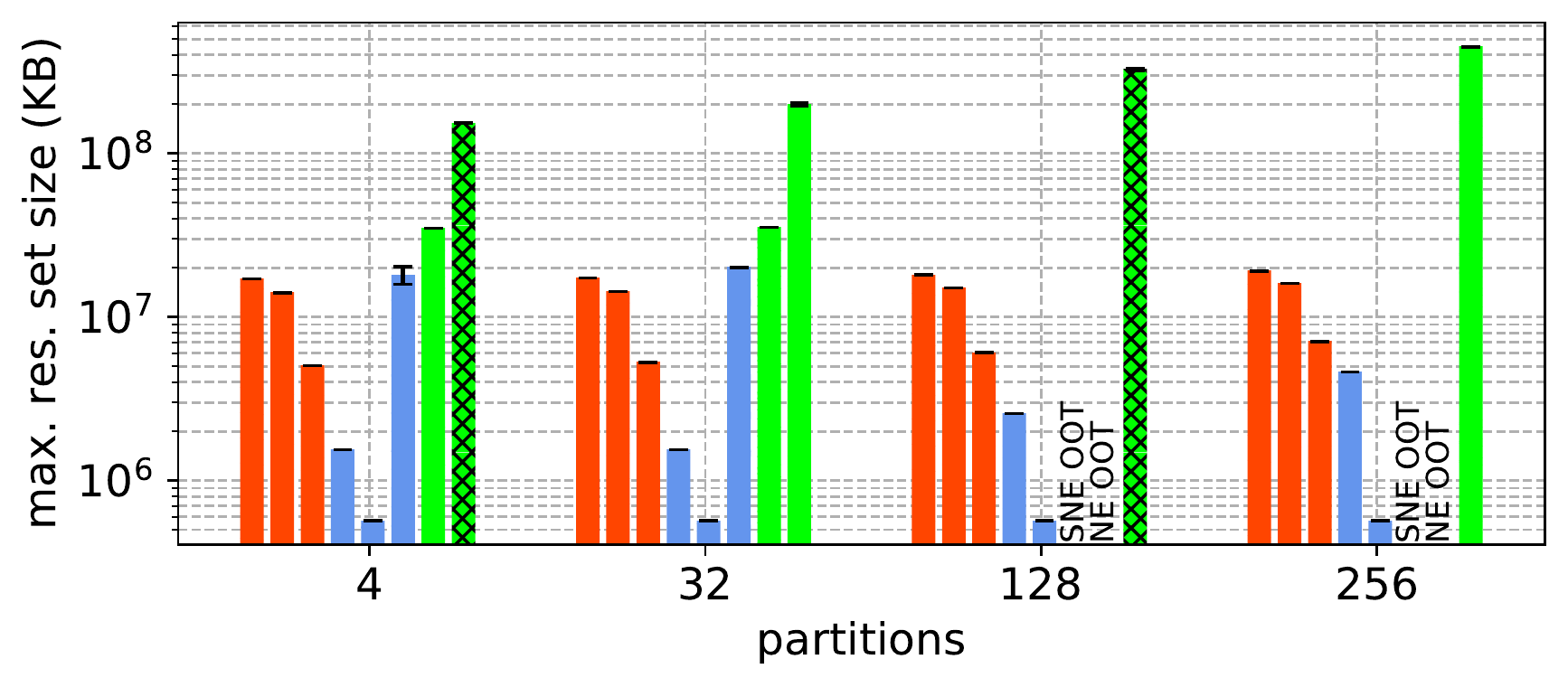}} \\
	\vspace{-0.35cm}
		\subfloat[UK: Replication factor.]{\label{a}   \includegraphics[width=0.31\textwidth]{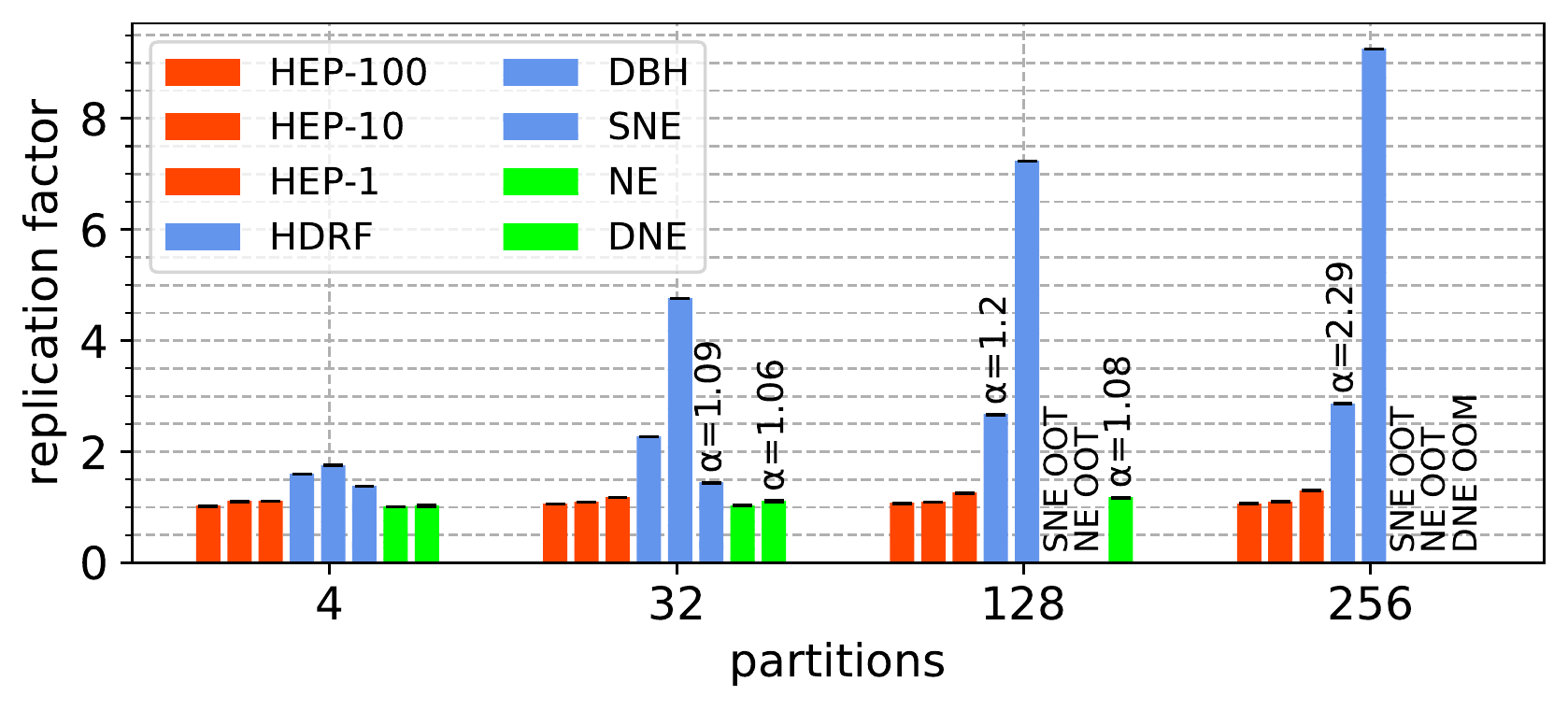}}
	\subfloat[UK: Run-time (logscale).]{\label{b}   \includegraphics[width=0.31\textwidth]{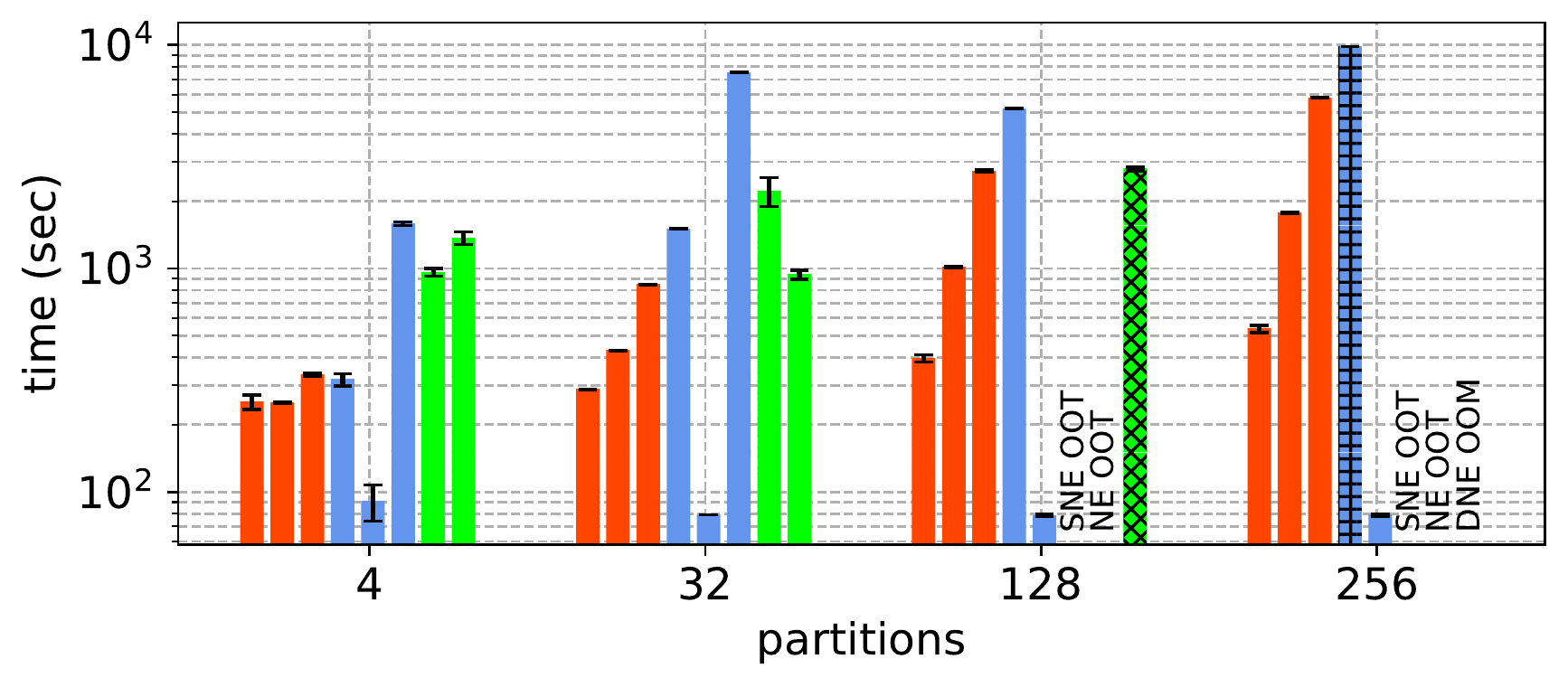}}
	\subfloat[UK: Memory overhead (logscale).]{\label{c}   \includegraphics[width=0.31\textwidth]{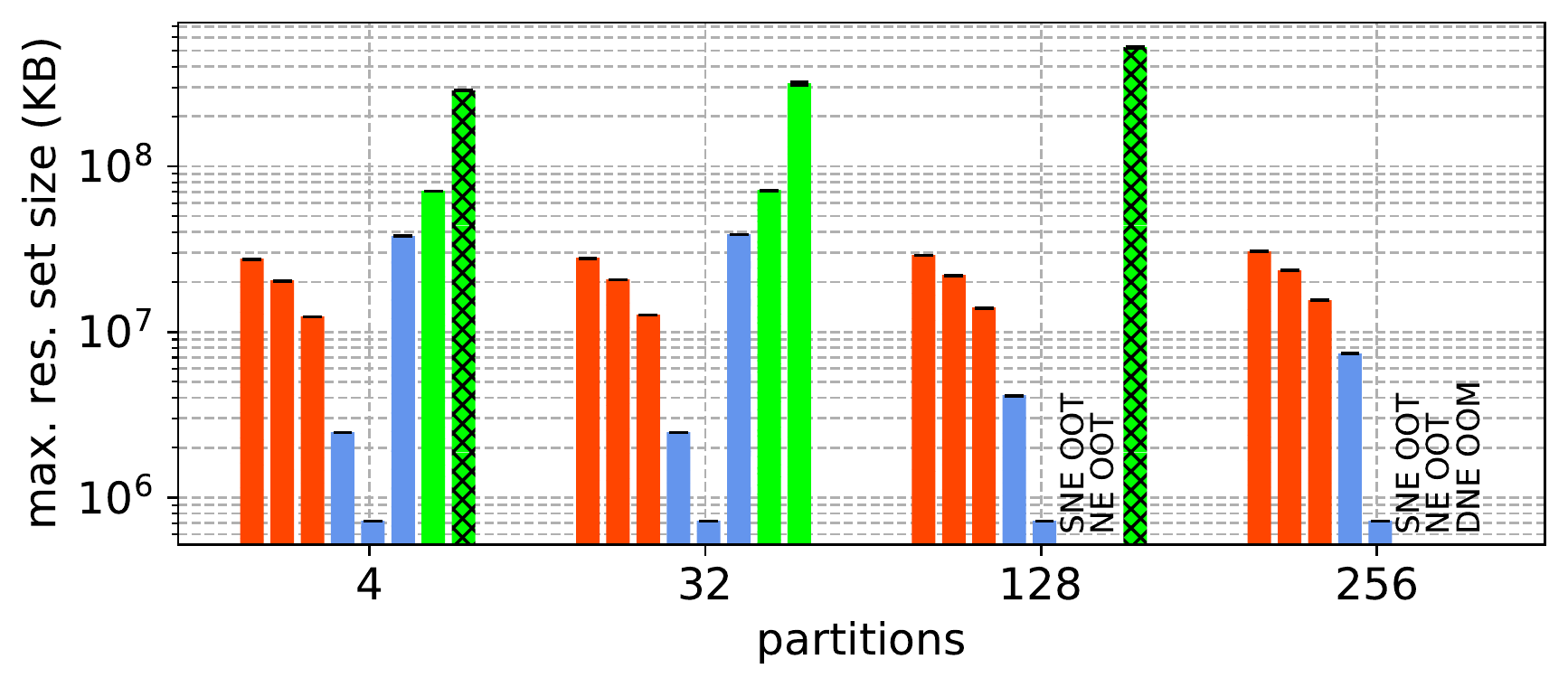}} \\
	\vspace{-0.35cm}
		\subfloat[GSH: Replication factor.]{\label{a}   \includegraphics[width=0.165\textwidth]{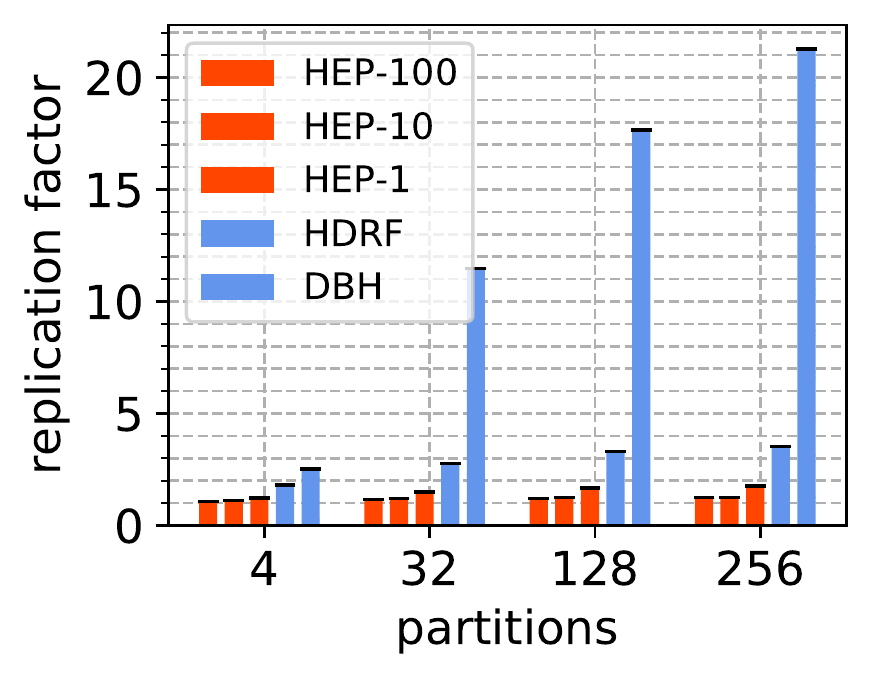}}
	\subfloat[GSH: Run-time (log).]{\label{b}   \includegraphics[width=0.165\textwidth]{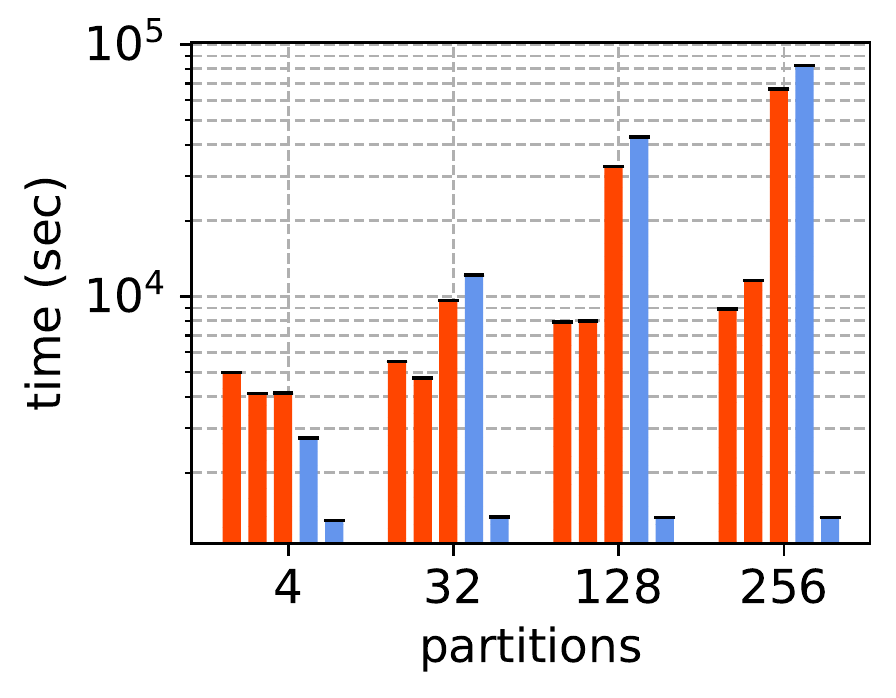}}
	\subfloat[GSH: Mem. overh. (log).]{\label{c}   \includegraphics[width=0.165\textwidth]{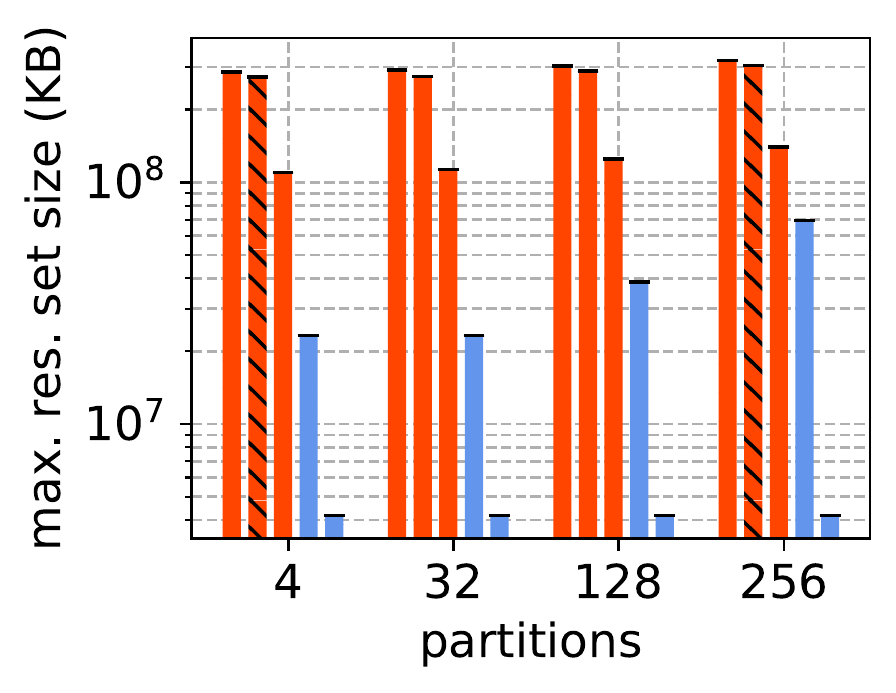}} 
		\subfloat[WDC: Replication factor.]{\label{a}   \includegraphics[width=0.165\textwidth]{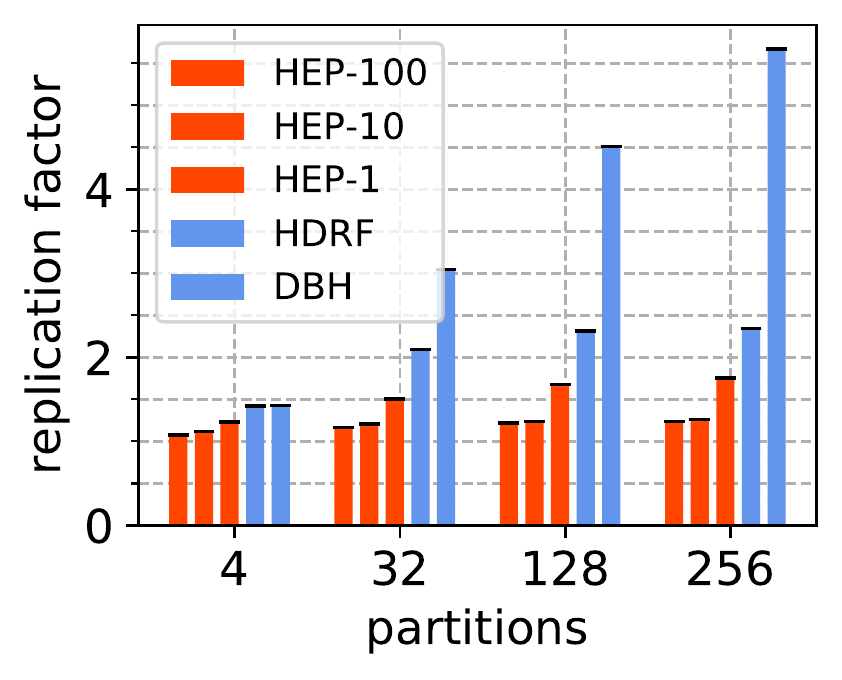}}
	\subfloat[WDC: Run-time (log).]{\label{b}   \includegraphics[width=0.165\textwidth]{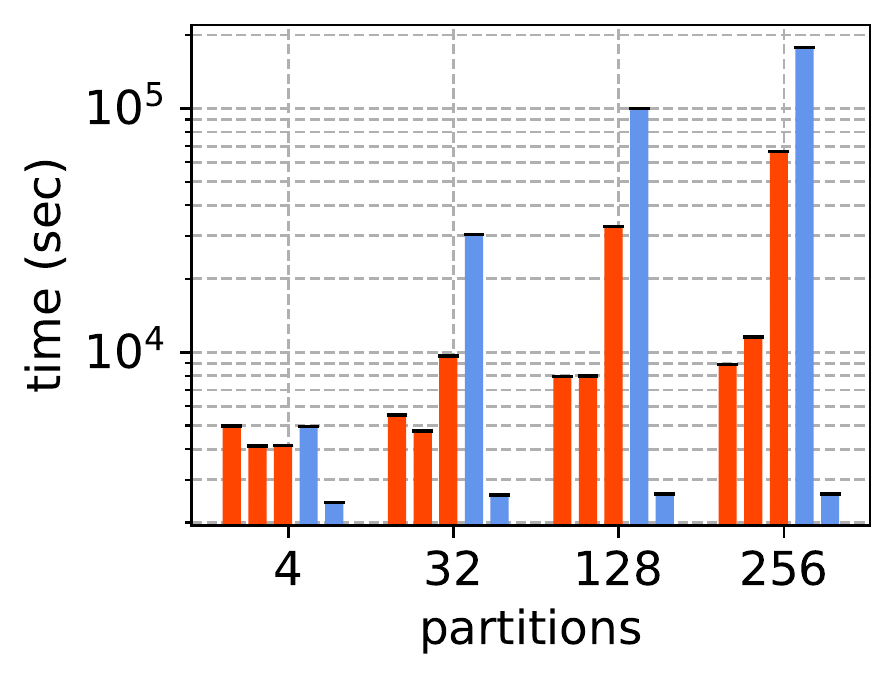}}
	\subfloat[WDC: Mem. overh. (log).]{\label{c}   \includegraphics[width=0.165\textwidth]{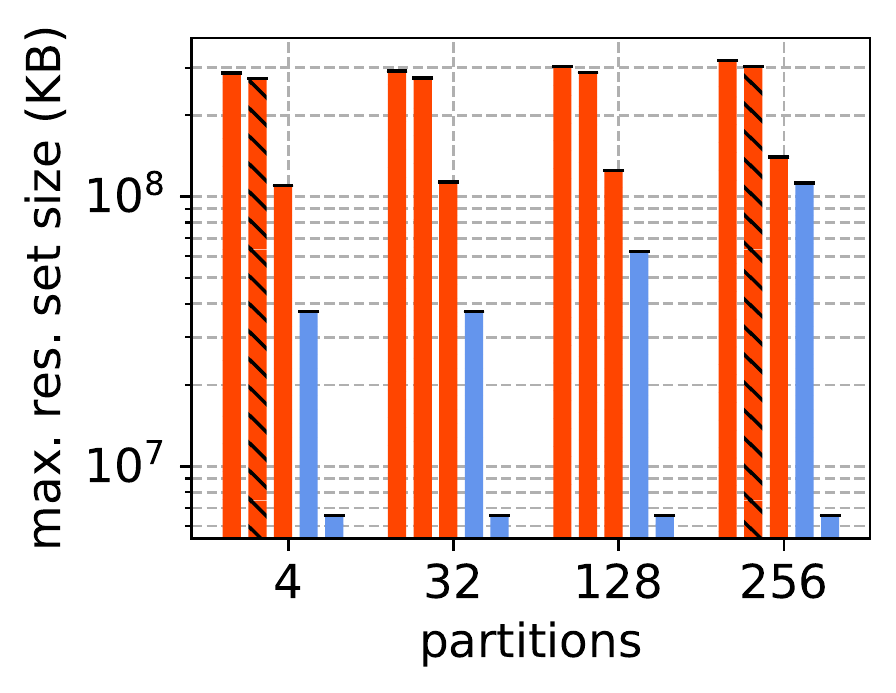}}
	\vspace{-5pt}
	\caption{Performance results on real-world graphs. HEP-$x$ indicates that HEP was run with a setting of $\tau = x$. ``OOM'': out-of-memory error, ``OOT'': excessive run-time, ``FAIL'': crashed for other reasons.}
	\label{eval:perf}
	\vspace{-10pt}
\end{figure*}

\textbf{Performance Metrics.}
We measure replication factor, run-time (including graph ingestion) and memory overhead (maximum resident set size). We also track edge balancing: Most partitioners kept the partitions balanced; in case partitioning was imbalanced, we report the balancing factor $\alpha$ in the plots.

\textbf{Experiments.} 
We perform systematic experiments with $k=\{4,32,128,256\}$ partitions. We configured HEP with the following values of $\tau$: 100, 10 and 1. For each graph, we first perform one warm-up run that does not count into the measurements (e.g., to warm up caches). We aborted experiments on smaller graphs (<~10~B edges) when they took more than 12 hours, as some partitioners (NE, SNE and METIS) would ``freeze'' in an unknown state for many hours. On the smaller graphs, we repeat every experiment 3 times and report the mean values with standard deviations; for GSH and WDC, we run experiments only once due to very large run-times. In Appendix~A, we discuss configurations and implementation of the baseline partitioners as well as the input formats of graph data.

\renewcommand{\arraystretch}{1.04}
\begin{table*}[h]
\parbox{.75\textwidth}{
{\footnotesize
	\begin{center}
		\begin{tabular}{|p{0.9cm}||l|l|l||l|l|l||l|l|l||l|l|l||l|l|l|}
			\hline
			\emph{Algor. /} &  \multicolumn{3}{c||}{Partitioning Time}  &  \multicolumn{3}{c||}{Rep. Factor} & \multicolumn{3}{c||}{PageRank}  &  \multicolumn{3}{c||}{BFS}   &  \multicolumn{3}{c|}{Conn. Comp.}     \\	
			\emph{Graph} & OK & IT & TW & OK & IT & TW & OK & IT & TW  & OK & IT & TW & OK & IT  & TW  \\	
			\hline \hline
			HEP-100 & 38 & 101 & 885 & 2.51 & 1.06 & 1.95 & \underline{122} & 628 & \textbf{1,239} &  \underline{\textbf{489}} & 2,675 & \textbf{10,396} & 38 & 244 & \textbf{382}  \\ \hline
			HEP-10 & 37 & 114 & 779 & 2.86 &  1.10 & 1.99 & 127 & 	\underline{570} & 1,242 & 503 & \underline{\textbf{2,508}} & 10,544 &  38 & 243 & \textbf{382} \\ \hline
			HEP-1 & 45 & 272 & 1,091 & 4.52  & 1.25 & 2.17 &144 & 	  \textbf{538} & 1,495 & 589 & 2,521  & 11,246  & 40  & \textbf{236} & 400 \\ \hline\hline
			NE & 88 & 467 & 3,553 &  2.50     & 1.04 & 1.92 & 	\textbf{117}  &	702 & 1,263 & 498 & 2,732 & 10,999  & \textbf{36} & 250 & 388 \\ \hline
			SNE & 110 & 2,488 & 3,149  &  4.57 & 1.31 & 2.80 & 148  &		729 & 1,608  & 572 & 2,732 & 12,083 & 45 & 307 & 458 \\ \hline
			HDRF & 52 & 441 & 758 & 10.78 & 2.18 & 3.61 & 159 & 			617 & 1,440 & 585 & 2,815  & 11,953 & 42 & 246 & 433 \\ \hline
			DBH & 6 & 31 &  63   & 12.41 & 5.04 & 3.76  & 184  & 			932 & \underline{1,381} & 633 & 3,342 & \underline{11,187} & \underline{45} & \underline{279} &  \underline{415} \\ \hline
		\end{tabular} 
	\end{center}
}
\caption{Processing time of distributed graph algorithms and partitioning time (sec). Lowest total time (partitioning + processing) is \underline{underlined}; lowest processing time is printed in \textbf{bold}.}
\label{tab:processing}
}\hfill
\parbox{.22\textwidth}{
{\footnotesize
	\begin{center}
	 \vspace{1.0cm}
		\begin{tabular}{|l|l|l|l|}
			\hline
			 & OK& IT & TW  \\	
			\hline
			HEP-100 & 0.184 & 0.425 & 0.320 \\
			HEP-10  & 0.168 & 0.376 & 0.222 \\
			HEP-1 & 0.124 & 0.196 & 0.216 \\	
			\hline
		\end{tabular}
	\end{center}
}
\caption{HEP: Vertex balancing (std. deviation / average number of vertex replicas per partition) at $k=32$.}
\label{tab:vertex_balance}
}
\vspace{-20pt}
\end{table*}

\subsection{Graph Partitioning}
\label{ref:Graph Partitioning}

Figure~\ref{eval:perf} shows all results. The main observations are as follows.

(1) HEP has a lower memory overhead than the in-memory partitioners NE, DNE and METIS (see Figure~\ref{eval:perf}(c, f, i, l, o)). Regardless of the setting of $\tau$, memory overheads of the in-memory partitioners are higher by up to an order of magnitude. At the same time, HEP can reach competitive replication factors to NE (especially, at settings of $\tau \geq 10$), which is the partitioner with the best replication factor throughout all experiments (cf. Figure~\ref{eval:perf}(a, d, g, i, m)). 

A major cause for the improvements compared to NE is the lazy edge removal method, which allowed us to remove an auxiliary data structure in NE, namely, an unsorted edge list, that keeps track of whether \emph{every edge} is still valid or not. Furthermore, the run-time of HEP is better, as we use more efficient data structures, namely, the CSR representation. Different from HEP, the unsorted edge list that NE internally employs causes a lot of random access and cache misses while performing the neighborhood expansion steps.

DNE is the state-of-the-art in scalable edge partitioning, being able to scale across large clusters of machines. However, this comes with a cost: DNE requires an order of magnitude more memory than HEP. Further, if the resources used for partitioning are limited (in our case to 64 hardware threads), the run-time of HEP is often competitive to DNE; in some cases, even lower. Furthermore, the distributed and parallel nature of DNE leads to a degradation of the yielded replication factors. HEP achieves better replication factors than DNE in all evaluated settings of $\tau$ on all graphs. Finally, DNE could not always keep the partitions balanced, while HEP achieved perfect balancing in all cases.

METIS is often considered the ``gold standard'' in graph partitioning quality and is commonly used as a baseline~\cite{Zhang:2017:GEP:3097983.3098033, Margo:2015:SDG:2824032.2824046}. While the replication factor obtained by METIS is relatively good in some cases (especially on the IT graph), METIS takes a lot of time, as it follows a multi-level partitioning approach with coarsening and refinement phases. HEP is faster, uses less memory, and achieves better replication factors than METIS. In some instances, the advantage in run-time of HEP is more than an order of magnitude, e.g., on the TW graph with $k=32$, it is 15 minutes for HEP-100 and 10 hours for METIS (Figure~\ref{eval:perf}(h)), while the replication factor of HEP-100 is 1.99 vs. METIS with 5.68 (Figure~\ref{eval:perf}(g)).

(2) Compared to streaming partitioners (ADWISE, HDRF, DBH, and SNE), HEP achieves much better replication factors. At the same time, HEP's memory overhead can be reduced such that it is close to streaming partitioning. By setting a lower value of $\tau$, on all graphs and all number of partitions, we observe a significant reduction in memory overhead of HEP. At $\tau = 1$, HEP's memory overhead comes close to streaming partitioning (Figure~\ref{eval:perf}(c, f, i, l, o, r, u)). This way, memory bounds on smaller machines can be kept. However, HEP achieves  a better replication factor than streaming partitioners (Figure~\ref{eval:perf}(a, d, g, j, m, p, s)). 

\textbf{Summary:} HEP combines the best of both worlds by integrating in-memory with streaming partitioning. This brings the following benefits to the end user: (1) Flexibility to adapt memory overheads to the graphs at hand and the existing hardware capabilities without needing to switch between partitioners. Instead of having to install or even integrate into another system many different partitioners, one can simply employ HEP and use the tuning knob (the parameter $\tau$) to fit the graph into memory. (2) In scenarios where in-memory partitioning is not feasible because the graph is too large, HEP yields better replication factors than streaming partitioners, especially on social network graphs. (3) In scenarios where in-memory partitioning would theoretically be feasible, using HEP with a high setting of $\tau$ yields a good replication factor and is much faster than the previous state-of-the-art partitioner NE.

\subsection{Distributed Graph Processing}

We evaluate the run-time of distributed graph processing under different graph partitionings. To this end, we set up a distributed Spark v3.0.0 cluster with 32 machines, each having 8 CPU cores and 20 GiB of RAM. They are connected by 10-GBit Ethernet. We perform a set of experiments, using three algorithms: PageRank (100 iterations), \underline{B}readth \underline{F}irst \underline{S}earch (subsequently from 10 different random seed vertices) and \underline{C}onnected \underline{C}omponents. PageRank is an algorithm that keeps all vertices active in every iteration; hence, it is very communication-intensive. BFS starts with all vertices but one seed vertex being inactive; from that seed vertex, the computation spreads through the graph. Finally, CC starts with all vertices being active, but in every iteration, more and more of the vertices become inactive. Hence, these three algorithms are characteristic for different sorts of workloads.  We compare the run-time of these graph processing algorithms on the Spark cluster using prepartitioned graphs with different partitioning algorithms. Please note that processing larger graphs than the TW graph led to out-of-memory errors in Spark. We report the mean result of three runs for each experiment. We did not perform partitioning with ADWISE or METIS, as these partitioners already take more run-time for partitioning than the entire subsequent graph processing job takes for processing. Further, DNE led to largely imbalanced partitions (cf. Section~\ref{ref:Graph Partitioning}) which can hinder good processing performance. Instead, we compare HEP to NE, SNE, HDRF and DBH.

\begin{figure*}
	\centering	
	\captionsetup[subfloat]{captionskip=-2pt}
	\subfloat[OK: Replication factor.]{\includegraphics[width=0.235\textwidth]{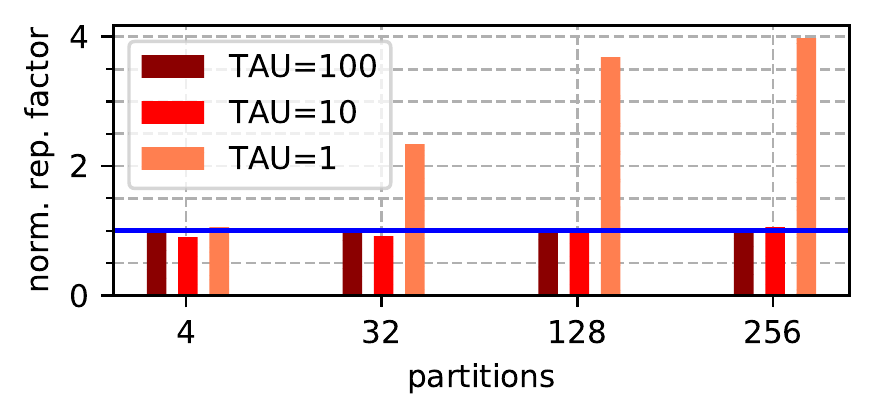}}
	\subfloat[OK: Run-time (logscale).]{\label{b}   \includegraphics[width=0.235\textwidth]{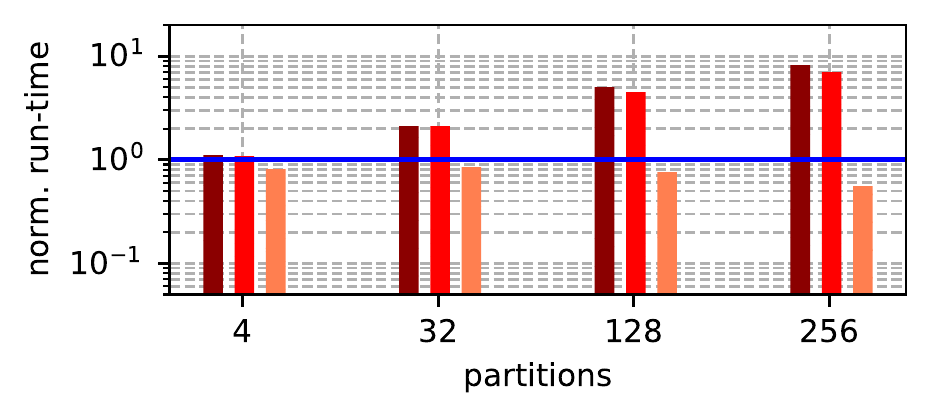}} 
	\subfloat[OK: Memory overhead.]{\label{c}   \includegraphics[width=0.235\textwidth]{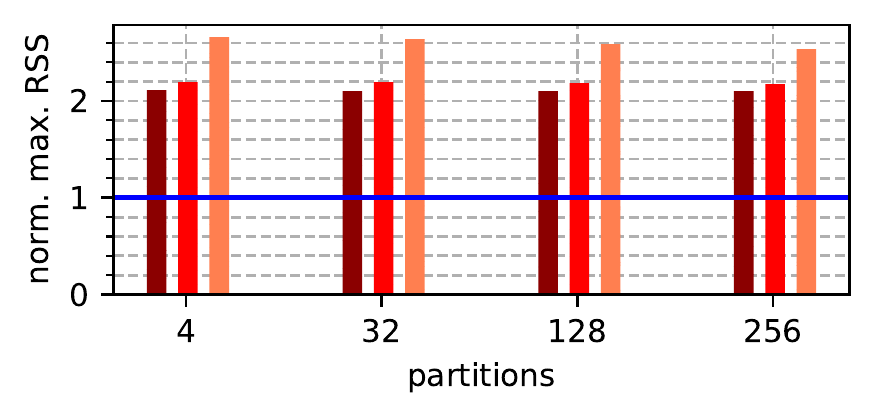}}
	\subfloat[OK: Edge Type Ratios.]{\label{c}   \includegraphics[width=0.18\textwidth]{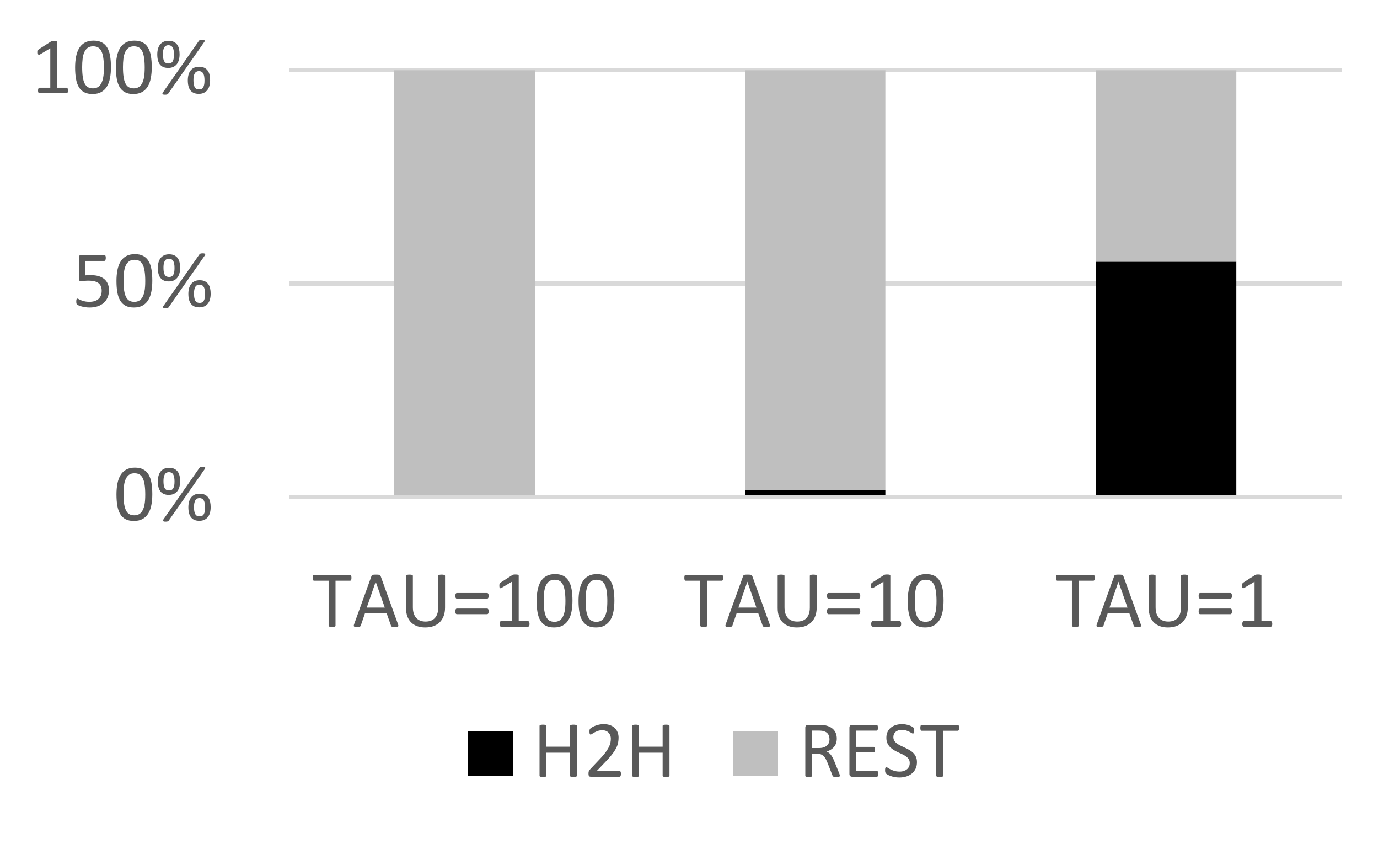}}\\
	\vspace{-0.41cm}
		\subfloat[IT: Replication factor.]{\includegraphics[width=0.235\textwidth]{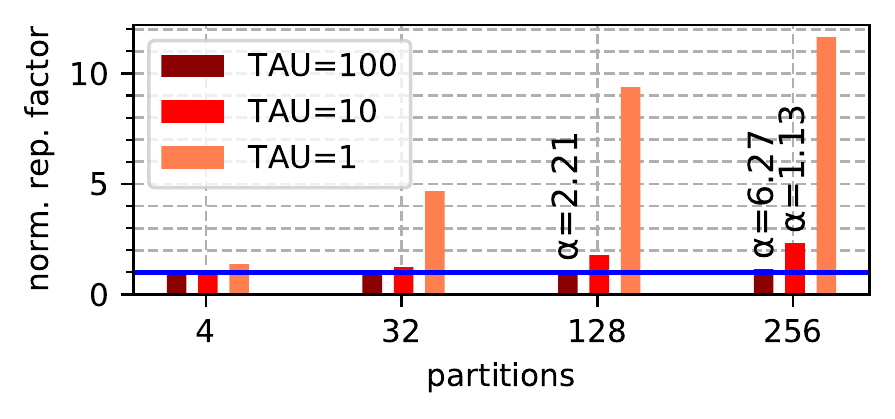}}
	\subfloat[IT: Run-time (logscale).]{\label{b}   \includegraphics[width=0.235\textwidth]{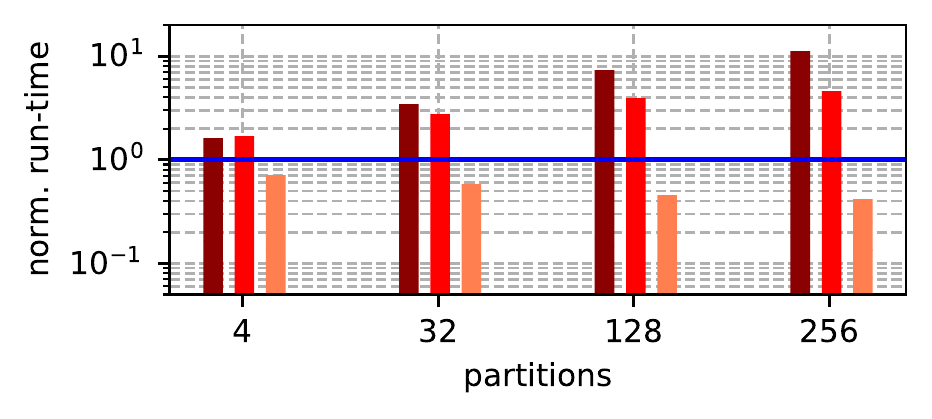}} 
	\subfloat[IT: Memory overhead.]{\label{c}   \includegraphics[width=0.235\textwidth]{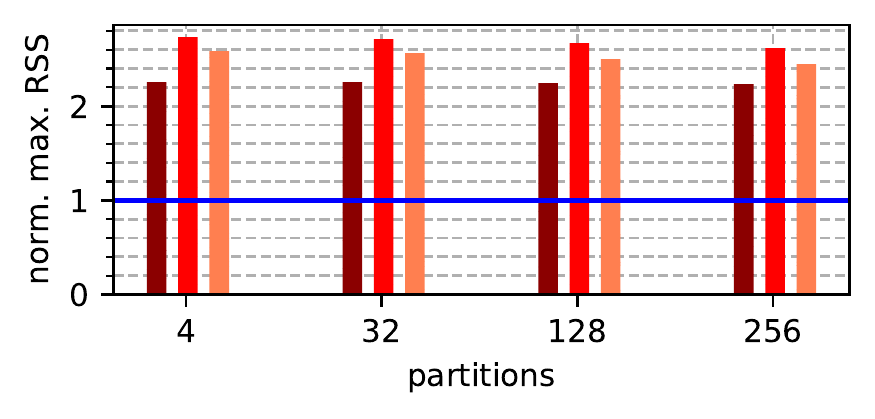}}
	\subfloat[IT: Edge Type Ratios.]{\label{c}   \includegraphics[width=0.18\textwidth]{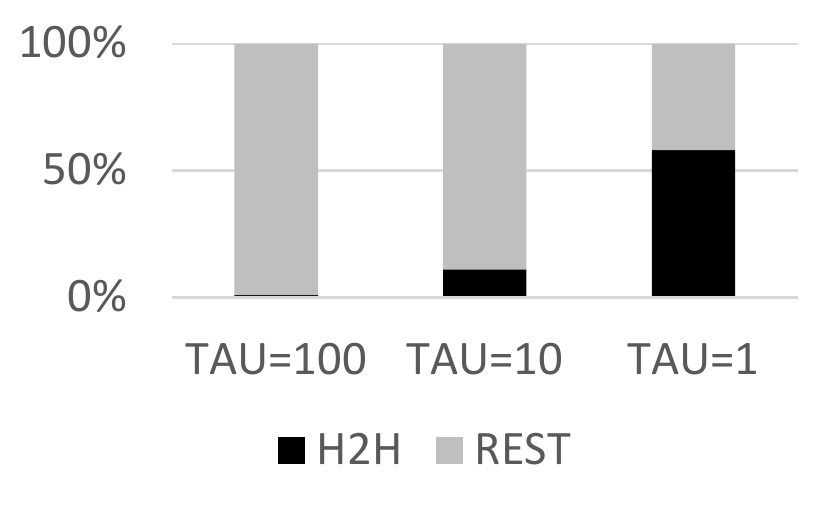}}\\
	\vspace{-0.41cm}
		\subfloat[TW: Replication factor.]{\includegraphics[width=0.235\textwidth]{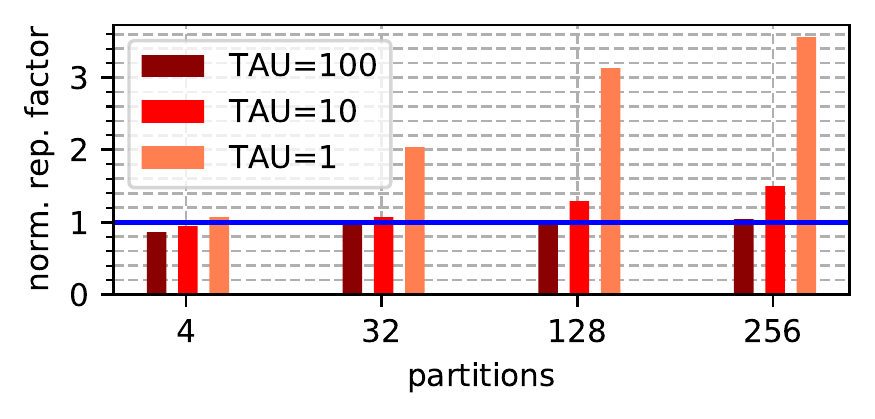}}
	\subfloat[TW: Run-time (logscale).]{\label{b}   \includegraphics[width=0.235\textwidth]{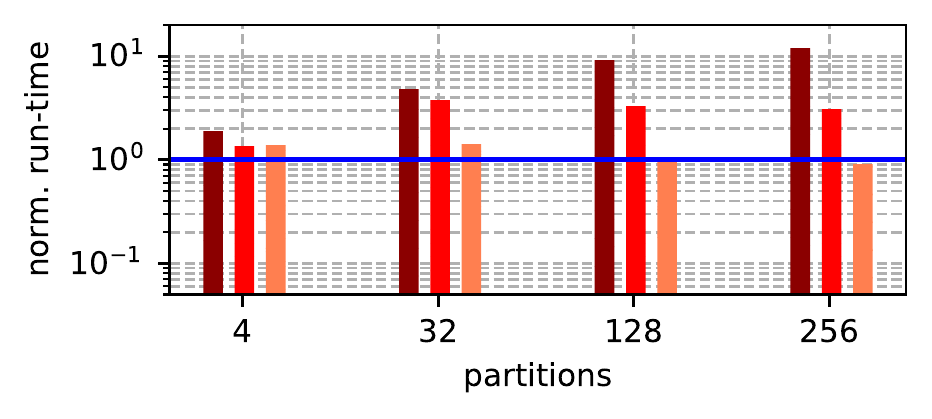}} 
	\subfloat[TW: Memory overhead.]{\label{c}   \includegraphics[width=0.235\textwidth]{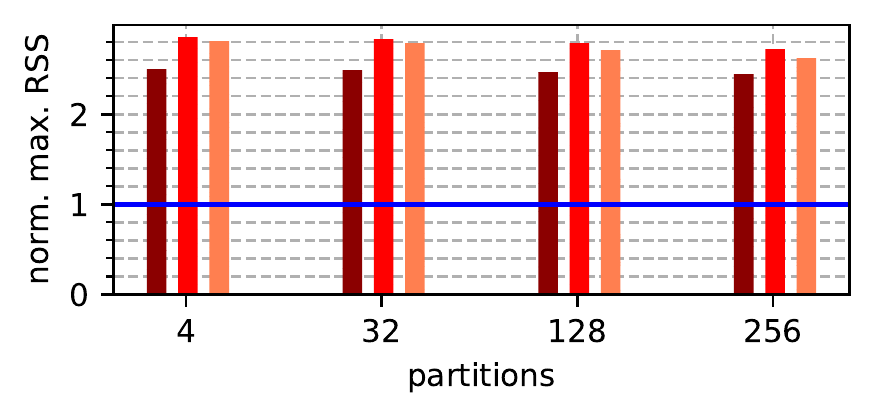}}
	\subfloat[TW: Edge Type Ratios.]{\label{c}   \includegraphics[width=0.18\textwidth]{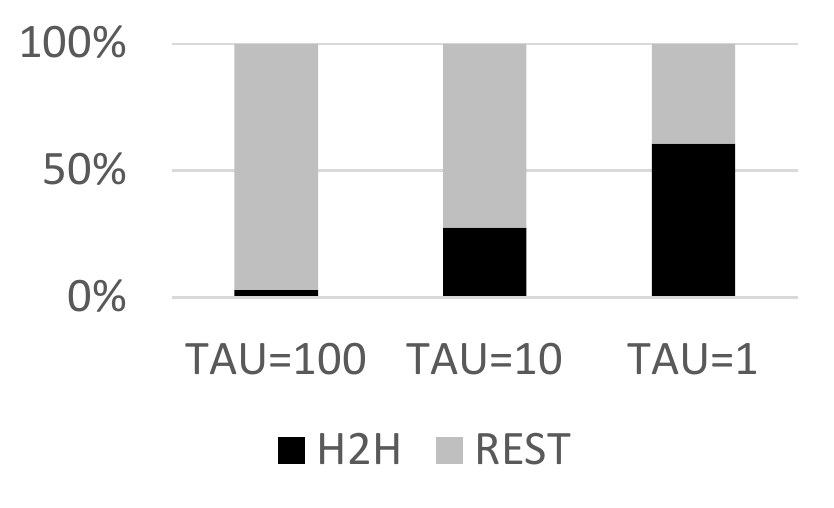}}\\
	\vspace{-0.41cm}
		\subfloat[FR: Replication factor.]{\includegraphics[width=0.235\textwidth]{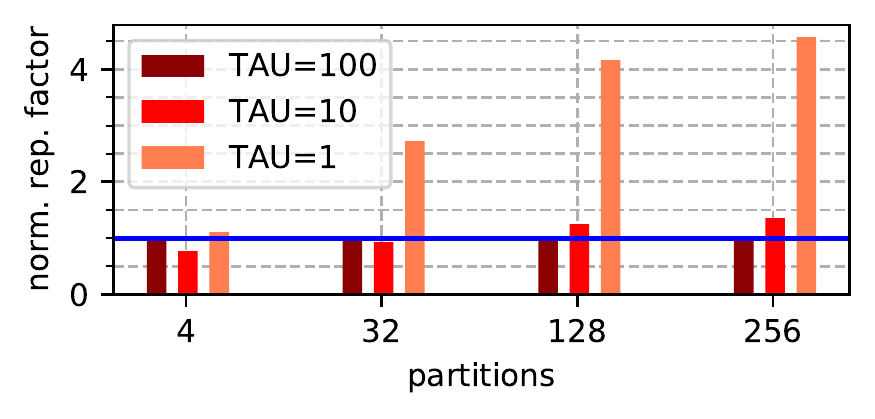}}
	\subfloat[FR: Run-time (logscale).]{\label{b}   \includegraphics[width=0.235\textwidth]{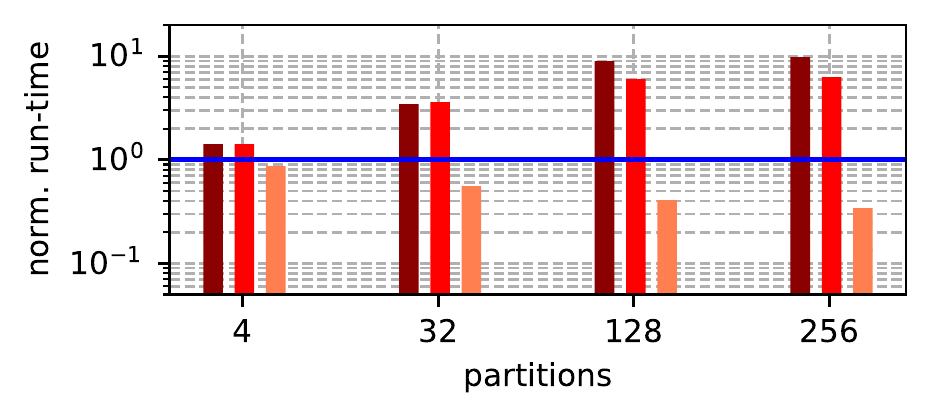}} 
	\subfloat[FR: Memory overhead.]{\label{c}   \includegraphics[width=0.235\textwidth]{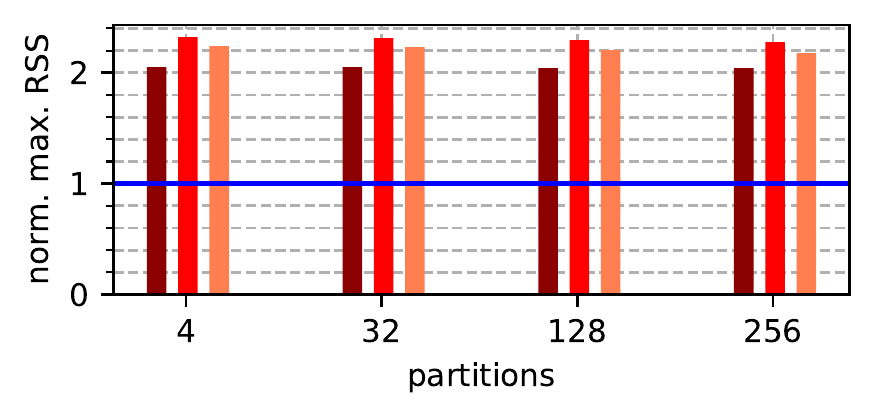}}
	\subfloat[FR: Edge Type Ratios.]{\label{c}   \includegraphics[width=0.18\textwidth]{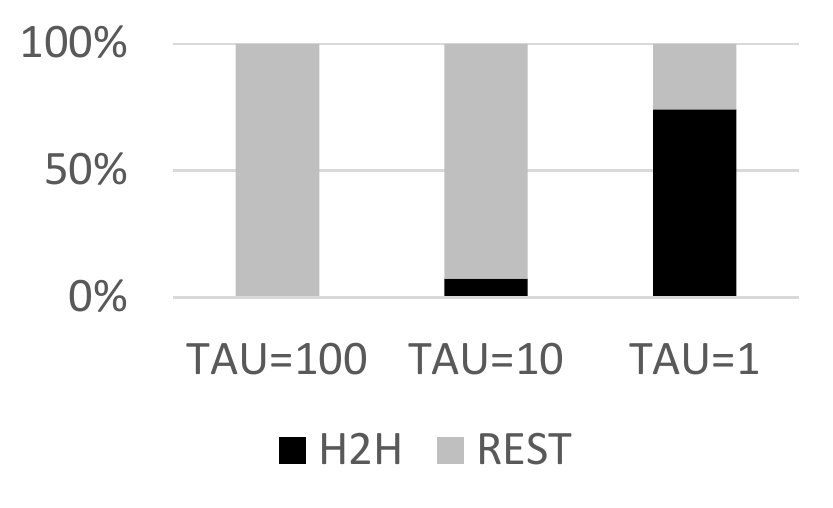}}\\
	\vspace{-0.41cm}
		\subfloat[UK: Replication factor.]{\includegraphics[width=0.235\textwidth]{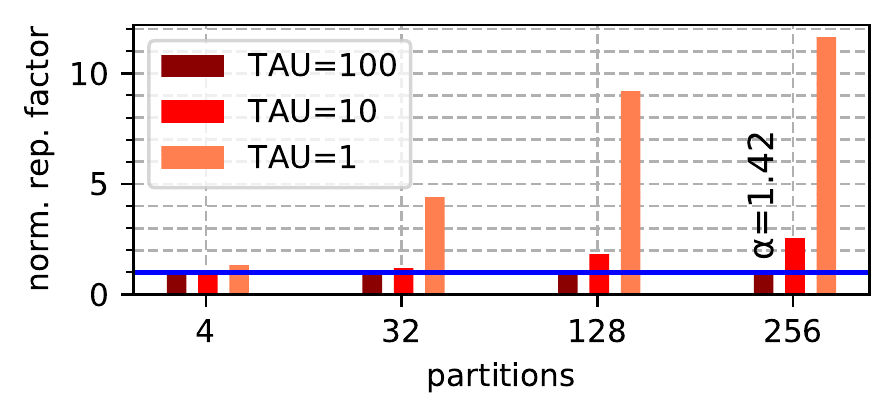}}
	\subfloat[UK: Run-time (logscale).]{\label{b}   \includegraphics[width=0.235\textwidth]{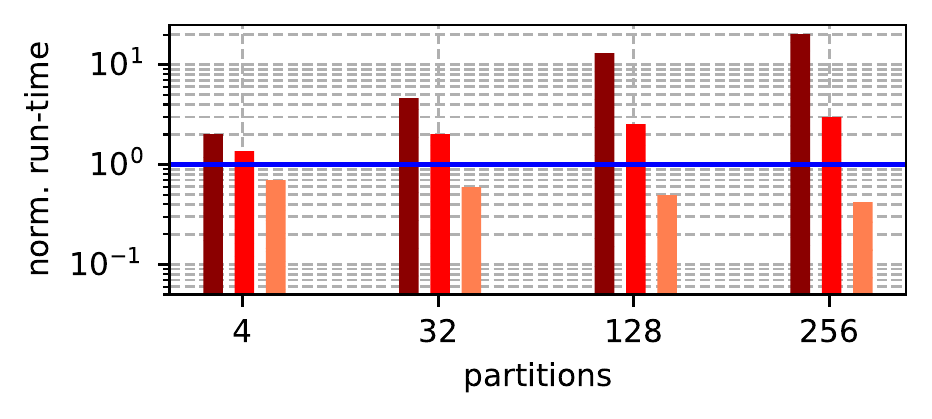}} 
	\subfloat[UK: Memory overhead.]{\label{c}   \includegraphics[width=0.235\textwidth]{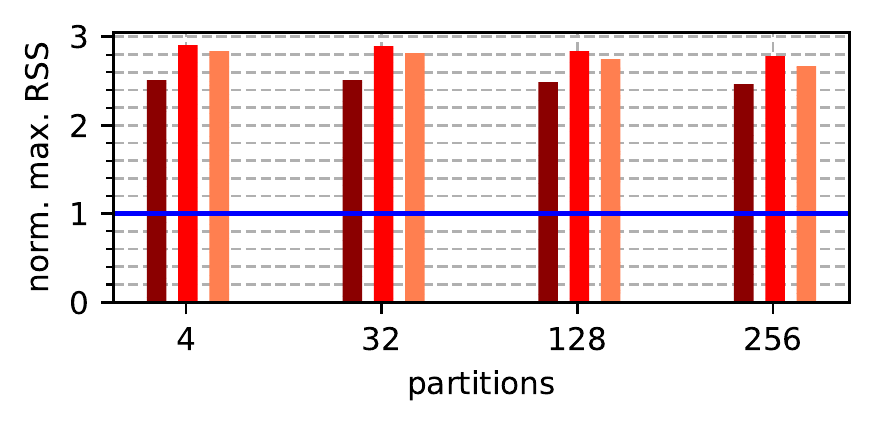}}
	\subfloat[UK: Edge Type Ratios.]{\label{c}   \includegraphics[width=0.18\textwidth]{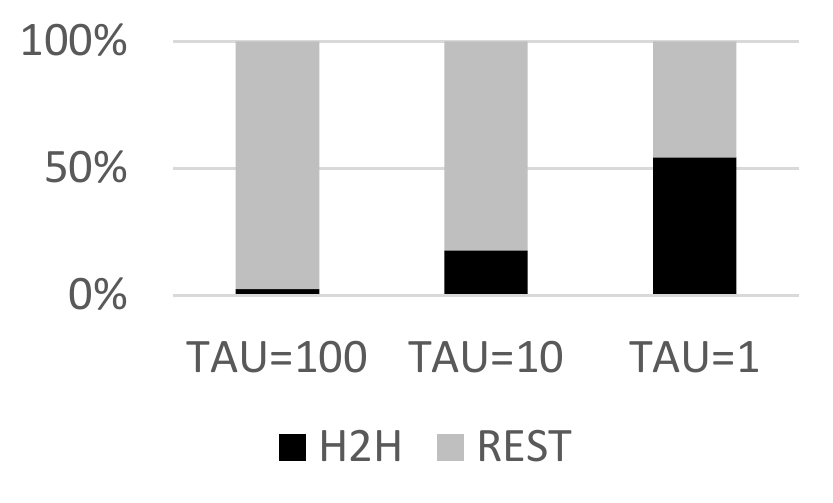}}\\
	\vspace{-10pt}
	\caption{Performance of simple hybrid strategy (NE + random streaming), normalized to the results of HEP (blue line).}
	\label{eval:comparison}
	\vspace{-8pt}
\end{figure*}

Table~\ref{tab:processing} shows the results. In many cases, the total run-time (partitioning + processing) is lowest when using HEP partitioning. There are two exceptions: First, if the processing task itself is very short, like in the Connected Components algorithm, it is better to perform fast partitioning with DBH, as despite of the higher run-time of graph processing, the total run-time is still lower. Second, different from the other graphs, on the TW graph, DBH performed reasonably well in terms of replication factor. Hence, although the graph processing run-time is faster when using HEP, the total run-time is still better with DBH. 

There are some surprising observations on the IT graph. This graph could be partitioned with very high quality by most of the partitioners (HEP, NE, SNE, and HDRF). As a result, communication overhead only plays a minor role and load balancing becomes more important. Here, HEP shows a hidden strength of its hybrid partitioning approach: It optimizes both vertex replication and \emph{vertex} balancing at the same time, as the streaming phase shows better vertex balancing than the neighborhood expansion phase (cf. Table~\ref{tab:vertex_balance}). Therefore, for all graph processing algorithms, the fastest processing time on the IT graph is obtained with HEP-1 or HEP-10, even though these do \underline{not} yield the lowest replication factor. 

We draw the following conclusions. First of all, the choice of partitioner largely depends on the run-time of the subsequent graph processing job. For very short processing jobs, simple hashing is still the best option, as it is the fastest way to partition a graph.  Second, for long-running graph processing jobs, like PageRank or multiple subsequent runs of Breadth First Search, it is in many cases beneficial to perform high-quality graph partitioning. Here, the optimal choice of partitioner depends on the replication factor achieved and the run-time of partitioning. Third, good partitioning quality, i.e., low communication overhead, matters a lot to the performance of graph processing. However, there is a saturation point where further improvements of partitioning quality do not lead to better processing performance, as the bottleneck shifts from communication overhead to balancing. While many partitioners try to balance the number of \emph{edges} across the partitions, we have shown that there are cases where vertex balancing also plays a role in the overall processing performance. HEP's hybrid approach shows to be advantageous in such a situation, leading to the best processing performance among all partitioners on the IT graph.

\textbf{Overall recommendation:} Use HEP for long-running graph processing jobs on graphs were it can achieve very low replication factors (e.g., on web graphs, cf. Figure~\ref{eval:perf}).

\subsection{Comparison to Simple Hybrid Partitioning}
How much of HEP's performance is due to its specific design (using NE++ and HDRF), as opposed to hybrid partitioning \emph{per se}? To answer this question, we compare HEP's performance to a simple hybrid partitioning baseline. We first divide the input graph $G$ into one subgraph $G_{\mathit{H2H}}$ that contains only edges incident to two high-degree vertices and another subgraph $G_{\textit{REST}}$ that contains the remaining edges. Then, $G_{\textit{REST}}$ is partitioned with NE and $G_{\mathit{H2H}}$ is partitioned with random streaming partitioning. The main observations (see Figure~\ref{eval:comparison}) are as follows.

(1) NE++ has up to $\mathtt{\sim}20\times$ lower run-time than NE (Figure~\ref{eval:comparison}(r), $\tau=100, k=256$) when both are run on the same set of edges. Only at a low value of $\tau$ ($\tau = 1.0$), the baseline sometimes runs faster than HEP (by up to $2.5\times$), as $G_{\mathit{H2H}}$ is larger than $G_{\textit{REST}}$ and partitioning is dominated by streaming (Figure~\ref{eval:comparison}(d,h,l,p,t)). Random streaming partitioning is faster than HDRF, because it has no scoring function; however, the replication factor is much worse (see (3)).

(2) NE++ has $2 - 3\times$ lower memory overhead than NE when both are run on the same set of edges (Figure~\ref{eval:comparison}(c,g,k,o,s)). Beyond the savings of hybrid partitioning, NE++ saves additional memory by pruning the adjacency lists of high-degree vertices and by avoiding auxiliary data structures to track which edges have already been assigned to partitions. 

(3) HDRF outperforms random streaming partitioning in terms of partitioning quality by up to $\mathtt{\sim}12\times$ (Figure~\ref{eval:comparison}(q), $\tau=1, k=256$). At $\tau = 1$, the amount of edges in $G_{\mathit{H2H}}$ is higher than in $G_{\textit{REST}}$, so that partitioning quality of the streaming phase matters a lot.

In summary, we have shown that the design of HEP realizes the idea of hybrid partitioning in an efficient and effective manner.

\renewcommand{\arraystretch}{1.1}
\begin{table}
{\footnotesize
	\begin{center}
		\begin{tabular}{|l||l|l|l|l|l|l|l|}
			\hline
			Mem. limit (MB) & 1,000 & 900 & 800 & 700 & 600  & 500   & 400 \\ \hline\hline
			Run-time (sec) & 42 & 65 & 116  & 205 & 374   &  587 & 1,736  \\ \hline
			Hard page faults & 61 K & 156 K & 365 K  & 688 K & 1.32 M   &  2.13 M  & 5.79 M \\ \hline
		\end{tabular}
	\end{center}
}
\caption{Performance of paging on OK graph.}
\label{tab:paging}
\vspace{-20pt}
\end{table}

\subsection{Comparison to Paging}
Instead of hybrid partitioning, one could exploit memory paging to partition large graphs with an in-memory algorithm and swap pages to disk that do not fit into memory. To evaluate this, we ran NE++ on the OK graph for $k=32$ partitions with different memory restrictions (using \texttt{cgroups}) employing an SSD for fast swapping. The run-time and hard page faults quickly increase when we restrict the process memory (Table~\ref{tab:paging}). For instance, at a memory restriction of 400 MB, NE++ takes 1,736 seconds and induces 5.79 million hard page faults, while with HEP at $\tau = 1$, we keep a maximum resident set size of 417 MB at a run-time of 45 seconds without any hard page faults. The advantage of paged NE++ is in a better replication factor (2.51 with NE++ compared to 4.52 with HEP at $\tau = 1$). Similar results were observed for other values of $k$.

\section{Related Work}
\label{sec:related}

Existing graph partitioners are either in-memory or streaming partitioners. The most prominent type of in-memory partitioners are \emph{multi-level} partitioners, such as Chaco~\cite{10.1145/224170.224228}, Jostle~\cite{doi:10.1137/S1064827598337373}, SCOTCH~\cite{10.5555/645560.658570}, METIS~\cite{Karypis:1998:FHQ:305219.305248, lasalle2013multi} or KaHiP~\cite{schlag2019scalable, sanders2011engineering}. They apply three subsequent steps: First, the graph is coarsened by combining neighboring vertices. Second, the coarsened graph is partitioned. Third, the partitioning of the coarsened graph is refined to the original graph. The main difference between the various multi-level partitioners is in which algorithms they use for each of these steps. These partitioners focus on partitioning quality, but come with a high overhead. In our experiments, KaHiP ran out of memory on large graphs, while METIS ran for days on some graphs. 

Sheep~\cite{Margo:2015:SDG:2824032.2824046} is a distributed edge partitioner that builds and partitions the elimination tree of the input graph. Spinner~\cite{spinner} and XtraPuLP~\cite{xtrapulp} are distributed and parallel vertex partitioners that build upon the label propagation algorithm for community detection. These partitioners are outperformed by DNE in terms of replication factor, run-time, memory efficiency and scalability~\cite{dne}. HEP performs better than DNE in terms of partitioning quality and memory overhead. NE~\cite{Zhang:2017:GEP:3097983.3098033} is a sequential partitioner that yields superior partitioning quality compared to other non-multi-level partitioners. In HEP, we improve the design of NE and overcome shortcomings that bloat its memory overhead and run-time. 

Some in-memory partitioners, like Sheep~\cite{Margo:2015:SDG:2824032.2824046} and Streaming NE (SNE)~\cite{Zhang:2017:GEP:3097983.3098033}, decrease their memory overhead by creating chunks of the input graph that are partitioned sequentially one at a time. However, this leads to longer run-times, and in case of SNE, also to worse partitioning quality. HEP also makes such trade-offs, but its performance is superior to prior approaches, as instead of applying the \emph{same} in-memory algorithm to \emph{subsequent} chunks of the input graph, it applies \emph{different} algorithms to \emph{selected} parts of the graph. This maximizes partitioning quality on the most crucial part of the graph, i.e., edges incident to low-degree vertices, while relaxing the partitioning quality only on a low number of high-degree vertices.  

We have evaluated a large range of streaming partitioners in Section~\ref{sec:evaluation}. They work either on hashing~\cite{dbh, grid} or they gather and exploit partitioning state while passing through the graph stream~\cite{Petroni:2015:HSP:2806416.2806424, 8416335, Tsourakakis:2014:FSG:2556195.2556213, Stanton:2012:SGP:2339530.2339722}). These partitioners are outperformed by HEP in terms of replication factor. Common variations of the streaming model are \emph{re-streaming}~\cite{Nishimura:2013:RGP:2487575.2487696}, i.e., performing multiple passes through the graph stream and refining the partitioning in each pass, and \emph{local reordering}~\cite{8416335}, i.e., buffering parts of the graph stream and reordering it to improve partitioning quality. TLP~\cite{TLP} builds on a similar neighbor expansion scheme as NE, but in a semi-streaming manner. The authors claim a low space complexity, but TLP requires the graph to be sorted and streamed multiple times in different breadth-first-search orderings. This overhead seems prohibitive to achieve competitive run-time.

Fan et al.~\cite{10.1145/3318464.3389745} propose an approach to refine a given vertex or edge partitioning to a hybrid partitioning that is tailored to the graph processing job. HEP can be used as a pre-partitioning algorithm for this. Further, Fan et al.~\cite{incremental} propose a method to incrementalize iterative vertex-cut partitioners, which is also applicable to NE++.
A similar problem to graph partitioning is to find ``vertex separators''~\cite{10.1145/1060590.1060674, vertex_separator}, i.e., vertices whose removal from the graph would split it into two or more disconnected components.

Compression techniques, such as SlashBurn~\cite{slashburn} or Frame of Reference~\cite{FOR}, can reduce the memory footprint of graph processing algorithms. However, the efficacy of compression depends on the graph data. Different from that, HEP can reduce the memory overhead in a predictable way by tuning $\tau$. Frame of Reference also benefits from sorted adjacency lists, which is at odds with NE++'s pruning mechanism that changes the ordering.

PowerLyra~\cite{powerlyra} is a graph processing system that follows a hybrid partitioning and processing strategy: It applies vertex partitioning to low-degree vertices and edge partitioning to edges incident to high-degree vertices. While this idea is related to ours in its nature of treating low-degree and high-degree vertices differently, PowerLyra is completely in-memory. Out-of-core graph processing systems, such as GraphChi \cite{graphchi}, X-Stream~\cite{Roy:2013:XEG:2517349.2522740}, Chaos~\cite{Roy:2015:CSG:2815400.2815408}, GridGraph~\cite{gridgraph} and Mosaic~\cite{Maass:2017:MPT:3064176.3064191} commonly pre-process the graph data to increase access locality. These pre-processing problems differ from the partitioning problem we target with HEP. It would be inefficient to implement the NE algorithm in out-of-core graph processing frameworks, as the number of iterations in NE is linear in the number of vertices, while each iteration in these frameworks requires a complete pass through the entire graph.

\section{Conclusions}
\label{sec:conclusion}
In this paper, we proposed HEP, a new hybrid system for edge partitioning. HEP separates the edge set into two sub-sets based on the vertex degrees and partitions them separately with a novel in-memory algorithm NE++ and subsequent informed stateful streaming partitioning. Evaluations show that HEP yields a state-of-the-art replication factor faster and with less memory overhead than pure in-memory systems, while yielding a better replication factor than streaming partitioning. Finally, graph partitioning with HEP leads to a faster run-time of graph processing jobs.

In future work, we aim to further improve the performance of HEP by focusing on parallelism and distribution. Beyond that, we aim to explore the extension of the hybrid in-memory and streaming partitioning paradigm to hypergraphs~\cite{8621968, Alistarh:2015:SMH:2969442.2969452}.

\section*{Appendix A: Configurations}
\label{sec:appendixA}

\textit{System Parameters}. We follow the author recommendations to set all system parameters. For HDRF, we set $\lambda = 1.1$~\cite{Petroni:2015:HSP:2806416.2806424}. For SNE, we set a sample size of 2~\cite{Zhang:2017:GEP:3097983.3098033}. We compiled DNE with 32-bit vertex IDs, which is optimal for the graphs we partitioned (having less than $2^{32}$ vertices) and set an expansion ratio of 0.1 and a balance factor of 1.05. DNE always starts one separate process per partition. As our machine has 64 hardware threads, we use $\lceil\frac{64}{k}\rceil$ threads per process. We ran METIS~\cite{Karypis:1998:FHQ:305219.305248} using direct $k$-way partitioning and set \texttt{-objtype=vol} to optimize for communication volume. As METIS is a vertex partitioner, we weighted each vertex with its degree before partitioning and then assigned each edge $e = (u,v)$ randomly either to the partition of $u$ or of $v$ (see also~\cite{Zhang:2017:GEP:3097983.3098033, Bourse:2014:BGE:2623330.2623660}). We did not count the time for this conversion step into the METIS run-time.

\textit{Implementation}. When possible, we used the reference implementation provided by the respective authors~\cite{SNE-code, DNE-code, METIS-code}. The stand-alone implementation of HDRF~\cite{HDRF-code} does not work for large graphs and shows excessive run-time, while the version integrated into PowerGraph~\cite{POWERGRAPH-code} cannot be used stand-alone. We, hence, re-implemented the HDRF algorithm in C++. DBH has no public reference implementation, so that we also re-implemented it in C++.

\textit{Input Formats}.
For HEP, HDRF, DBH, NE, and SNE, the input graph is provided as binary edge list with 32-bit vertex ids. Other partitioners require different input formats. We do not include the transformation time of the graph data into our run-time results.

\section*{Appendix B: Proof of Theorem 1}
\label{sec:appendixB}
\begin{proof}
We first analyze when in the NE algorithm (Algorithm~1) the adjacency list of a vertex is accessed. 
First, when a vertex $v$ is moved to the core set $C$, its neighbors which are neither in $C$ nor in $S_i$ are moved to $S_i$ (lines 14--15). Second, when a vertex $v$ is moved to the secondary set $S_i$, it is checked for each neighbor $u$ of $v$ whether $u$ is in $C$ or $S_i$, in order to update the external degree of $v$ and to assign corresponding edges to partition $p_i$ (lines 20--28). 

After a vertex $v$ is moved to $C$ in the expansion phase of partition $p_i$, it can neither be moved to $C$ another time nor be moved to $S_j$ in the expansion phase of a subsequent partition $p_j, j>i$. In other words, once moved to $C$, a vertex stays in $C$. Therefore, once $v$ is moved to $C$, the adjacency list of $v$ will not be accessed again. On the other hand, if $v$ is not moved to $C$, it remains in $S_i$ at the end of the expansion phase of $p_i$. This proves the theorem.
\end{proof}

\paragraph*{Acknowledgements}
This work is funded in part by the Deutsche Forschungsgemeinschaft (DFG, German Research Foundation) - 438107855.


\bibliographystyle{ACM-Reference-Format}
\bibliography{paper}


\begin{thebibliography}{67}


\ifx \showCODEN    \undefined \def \showCODEN     #1{\unskip}     \fi
\ifx \showDOI      \undefined \def \showDOI       #1{#1}\fi
\ifx \showISBNx    \undefined \def \showISBNx     #1{\unskip}     \fi
\ifx \showISBNxiii \undefined \def \showISBNxiii  #1{\unskip}     \fi
\ifx \showISSN     \undefined \def \showISSN      #1{\unskip}     \fi
\ifx \showLCCN     \undefined \def \showLCCN      #1{\unskip}     \fi
\ifx \shownote     \undefined \def \shownote      #1{#1}          \fi
\ifx \showarticletitle \undefined \def \showarticletitle #1{#1}   \fi
\ifx \showURL      \undefined \def \showURL       {\relax}        \fi
\providecommand\bibfield[2]{#2}
\providecommand\bibinfo[2]{#2}
\providecommand\natexlab[1]{#1}
\providecommand\showeprint[2][]{arXiv:#2}

\bibitem[\protect\citeauthoryear{??}{gir}{[n.d.]}]%
        {giraph}
 \bibinfo{year}{[n.d.]}\natexlab{}.
\newblock \bibinfo{title}{{Apache Giraph}}.
\newblock
\newblock
\urldef\tempurl%
\url{https://giraph.apache.org/}
\showURL{%
\tempurl}


\bibitem[\protect\citeauthoryear{??}{DNE}{[n.d.]}]%
        {DNE-code}
 \bibinfo{year}{[n.d.]}\natexlab{}.
\newblock \bibinfo{title}{{DNE}}.
\newblock
\newblock
\urldef\tempurl%
\url{https://github.com/masatoshihanai/DistributedNE}
\showURL{%
\tempurl}


\bibitem[\protect\citeauthoryear{??}{fri}{[n.d.]}]%
        {friendster}
 \bibinfo{year}{[n.d.]}\natexlab{}.
\newblock \bibinfo{title}{{FR} Graph}.
\newblock
\newblock
\urldef\tempurl%
\url{https://snap.stanford.edu/data/com-Friendster.html}
\showURL{%
\tempurl}


\bibitem[\protect\citeauthoryear{??}{gsh}{[n.d.]}]%
        {gsh}
 \bibinfo{year}{[n.d.]}\natexlab{}.
\newblock \bibinfo{title}{{GSH} Graph}.
\newblock
\newblock
\urldef\tempurl%
\url{http://law.di.unimi.it/webdata/gsh-2015/}
\showURL{%
\tempurl}


\bibitem[\protect\citeauthoryear{??}{HDR}{[n.d.]}]%
        {HDRF-code}
 \bibinfo{year}{[n.d.]}\natexlab{}.
\newblock \bibinfo{title}{{HDRF}}.
\newblock
\newblock
\urldef\tempurl%
\url{https://github.com/fabiopetroni/VGP}
\showURL{%
\tempurl}


\bibitem[\protect\citeauthoryear{??}{it}{[n.d.]}]%
        {it}
 \bibinfo{year}{[n.d.]}\natexlab{}.
\newblock \bibinfo{title}{{IT} Graph}.
\newblock
\newblock
\urldef\tempurl%
\url{http://law.di.unimi.it/webdata/it-2004/}
\showURL{%
\tempurl}


\bibitem[\protect\citeauthoryear{??}{MET}{[n.d.]}]%
        {METIS-code}
 \bibinfo{year}{[n.d.]}\natexlab{}.
\newblock \bibinfo{title}{{METIS}}.
\newblock
\newblock
\urldef\tempurl%
\url{http://glaros.dtc.umn.edu/gkhome/metis/metis/download}
\showURL{%
\tempurl}


\bibitem[\protect\citeauthoryear{??}{ork}{[n.d.]}]%
        {orkut}
 \bibinfo{year}{[n.d.]}\natexlab{}.
\newblock \bibinfo{title}{{OK} Graph}.
\newblock
\newblock
\urldef\tempurl%
\url{https://snap.stanford.edu/data/com-Orkut.html}
\showURL{%
\tempurl}


\bibitem[\protect\citeauthoryear{??}{POW}{[n.d.]}]%
        {POWERGRAPH-code}
 \bibinfo{year}{[n.d.]}\natexlab{}.
\newblock \bibinfo{title}{{PowerGraph}}.
\newblock
\newblock
\urldef\tempurl%
\url{https://github.com/jegonzal/PowerGraph}
\showURL{%
\tempurl}


\bibitem[\protect\citeauthoryear{??}{SNE}{[n.d.]}]%
        {SNE-code}
 \bibinfo{year}{[n.d.]}\natexlab{}.
\newblock \bibinfo{title}{{(S)NE}}.
\newblock
\newblock
\urldef\tempurl%
\url{https://github.com/ansrlab/edgepart}
\showURL{%
\tempurl}


\bibitem[\protect\citeauthoryear{??}{twi}{[n.d.]}]%
        {twitter}
 \bibinfo{year}{[n.d.]}\natexlab{}.
\newblock \bibinfo{title}{{TW} Graph}.
\newblock
\newblock
\urldef\tempurl%
\url{https://snap.stanford.edu/data/twitter-2010.html}
\showURL{%
\tempurl}


\bibitem[\protect\citeauthoryear{??}{uk}{[n.d.]}]%
        {uk}
 \bibinfo{year}{[n.d.]}\natexlab{}.
\newblock \bibinfo{title}{{UK} Graph}.
\newblock
\newblock
\urldef\tempurl%
\url{http://law.di.unimi.it/webdata/uk-2007-05/}
\showURL{%
\tempurl}


\bibitem[\protect\citeauthoryear{??}{wdc}{[n.d.]}]%
        {wdc}
 \bibinfo{year}{[n.d.]}\natexlab{}.
\newblock \bibinfo{title}{{WDC} Graph}.
\newblock
\newblock
\urldef\tempurl%
\url{http://webdatacommons.org/hyperlinkgraph/}
\showURL{%
\tempurl}


\bibitem[\protect\citeauthoryear{Albert and Barab\'asi}{Albert and
  Barab\'asi}{2002}]%
        {RevModPhys.74.47}
\bibfield{author}{\bibinfo{person}{R\'eka Albert} {and}
  \bibinfo{person}{Albert-L\'aszl\'o Barab\'asi}.}
  \bibinfo{year}{2002}\natexlab{}.
\newblock \showarticletitle{Statistical mechanics of complex networks}.
\newblock \bibinfo{journal}{\emph{Rev. Mod. Phys.}}  \bibinfo{volume}{74}
  (\bibinfo{date}{Jan} \bibinfo{year}{2002}), \bibinfo{pages}{47--97}.
\newblock
Issue 1.
\urldef\tempurl%
\url{https://doi.org/10.1103/RevModPhys.74.47}
\showDOI{\tempurl}


\bibitem[\protect\citeauthoryear{Alistarh, Iglesias, and Vojnovic}{Alistarh
  et~al\mbox{.}}{2015}]%
        {Alistarh:2015:SMH:2969442.2969452}
\bibfield{author}{\bibinfo{person}{Dan Alistarh}, \bibinfo{person}{Jennifer
  Iglesias}, {and} \bibinfo{person}{Milan Vojnovic}.}
  \bibinfo{year}{2015}\natexlab{}.
\newblock \showarticletitle{Streaming Min-max Hypergraph Partitioning}. In
  \bibinfo{booktitle}{\emph{Proceedings of the 28th International Conference on
  Neural Information Processing Systems - Volume 2}}
  \emph{(\bibinfo{series}{NIPS'15})}. \bibinfo{publisher}{MIT Press},
  \bibinfo{address}{Cambridge, MA, USA}, \bibinfo{pages}{1900--1908}.
\newblock
\urldef\tempurl%
\url{http://dl.acm.org/citation.cfm?id=2969442.2969452}
\showURL{%
\tempurl}


\bibitem[\protect\citeauthoryear{Boldi, Marino, Santini, and Vigna}{Boldi
  et~al\mbox{.}}{2014}]%
        {BMSB}
\bibfield{author}{\bibinfo{person}{Paolo Boldi}, \bibinfo{person}{Andrea
  Marino}, \bibinfo{person}{Massimo Santini}, {and} \bibinfo{person}{Sebastiano
  Vigna}.} \bibinfo{year}{2014}\natexlab{}.
\newblock \showarticletitle{{BUbiNG}: Massive Crawling for the Masses}. In
  \bibinfo{booktitle}{\emph{Proceedings of the Companion Publication of the
  23rd International Conference on World Wide Web}}.
  \bibinfo{publisher}{International World Wide Web Conferences Steering
  Committee}, \bibinfo{pages}{227--228}.
\newblock


\bibitem[\protect\citeauthoryear{Boldi, Rosa, Santini, and Vigna}{Boldi
  et~al\mbox{.}}{2011}]%
        {BRSLLP}
\bibfield{author}{\bibinfo{person}{Paolo Boldi}, \bibinfo{person}{Marco Rosa},
  \bibinfo{person}{Massimo Santini}, {and} \bibinfo{person}{Sebastiano Vigna}.}
  \bibinfo{year}{2011}\natexlab{}.
\newblock \showarticletitle{Layered Label Propagation: A MultiResolution
  Coordinate-Free Ordering for Compressing Social Networks}. In
  \bibinfo{booktitle}{\emph{Proceedings of the 20th International Conference on
  World Wide Web}}, \bibfield{editor}{\bibinfo{person}{Sadagopan Srinivasan},
  \bibinfo{person}{Krithi Ramamritham}, \bibinfo{person}{Arun Kumar},
  \bibinfo{person}{M.~P. Ravindra}, \bibinfo{person}{Elisa Bertino}, {and}
  \bibinfo{person}{Ravi Kumar}} (Eds.). \bibinfo{publisher}{ACM Press},
  \bibinfo{pages}{587--596}.
\newblock


\bibitem[\protect\citeauthoryear{Boldi and Vigna}{Boldi and Vigna}{2004}]%
        {BoVWFI}
\bibfield{author}{\bibinfo{person}{Paolo Boldi} {and}
  \bibinfo{person}{Sebastiano Vigna}.} \bibinfo{year}{2004}\natexlab{}.
\newblock \showarticletitle{The {W}eb{G}raph Framework {I}: {C}ompression
  Techniques}. In \bibinfo{booktitle}{\emph{Proc. of the Thirteenth
  International World Wide Web Conference (WWW 2004)}}.
  \bibinfo{publisher}{ACM}, \bibinfo{address}{Manhattan, USA},
  \bibinfo{pages}{595--601}.
\newblock


\bibitem[\protect\citeauthoryear{Bourse, Lelarge, and Vojnovic}{Bourse
  et~al\mbox{.}}{2014}]%
        {Bourse:2014:BGE:2623330.2623660}
\bibfield{author}{\bibinfo{person}{Florian Bourse}, \bibinfo{person}{Marc
  Lelarge}, {and} \bibinfo{person}{Milan Vojnovic}.}
  \bibinfo{year}{2014}\natexlab{}.
\newblock \showarticletitle{Balanced Graph Edge Partition}. In
  \bibinfo{booktitle}{\emph{Proceedings of the 20th ACM SIGKDD International
  Conference on Knowledge Discovery and Data Mining}}
  \emph{(\bibinfo{series}{KDD '14})}. \bibinfo{publisher}{ACM},
  \bibinfo{address}{New York, NY, USA}, \bibinfo{pages}{1456--1465}.
\newblock
\showISBNx{978-1-4503-2956-9}
\urldef\tempurl%
\url{https://doi.org/10.1145/2623330.2623660}
\showDOI{\tempurl}


\bibitem[\protect\citeauthoryear{Brandt and Wattenhofer}{Brandt and
  Wattenhofer}{2017}]%
        {vertex_separator}
\bibfield{author}{\bibinfo{person}{Sebastian Brandt} {and}
  \bibinfo{person}{Roger Wattenhofer}.} \bibinfo{year}{2017}\natexlab{}.
\newblock \showarticletitle{Approximating Small Balanced Vertex Separators in
  Almost Linear Time}. In \bibinfo{booktitle}{\emph{Algorithms and Data
  Structures}}, \bibfield{editor}{\bibinfo{person}{Faith Ellen},
  \bibinfo{person}{Antonina Kolokolova}, {and}
  \bibinfo{person}{J{\"o}rg-R{\"u}diger Sack}} (Eds.).
  \bibinfo{publisher}{Springer International Publishing},
  \bibinfo{address}{Cham}, \bibinfo{pages}{229--240}.
\newblock
\showISBNx{978-3-319-62127-2}


\bibitem[\protect\citeauthoryear{Bulu{\c{c}}, Meyerhenke, Safro, Sanders, and
  Schulz}{Bulu{\c{c}} et~al\mbox{.}}{2016}]%
        {gp-survey}
\bibfield{author}{\bibinfo{person}{Ayd{\i}n Bulu{\c{c}}},
  \bibinfo{person}{Henning Meyerhenke}, \bibinfo{person}{Ilya Safro},
  \bibinfo{person}{Peter Sanders}, {and} \bibinfo{person}{Christian Schulz}.}
  \bibinfo{year}{2016}\natexlab{}.
\newblock \showarticletitle{Recent Advances in Graph Partitioning}. In
  \bibinfo{booktitle}{\emph{Algorithm Engineering: Selected Results and
  Surveys}}, \bibfield{editor}{\bibinfo{person}{Lasse Kliemann} {and}
  \bibinfo{person}{Peter Sanders}} (Eds.). \bibinfo{publisher}{Springer
  International Publishing}, \bibinfo{address}{Cham},
  \bibinfo{pages}{117--158}.
\newblock
\showISBNx{978-3-319-49487-6}
\urldef\tempurl%
\url{https://doi.org/10.1007/978-3-319-49487-6_4}
\showDOI{\tempurl}


\bibitem[\protect\citeauthoryear{Bulu\c{c}, Fineman, Frigo, Gilbert, and
  Leiserson}{Bulu\c{c} et~al\mbox{.}}{2009}]%
        {10.1145/1583991.1584053}
\bibfield{author}{\bibinfo{person}{Aydin Bulu\c{c}}, \bibinfo{person}{Jeremy~T.
  Fineman}, \bibinfo{person}{Matteo Frigo}, \bibinfo{person}{John~R. Gilbert},
  {and} \bibinfo{person}{Charles~E. Leiserson}.}
  \bibinfo{year}{2009}\natexlab{}.
\newblock \showarticletitle{Parallel Sparse Matrix-Vector and
  Matrix-Transpose-Vector Multiplication Using Compressed Sparse Blocks}. In
  \bibinfo{booktitle}{\emph{Proceedings of the Twenty-First Annual Symposium on
  Parallelism in Algorithms and Architectures}} \emph{(\bibinfo{series}{SPAA
  ’09})}. \bibinfo{publisher}{Association for Computing Machinery},
  \bibinfo{address}{New York, NY, USA}, \bibinfo{pages}{233–244}.
\newblock
\showISBNx{9781605586069}
\urldef\tempurl%
\url{https://doi.org/10.1145/1583991.1584053}
\showDOI{\tempurl}


\bibitem[\protect\citeauthoryear{Chen, Shi, Chen, and Chen}{Chen
  et~al\mbox{.}}{2015}]%
        {powerlyra}
\bibfield{author}{\bibinfo{person}{Rong Chen}, \bibinfo{person}{Jiaxin Shi},
  \bibinfo{person}{Yanzhe Chen}, {and} \bibinfo{person}{Haibo Chen}.}
  \bibinfo{year}{2015}\natexlab{}.
\newblock \showarticletitle{Power{L}yra: Differentiated Graph Computation and
  Partitioning on Skewed Graphs}. In \bibinfo{booktitle}{\emph{Proceedings of
  the Tenth European Conference on Computer Systems}}
  \emph{(\bibinfo{series}{EuroSys '15})}. \bibinfo{publisher}{ACM},
  \bibinfo{address}{New York, NY, USA}, Article \bibinfo{articleno}{1},
  \bibinfo{numpages}{15}~pages.
\newblock
\showISBNx{978-1-4503-3238-5}
\urldef\tempurl%
\url{https://doi.org/10.1145/2741948.2741970}
\showDOI{\tempurl}


\bibitem[\protect\citeauthoryear{Fan, Jin, Liu, Lu, Luo, Xu, Yin, Yu, and
  Zhou}{Fan et~al\mbox{.}}{2020a}]%
        {10.1145/3318464.3389745}
\bibfield{author}{\bibinfo{person}{Wenfei Fan}, \bibinfo{person}{Ruochun Jin},
  \bibinfo{person}{Muyang Liu}, \bibinfo{person}{Ping Lu},
  \bibinfo{person}{Xiaojian Luo}, \bibinfo{person}{Ruiqi Xu},
  \bibinfo{person}{Qiang Yin}, \bibinfo{person}{Wenyuan Yu}, {and}
  \bibinfo{person}{Jingren Zhou}.} \bibinfo{year}{2020}\natexlab{a}.
\newblock \showarticletitle{Application Driven Graph Partitioning}. In
  \bibinfo{booktitle}{\emph{Proceedings of the 2020 ACM SIGMOD International
  Conference on Management of Data}} \emph{(\bibinfo{series}{SIGMOD '20})}.
  \bibinfo{publisher}{Association for Computing Machinery},
  \bibinfo{address}{New York, NY, USA}, \bibinfo{pages}{1765–1779}.
\newblock
\showISBNx{9781450367356}
\urldef\tempurl%
\url{https://doi.org/10.1145/3318464.3389745}
\showDOI{\tempurl}


\bibitem[\protect\citeauthoryear{Fan, Liu, Tian, Xu, and Zhou}{Fan
  et~al\mbox{.}}{2020b}]%
        {incremental}
\bibfield{author}{\bibinfo{person}{Wenfei Fan}, \bibinfo{person}{Muyang Liu},
  \bibinfo{person}{Chao Tian}, \bibinfo{person}{Ruiqi Xu}, {and}
  \bibinfo{person}{Jingren Zhou}.} \bibinfo{year}{2020}\natexlab{b}.
\newblock \showarticletitle{Incrementalization of Graph Partitioning
  Algorithms}.
\newblock \bibinfo{journal}{\emph{Proc. VLDB Endow.}} \bibinfo{volume}{13},
  \bibinfo{number}{8} (\bibinfo{year}{2020}), 14.
\newblock


\bibitem[\protect\citeauthoryear{Feige, Hajiaghayi, and Lee}{Feige
  et~al\mbox{.}}{2005}]%
        {10.1145/1060590.1060674}
\bibfield{author}{\bibinfo{person}{Uriel Feige},
  \bibinfo{person}{Mohammad~Taghi Hajiaghayi}, {and} \bibinfo{person}{James~R.
  Lee}.} \bibinfo{year}{2005}\natexlab{}.
\newblock \showarticletitle{Improved Approximation Algorithms for
  Minimum-Weight Vertex Separators}. In \bibinfo{booktitle}{\emph{Proceedings
  of the Thirty-Seventh Annual ACM Symposium on Theory of Computing}}
  \emph{(\bibinfo{series}{STOC ’05})}. \bibinfo{publisher}{Association for
  Computing Machinery}, \bibinfo{address}{New York, NY, USA},
  \bibinfo{pages}{563–572}.
\newblock
\showISBNx{1581139608}
\urldef\tempurl%
\url{https://doi.org/10.1145/1060590.1060674}
\showDOI{\tempurl}


\bibitem[\protect\citeauthoryear{{Goldstein}, {Ramakrishnan}, and
  {Shaft}}{{Goldstein} et~al\mbox{.}}{1998}]%
        {FOR}
\bibfield{author}{\bibinfo{person}{Jonathan {Goldstein}},
  \bibinfo{person}{Raghu {Ramakrishnan}}, {and} \bibinfo{person}{Uri {Shaft}}.}
  \bibinfo{year}{1998}\natexlab{}.
\newblock \showarticletitle{Compressing relations and indexes}. In
  \bibinfo{booktitle}{\emph{Proceedings 14th International Conference on Data
  Engineering}}. \bibinfo{pages}{370--379}.
\newblock
\urldef\tempurl%
\url{https://doi.org/10.1109/ICDE.1998.655800}
\showDOI{\tempurl}


\bibitem[\protect\citeauthoryear{Gonzalez, Low, Gu, Bickson, and
  Guestrin}{Gonzalez et~al\mbox{.}}{2012}]%
        {powergraph}
\bibfield{author}{\bibinfo{person}{Joseph~E. Gonzalez},
  \bibinfo{person}{Yucheng Low}, \bibinfo{person}{Haijie Gu},
  \bibinfo{person}{Danny Bickson}, {and} \bibinfo{person}{Carlos Guestrin}.}
  \bibinfo{year}{2012}\natexlab{}.
\newblock \showarticletitle{Power{G}raph: Distributed Graph-Parallel
  Computation on Natural Graphs}. In \bibinfo{booktitle}{\emph{10th {USENIX}
  Symposium on Operating Systems Design and Implementation ({OSDI} 12)}}.
  \bibinfo{publisher}{{USENIX}}, \bibinfo{pages}{17--30}.
\newblock
\showISBNx{978-1-931971-96-6}
\urldef\tempurl%
\url{https://www.usenix.org/conference/osdi12/technical-sessions/presentation/gonzalez}
\showURL{%
\tempurl}


\bibitem[\protect\citeauthoryear{Gonzalez, Xin, Dave, Crankshaw, Franklin, and
  Stoica}{Gonzalez et~al\mbox{.}}{2014}]%
        {graphx}
\bibfield{author}{\bibinfo{person}{Joseph~E. Gonzalez},
  \bibinfo{person}{Reynold~S. Xin}, \bibinfo{person}{Ankur Dave},
  \bibinfo{person}{Daniel Crankshaw}, \bibinfo{person}{Michael~J. Franklin},
  {and} \bibinfo{person}{Ion Stoica}.} \bibinfo{year}{2014}\natexlab{}.
\newblock \showarticletitle{Graph{X}: Graph Processing in a Distributed
  Dataflow Framework}. In \bibinfo{booktitle}{\emph{11th {USENIX} Symposium on
  Operating Systems Design and Implementation ({OSDI} 14)}}.
  \bibinfo{publisher}{{USENIX}}, \bibinfo{pages}{599--613}.
\newblock
\showISBNx{978-1-931971-16-4}
\urldef\tempurl%
\url{https://www.usenix.org/conference/osdi14/technical-sessions/presentation/gonzalez}
\showURL{%
\tempurl}


\bibitem[\protect\citeauthoryear{Hanai, Suzumura, Tan, Liu, Theodoropoulos, and
  Cai}{Hanai et~al\mbox{.}}{2019}]%
        {dne}
\bibfield{author}{\bibinfo{person}{Masatoshi Hanai}, \bibinfo{person}{Toyotaro
  Suzumura}, \bibinfo{person}{Wen~Jun Tan}, \bibinfo{person}{Elvis Liu},
  \bibinfo{person}{Georgios Theodoropoulos}, {and} \bibinfo{person}{Wentong
  Cai}.} \bibinfo{year}{2019}\natexlab{}.
\newblock \showarticletitle{Distributed Edge Partitioning for Trillion-Edge
  Graphs}.
\newblock \bibinfo{journal}{\emph{Proc. VLDB Endow.}} \bibinfo{volume}{12},
  \bibinfo{number}{13} (\bibinfo{date}{Sept.} \bibinfo{year}{2019}),
  \bibinfo{pages}{2379–2392}.
\newblock
\showISSN{2150-8097}
\urldef\tempurl%
\url{https://doi.org/10.14778/3358701.3358706}
\showDOI{\tempurl}


\bibitem[\protect\citeauthoryear{Hendrickson and Leland}{Hendrickson and
  Leland}{1995}]%
        {10.1145/224170.224228}
\bibfield{author}{\bibinfo{person}{Bruce Hendrickson} {and}
  \bibinfo{person}{Robert Leland}.} \bibinfo{year}{1995}\natexlab{}.
\newblock \showarticletitle{A Multilevel Algorithm for Partitioning Graphs}. In
  \bibinfo{booktitle}{\emph{Proceedings of the 1995 ACM/IEEE Conference on
  Supercomputing}} \emph{(\bibinfo{series}{Supercomputing ’95})}.
  \bibinfo{publisher}{Association for Computing Machinery},
  \bibinfo{address}{New York, NY, USA}, \bibinfo{pages}{28–es}.
\newblock
\showISBNx{0897918169}
\urldef\tempurl%
\url{https://doi.org/10.1145/224170.224228}
\showDOI{\tempurl}


\bibitem[\protect\citeauthoryear{Jain, Liao, and Willke}{Jain
  et~al\mbox{.}}{2013}]%
        {grid}
\bibfield{author}{\bibinfo{person}{Nilesh Jain}, \bibinfo{person}{Guangdeng
  Liao}, {and} \bibinfo{person}{Theodore~L. Willke}.}
  \bibinfo{year}{2013}\natexlab{}.
\newblock \showarticletitle{Graph{B}uilder: Scalable Graph {ETL} Framework}. In
  \bibinfo{booktitle}{\emph{First International Workshop on Graph Data
  Management Experiences and Systems}} \emph{(\bibinfo{series}{GRADES '13})}.
  \bibinfo{publisher}{ACM}, \bibinfo{address}{New York, NY, USA}, Article
  \bibinfo{articleno}{4}, \bibinfo{numpages}{6}~pages.
\newblock
\showISBNx{978-1-4503-2188-4}
\urldef\tempurl%
\url{https://doi.org/10.1145/2484425.2484429}
\showDOI{\tempurl}


\bibitem[\protect\citeauthoryear{Ji, Bu, Li, and Wu}{Ji et~al\mbox{.}}{2019}]%
        {TLP}
\bibfield{author}{\bibinfo{person}{Shengwei Ji}, \bibinfo{person}{Chenyang Bu},
  \bibinfo{person}{Lei Li}, {and} \bibinfo{person}{Xindong Wu}.}
  \bibinfo{year}{2019}\natexlab{}.
\newblock \showarticletitle{Local Graph Edge Partitioning with a Two-Stage
  Heuristic Method}. In \bibinfo{booktitle}{\emph{2019 IEEE 39th International
  Conference on Distributed Computing Systems (ICDCS)}}.
  \bibinfo{pages}{228--237}.
\newblock


\bibitem[\protect\citeauthoryear{Karypis and Kumar}{Karypis and Kumar}{1998}]%
        {Karypis:1998:FHQ:305219.305248}
\bibfield{author}{\bibinfo{person}{George Karypis} {and} \bibinfo{person}{Vipin
  Kumar}.} \bibinfo{year}{1998}\natexlab{}.
\newblock \showarticletitle{A Fast and High Quality Multilevel Scheme for
  Partitioning Irregular Graphs}.
\newblock \bibinfo{journal}{\emph{SIAM J. Sci. Comput.}} \bibinfo{volume}{20},
  \bibinfo{number}{1} (\bibinfo{date}{Dec.} \bibinfo{year}{1998}),
  \bibinfo{pages}{359--392}.
\newblock
\showISSN{1064-8275}
\urldef\tempurl%
\url{https://doi.org/10.1137/S1064827595287997}
\showDOI{\tempurl}


\bibitem[\protect\citeauthoryear{{Kernighan} and {Lin}}{{Kernighan} and
  {Lin}}{1970}]%
        {kernighan70}
\bibfield{author}{\bibinfo{person}{Brian~W. {Kernighan}} {and}
  \bibinfo{person}{Shen {Lin}}.} \bibinfo{year}{1970}\natexlab{}.
\newblock \showarticletitle{An efficient heuristic procedure for partitioning
  graphs}.
\newblock \bibinfo{journal}{\emph{The Bell System Technical Journal}}
  \bibinfo{volume}{49}, \bibinfo{number}{2} (\bibinfo{year}{1970}),
  \bibinfo{pages}{291--307}.
\newblock


\bibitem[\protect\citeauthoryear{Kwak, Lee, Park, and Moon}{Kwak
  et~al\mbox{.}}{2010}]%
        {Kwak:2010:TSN:1772690.1772751}
\bibfield{author}{\bibinfo{person}{Haewoon Kwak}, \bibinfo{person}{Changhyun
  Lee}, \bibinfo{person}{Hosung Park}, {and} \bibinfo{person}{Sue Moon}.}
  \bibinfo{year}{2010}\natexlab{}.
\newblock \showarticletitle{What is Twitter, a Social Network or a News
  Media?}. In \bibinfo{booktitle}{\emph{Proceedings of the 19th International
  Conference on World Wide Web}} \emph{(\bibinfo{series}{WWW '10})}.
  \bibinfo{publisher}{ACM}, \bibinfo{address}{New York, NY, USA},
  \bibinfo{pages}{591--600}.
\newblock
\showISBNx{978-1-60558-799-8}
\urldef\tempurl%
\url{https://doi.org/10.1145/1772690.1772751}
\showDOI{\tempurl}


\bibitem[\protect\citeauthoryear{Kyrola, Blelloch, and Guestrin}{Kyrola
  et~al\mbox{.}}{2012}]%
        {graphchi}
\bibfield{author}{\bibinfo{person}{Aapo Kyrola}, \bibinfo{person}{Guy
  Blelloch}, {and} \bibinfo{person}{Carlos Guestrin}.}
  \bibinfo{year}{2012}\natexlab{}.
\newblock \showarticletitle{Graph{C}hi: Large-Scale Graph Computation on Just a
  {PC}}. In \bibinfo{booktitle}{\emph{10th {USENIX} Symposium on Operating
  Systems Design and Implementation ({OSDI} 12)}}.
  \bibinfo{publisher}{{USENIX}}, \bibinfo{pages}{31--46}.
\newblock
\showISBNx{978-1-931971-96-6}
\urldef\tempurl%
\url{https://www.usenix.org/conference/osdi12/technical-sessions/presentation/kyrola}
\showURL{%
\tempurl}


\bibitem[\protect\citeauthoryear{LaSalle and Karypis}{LaSalle and
  Karypis}{2013}]%
        {lasalle2013multi}
\bibfield{author}{\bibinfo{person}{Dominique LaSalle} {and}
  \bibinfo{person}{George Karypis}.} \bibinfo{year}{2013}\natexlab{}.
\newblock \showarticletitle{Multi-threaded graph partitioning}. In
  \bibinfo{booktitle}{\emph{2013 IEEE 27th International Symposium on Parallel
  and Distributed Processing}}. IEEE, \bibinfo{pages}{225--236}.
\newblock


\bibitem[\protect\citeauthoryear{Leskovec and Krevl}{Leskovec and
  Krevl}{2014}]%
        {snapnets}
\bibfield{author}{\bibinfo{person}{Jure Leskovec} {and} \bibinfo{person}{Andrej
  Krevl}.} \bibinfo{year}{2014}\natexlab{}.
\newblock \bibinfo{title}{{SNAP Datasets}: {Stanford} Large Network Dataset
  Collection}.
\newblock \bibinfo{howpublished}{http://snap.stanford.edu/data}.
\newblock


\bibitem[\protect\citeauthoryear{{Lim}, {Kang}, and {Faloutsos}}{{Lim}
  et~al\mbox{.}}{2014}]%
        {slashburn}
\bibfield{author}{\bibinfo{person}{Yongsub {Lim}}, \bibinfo{person}{U {Kang}},
  {and} \bibinfo{person}{Christos {Faloutsos}}.}
  \bibinfo{year}{2014}\natexlab{}.
\newblock \showarticletitle{SlashBurn: Graph Compression and Mining beyond
  Caveman Communities}.
\newblock \bibinfo{journal}{\emph{IEEE Transactions on Knowledge and Data
  Engineering}} \bibinfo{volume}{26}, \bibinfo{number}{12}
  (\bibinfo{year}{2014}), \bibinfo{pages}{3077--3089}.
\newblock
\urldef\tempurl%
\url{https://doi.org/10.1109/TKDE.2014.2320716}
\showDOI{\tempurl}


\bibitem[\protect\citeauthoryear{Low, Bickson, Gonzalez, Guestrin, Kyrola, and
  Hellerstein}{Low et~al\mbox{.}}{2012}]%
        {10.14778/2212351.2212354}
\bibfield{author}{\bibinfo{person}{Yucheng Low}, \bibinfo{person}{Danny
  Bickson}, \bibinfo{person}{Joseph Gonzalez}, \bibinfo{person}{Carlos
  Guestrin}, \bibinfo{person}{Aapo Kyrola}, {and} \bibinfo{person}{Joseph~M.
  Hellerstein}.} \bibinfo{year}{2012}\natexlab{}.
\newblock \showarticletitle{Distributed {G}raph{L}ab: A Framework for Machine
  Learning and Data Mining in the Cloud}.
\newblock \bibinfo{journal}{\emph{Proc. VLDB Endow.}} \bibinfo{volume}{5},
  \bibinfo{number}{8} (\bibinfo{date}{April} \bibinfo{year}{2012}),
  \bibinfo{pages}{716–727}.
\newblock
\showISSN{2150-8097}
\urldef\tempurl%
\url{https://doi.org/10.14778/2212351.2212354}
\showDOI{\tempurl}


\bibitem[\protect\citeauthoryear{Maass, Min, Kashyap, Kang, Kumar, and
  Kim}{Maass et~al\mbox{.}}{2017}]%
        {Maass:2017:MPT:3064176.3064191}
\bibfield{author}{\bibinfo{person}{Steffen Maass}, \bibinfo{person}{Changwoo
  Min}, \bibinfo{person}{Sanidhya Kashyap}, \bibinfo{person}{Woonhak Kang},
  \bibinfo{person}{Mohan Kumar}, {and} \bibinfo{person}{Taesoo Kim}.}
  \bibinfo{year}{2017}\natexlab{}.
\newblock \showarticletitle{Mosaic: Processing a Trillion-Edge Graph on a
  Single Machine}. In \bibinfo{booktitle}{\emph{Proceedings of the Twelfth
  European Conference on Computer Systems}} \emph{(\bibinfo{series}{EuroSys
  '17})}. \bibinfo{publisher}{ACM}, \bibinfo{address}{New York, NY, USA},
  \bibinfo{pages}{527--543}.
\newblock
\showISBNx{978-1-4503-4938-3}
\urldef\tempurl%
\url{https://doi.org/10.1145/3064176.3064191}
\showDOI{\tempurl}


\bibitem[\protect\citeauthoryear{Malewicz, Austern, Bik, Dehnert, Horn, Leiser,
  and Czajkowski}{Malewicz et~al\mbox{.}}{2010}]%
        {pregel}
\bibfield{author}{\bibinfo{person}{Grzegorz Malewicz},
  \bibinfo{person}{Matthew~H. Austern}, \bibinfo{person}{Aart~J.C Bik},
  \bibinfo{person}{James~C. Dehnert}, \bibinfo{person}{Ilan Horn},
  \bibinfo{person}{Naty Leiser}, {and} \bibinfo{person}{Grzegorz Czajkowski}.}
  \bibinfo{year}{2010}\natexlab{}.
\newblock \showarticletitle{Pregel: A System for Large-scale Graph Processing}.
  In \bibinfo{booktitle}{\emph{Proceedings of the 2010 ACM SIGMOD International
  Conference on Management of Data}} \emph{(\bibinfo{series}{SIGMOD '10})}.
  \bibinfo{publisher}{ACM}, \bibinfo{address}{New York, NY, USA},
  \bibinfo{pages}{135--146}.
\newblock
\showISBNx{978-1-4503-0032-2}
\urldef\tempurl%
\url{https://doi.org/10.1145/1807167.1807184}
\showDOI{\tempurl}


\bibitem[\protect\citeauthoryear{Margo and Seltzer}{Margo and Seltzer}{2015}]%
        {Margo:2015:SDG:2824032.2824046}
\bibfield{author}{\bibinfo{person}{Daniel Margo} {and} \bibinfo{person}{Margo
  Seltzer}.} \bibinfo{year}{2015}\natexlab{}.
\newblock \showarticletitle{A Scalable Distributed Graph Partitioner}.
\newblock \bibinfo{journal}{\emph{Proc. VLDB Endow.}} \bibinfo{volume}{8},
  \bibinfo{number}{12} (\bibinfo{date}{Aug.} \bibinfo{year}{2015}),
  \bibinfo{pages}{1478--1489}.
\newblock
\showISSN{2150-8097}
\urldef\tempurl%
\url{https://doi.org/10.14778/2824032.2824046}
\showDOI{\tempurl}


\bibitem[\protect\citeauthoryear{Martella, Logothetis, Loukas, and
  Siganos}{Martella et~al\mbox{.}}{2017}]%
        {spinner}
\bibfield{author}{\bibinfo{person}{Claudio Martella},
  \bibinfo{person}{Dionysios Logothetis}, \bibinfo{person}{Andreas Loukas},
  {and} \bibinfo{person}{Georgos Siganos}.} \bibinfo{year}{2017}\natexlab{}.
\newblock \showarticletitle{Spinner: Scalable Graph Partitioning in the Cloud}.
  In \bibinfo{booktitle}{\emph{2017 IEEE 33rd International Conference on Data
  Engineering (ICDE)}}. \bibinfo{pages}{1083--1094}.
\newblock
\showISSN{2375-026X}
\urldef\tempurl%
\url{https://doi.org/10.1109/ICDE.2017.153}
\showDOI{\tempurl}


\bibitem[\protect\citeauthoryear{Mayer, Mayer, Bhowmik, Epple, and
  Rothermel}{Mayer et~al\mbox{.}}{2018a}]%
        {8621968}
\bibfield{author}{\bibinfo{person}{Christian Mayer}, \bibinfo{person}{Ruben
  Mayer}, \bibinfo{person}{Sukanya Bhowmik}, \bibinfo{person}{Lukas Epple},
  {and} \bibinfo{person}{Kurt Rothermel}.} \bibinfo{year}{2018}\natexlab{a}.
\newblock \showarticletitle{HYPE: Massive Hypergraph Partitioning with
  Neighborhood Expansion}. In \bibinfo{booktitle}{\emph{2018 IEEE International
  Conference on Big Data (Big Data)}}. \bibinfo{pages}{458--467}.
\newblock
\urldef\tempurl%
\url{https://doi.org/10.1109/BigData.2018.8621968}
\showDOI{\tempurl}


\bibitem[\protect\citeauthoryear{Mayer, Mayer, Tariq, Geppert, Laich, Rieger,
  and Rothermel}{Mayer et~al\mbox{.}}{2018b}]%
        {8416335}
\bibfield{author}{\bibinfo{person}{Christian Mayer}, \bibinfo{person}{Ruben
  Mayer}, \bibinfo{person}{Muhammad~Adnan Tariq}, \bibinfo{person}{Heiko
  Geppert}, \bibinfo{person}{Larissa Laich}, \bibinfo{person}{Lukas Rieger},
  {and} \bibinfo{person}{Kurt Rothermel}.} \bibinfo{year}{2018}\natexlab{b}.
\newblock \showarticletitle{{ADWISE}: Adaptive Window-Based Streaming Edge
  Partitioning for High-Speed Graph Processing}. In
  \bibinfo{booktitle}{\emph{2018 IEEE 38th International Conference on
  Distributed Computing Systems (ICDCS)}}. \bibinfo{pages}{685--695}.
\newblock
\showISSN{2575-8411}
\urldef\tempurl%
\url{https://doi.org/10.1109/ICDCS.2018.00072}
\showDOI{\tempurl}


\bibitem[\protect\citeauthoryear{Nishimura and Ugander}{Nishimura and
  Ugander}{2013}]%
        {Nishimura:2013:RGP:2487575.2487696}
\bibfield{author}{\bibinfo{person}{Joel Nishimura} {and} \bibinfo{person}{Johan
  Ugander}.} \bibinfo{year}{2013}\natexlab{}.
\newblock \showarticletitle{Restreaming Graph Partitioning: Simple Versatile
  Algorithms for Advanced Balancing}. In \bibinfo{booktitle}{\emph{Proceedings
  of the 19th ACM SIGKDD International Conference on Knowledge Discovery and
  Data Mining}} \emph{(\bibinfo{series}{KDD '13})}. \bibinfo{publisher}{ACM},
  \bibinfo{address}{New York, NY, USA}, \bibinfo{pages}{1106--1114}.
\newblock
\showISBNx{978-1-4503-2174-7}
\urldef\tempurl%
\url{https://doi.org/10.1145/2487575.2487696}
\showDOI{\tempurl}


\bibitem[\protect\citeauthoryear{Pacaci and \"{O}zsu}{Pacaci and
  \"{O}zsu}{2019}]%
        {Pacaci:2019:EAS:3299869.3300076}
\bibfield{author}{\bibinfo{person}{Anil Pacaci} {and} \bibinfo{person}{M.~Tamer
  \"{O}zsu}.} \bibinfo{year}{2019}\natexlab{}.
\newblock \showarticletitle{Experimental Analysis of Streaming Algorithms for
  Graph Partitioning}. In \bibinfo{booktitle}{\emph{Proceedings of the 2019
  International Conference on Management of Data}}
  \emph{(\bibinfo{series}{SIGMOD '19})}. \bibinfo{publisher}{ACM},
  \bibinfo{address}{New York, NY, USA}, \bibinfo{pages}{1375--1392}.
\newblock
\showISBNx{978-1-4503-5643-5}
\urldef\tempurl%
\url{https://doi.org/10.1145/3299869.3300076}
\showDOI{\tempurl}


\bibitem[\protect\citeauthoryear{Pellegrini and Roman}{Pellegrini and
  Roman}{1996}]%
        {10.5555/645560.658570}
\bibfield{author}{\bibinfo{person}{Fran\c{c}ois Pellegrini} {and}
  \bibinfo{person}{Jean Roman}.} \bibinfo{year}{1996}\natexlab{}.
\newblock \showarticletitle{SCOTCH: A Software Package for Static Mapping by
  Dual Recursive Bipartitioning of Process and Architecture Graphs}. In
  \bibinfo{booktitle}{\emph{Proceedings of the International Conference and
  Exhibition on High-Performance Computing and Networking}}
  \emph{(\bibinfo{series}{HPCN Europe 1996})}.
  \bibinfo{publisher}{Springer-Verlag}, \bibinfo{address}{Berlin, Heidelberg},
  \bibinfo{pages}{493–498}.
\newblock
\showISBNx{3540611428}


\bibitem[\protect\citeauthoryear{Petroni, Querzoni, Daudjee, Kamali, and
  Iacoboni}{Petroni et~al\mbox{.}}{2015}]%
        {Petroni:2015:HSP:2806416.2806424}
\bibfield{author}{\bibinfo{person}{Fabio Petroni}, \bibinfo{person}{Leonardo
  Querzoni}, \bibinfo{person}{Khuzaima Daudjee}, \bibinfo{person}{Shahin
  Kamali}, {and} \bibinfo{person}{Giorgio Iacoboni}.}
  \bibinfo{year}{2015}\natexlab{}.
\newblock \showarticletitle{{HDRF}: Stream-Based Partitioning for Power-Law
  Graphs}. In \bibinfo{booktitle}{\emph{Proceedings of the 24th ACM
  International Conference on Information and Knowledge Management}}
  \emph{(\bibinfo{series}{CIKM '15})}. \bibinfo{publisher}{ACM},
  \bibinfo{address}{New York, NY, USA}, \bibinfo{pages}{243--252}.
\newblock
\showISBNx{978-1-4503-3794-6}
\urldef\tempurl%
\url{https://doi.org/10.1145/2806416.2806424}
\showDOI{\tempurl}


\bibitem[\protect\citeauthoryear{Roy, Bindschaedler, Malicevic, and
  Zwaenepoel}{Roy et~al\mbox{.}}{2015}]%
        {Roy:2015:CSG:2815400.2815408}
\bibfield{author}{\bibinfo{person}{Amitabha Roy}, \bibinfo{person}{Laurent
  Bindschaedler}, \bibinfo{person}{Jasmina Malicevic}, {and}
  \bibinfo{person}{Willy Zwaenepoel}.} \bibinfo{year}{2015}\natexlab{}.
\newblock \showarticletitle{Chaos: Scale-out Graph Processing from Secondary
  Storage}. In \bibinfo{booktitle}{\emph{Proceedings of the 25th Symposium on
  Operating Systems Principles}} \emph{(\bibinfo{series}{SOSP '15})}.
  \bibinfo{publisher}{ACM}, \bibinfo{address}{New York, NY, USA},
  \bibinfo{pages}{410--424}.
\newblock
\showISBNx{978-1-4503-3834-9}
\urldef\tempurl%
\url{https://doi.org/10.1145/2815400.2815408}
\showDOI{\tempurl}


\bibitem[\protect\citeauthoryear{Roy, Mihailovic, and Zwaenepoel}{Roy
  et~al\mbox{.}}{2013}]%
        {Roy:2013:XEG:2517349.2522740}
\bibfield{author}{\bibinfo{person}{Amitabha Roy}, \bibinfo{person}{Ivo
  Mihailovic}, {and} \bibinfo{person}{Willy Zwaenepoel}.}
  \bibinfo{year}{2013}\natexlab{}.
\newblock \showarticletitle{X-{S}tream: Edge-centric Graph Processing Using
  Streaming Partitions}. In \bibinfo{booktitle}{\emph{Proceedings of the
  Twenty-Fourth ACM Symposium on Operating Systems Principles}}
  \emph{(\bibinfo{series}{SOSP '13})}. \bibinfo{publisher}{ACM},
  \bibinfo{address}{New York, NY, USA}, \bibinfo{pages}{472--488}.
\newblock
\showISBNx{978-1-4503-2388-8}
\urldef\tempurl%
\url{https://doi.org/10.1145/2517349.2522740}
\showDOI{\tempurl}


\bibitem[\protect\citeauthoryear{Sanders and Schulz}{Sanders and
  Schulz}{2011}]%
        {sanders2011engineering}
\bibfield{author}{\bibinfo{person}{Peter Sanders} {and}
  \bibinfo{person}{Christian Schulz}.} \bibinfo{year}{2011}\natexlab{}.
\newblock \showarticletitle{Engineering multilevel graph partitioning
  algorithms}. In \bibinfo{booktitle}{\emph{European Symposium on Algorithms}}.
  Springer, \bibinfo{pages}{469--480}.
\newblock


\bibitem[\protect\citeauthoryear{Schlag, Schulz, Seemaier, and Strash}{Schlag
  et~al\mbox{.}}{2019}]%
        {schlag2019scalable}
\bibfield{author}{\bibinfo{person}{Sebastian Schlag},
  \bibinfo{person}{Christian Schulz}, \bibinfo{person}{Daniel Seemaier}, {and}
  \bibinfo{person}{Darren Strash}.} \bibinfo{year}{2019}\natexlab{}.
\newblock \showarticletitle{Scalable Edge Partitioning}. In
  \bibinfo{booktitle}{\emph{2019 Proceedings of the Twenty-First Workshop on
  Algorithm Engineering and Experiments (ALENEX)}}. SIAM,
  \bibinfo{pages}{211--225}.
\newblock


\bibitem[\protect\citeauthoryear{Shao, Wang, and Li}{Shao
  et~al\mbox{.}}{2013}]%
        {10.1145/2463676.2467799}
\bibfield{author}{\bibinfo{person}{Bin Shao}, \bibinfo{person}{Haixun Wang},
  {and} \bibinfo{person}{Yatao Li}.} \bibinfo{year}{2013}\natexlab{}.
\newblock \showarticletitle{Trinity: A Distributed Graph Engine on a Memory
  Cloud}. In \bibinfo{booktitle}{\emph{Proceedings of the 2013 ACM SIGMOD
  International Conference on Management of Data}}
  \emph{(\bibinfo{series}{SIGMOD ’13})}. \bibinfo{publisher}{ACM},
  \bibinfo{address}{New York, NY, USA}, \bibinfo{pages}{505–516}.
\newblock
\showISBNx{9781450320375}
\urldef\tempurl%
\url{https://doi.org/10.1145/2463676.2467799}
\showDOI{\tempurl}


\bibitem[\protect\citeauthoryear{Slota, Rajamanickam, Devine, and
  Madduri}{Slota et~al\mbox{.}}{2017}]%
        {xtrapulp}
\bibfield{author}{\bibinfo{person}{George~M. Slota},
  \bibinfo{person}{Sivasankaran Rajamanickam}, \bibinfo{person}{Karen Devine},
  {and} \bibinfo{person}{Kamesh Madduri}.} \bibinfo{year}{2017}\natexlab{}.
\newblock \showarticletitle{Partitioning Trillion-Edge Graphs in Minutes}. In
  \bibinfo{booktitle}{\emph{2017 IEEE International Parallel and Distributed
  Processing Symposium (IPDPS)}}. \bibinfo{pages}{646--655}.
\newblock
\showISSN{1530-2075}
\urldef\tempurl%
\url{https://doi.org/10.1109/IPDPS.2017.95}
\showDOI{\tempurl}


\bibitem[\protect\citeauthoryear{Stanton and Kliot}{Stanton and Kliot}{2012}]%
        {Stanton:2012:SGP:2339530.2339722}
\bibfield{author}{\bibinfo{person}{Isabelle Stanton} {and}
  \bibinfo{person}{Gabriel Kliot}.} \bibinfo{year}{2012}\natexlab{}.
\newblock \showarticletitle{Streaming Graph Partitioning for Large Distributed
  Graphs}. In \bibinfo{booktitle}{\emph{Proceedings of the 18th ACM SIGKDD
  International Conference on Knowledge Discovery and Data Mining}}
  \emph{(\bibinfo{series}{KDD '12})}. \bibinfo{publisher}{ACM},
  \bibinfo{address}{New York, NY, USA}, \bibinfo{pages}{1222--1230}.
\newblock
\showISBNx{978-1-4503-1462-6}
\urldef\tempurl%
\url{https://doi.org/10.1145/2339530.2339722}
\showDOI{\tempurl}


\bibitem[\protect\citeauthoryear{{Tinney} and {Walker}}{{Tinney} and
  {Walker}}{1967}]%
        {1447941}
\bibfield{author}{\bibinfo{person}{William~F. {Tinney}} {and}
  \bibinfo{person}{John~W. {Walker}}.} \bibinfo{year}{1967}\natexlab{}.
\newblock \showarticletitle{Direct solutions of sparse network equations by
  optimally ordered triangular factorization}.
\newblock \bibinfo{journal}{\emph{Proc. IEEE}} \bibinfo{volume}{55},
  \bibinfo{number}{11} (\bibinfo{year}{1967}), \bibinfo{pages}{1801--1809}.
\newblock


\bibitem[\protect\citeauthoryear{Tsourakakis, Gkantsidis, Radunovic, and
  Vojnovic}{Tsourakakis et~al\mbox{.}}{2014}]%
        {Tsourakakis:2014:FSG:2556195.2556213}
\bibfield{author}{\bibinfo{person}{Charalampos Tsourakakis},
  \bibinfo{person}{Christos Gkantsidis}, \bibinfo{person}{Bozidar Radunovic},
  {and} \bibinfo{person}{Milan Vojnovic}.} \bibinfo{year}{2014}\natexlab{}.
\newblock \showarticletitle{{FENNEL}: Streaming Graph Partitioning for Massive
  Scale Graphs}. In \bibinfo{booktitle}{\emph{Proceedings of the 7th ACM
  International Conference on Web Search and Data Mining}}
  \emph{(\bibinfo{series}{WSDM '14})}. \bibinfo{publisher}{ACM},
  \bibinfo{address}{New York, NY, USA}, \bibinfo{pages}{333--342}.
\newblock
\showISBNx{978-1-4503-2351-2}
\urldef\tempurl%
\url{https://doi.org/10.1145/2556195.2556213}
\showDOI{\tempurl}


\bibitem[\protect\citeauthoryear{Verma, Leslie, Shin, and Gupta}{Verma
  et~al\mbox{.}}{2017}]%
        {verma-vldb}
\bibfield{author}{\bibinfo{person}{Shiv Verma}, \bibinfo{person}{Luke~M.
  Leslie}, \bibinfo{person}{Yosub Shin}, {and} \bibinfo{person}{Indranil
  Gupta}.} \bibinfo{year}{2017}\natexlab{}.
\newblock \showarticletitle{An Experimental Comparison of Partitioning
  Strategies in Distributed Graph Processing}.
\newblock \bibinfo{journal}{\emph{Proc. VLDB Endow.}} \bibinfo{volume}{10},
  \bibinfo{number}{5} (\bibinfo{date}{Jan.} \bibinfo{year}{2017}),
  \bibinfo{pages}{493--504}.
\newblock
\showISSN{2150-8097}
\urldef\tempurl%
\url{https://doi.org/10.14778/3055540.3055543}
\showDOI{\tempurl}


\bibitem[\protect\citeauthoryear{Walshaw and Cross}{Walshaw and Cross}{2000}]%
        {doi:10.1137/S1064827598337373}
\bibfield{author}{\bibinfo{person}{Chris Walshaw} {and} \bibinfo{person}{Mark
  Cross}.} \bibinfo{year}{2000}\natexlab{}.
\newblock \showarticletitle{Mesh Partitioning: A Multilevel Balancing and
  Refinement Algorithm}.
\newblock \bibinfo{journal}{\emph{SIAM Journal on Scientific Computing}}
  \bibinfo{volume}{22}, \bibinfo{number}{1} (\bibinfo{year}{2000}),
  \bibinfo{pages}{63--80}.
\newblock
\urldef\tempurl%
\url{https://doi.org/10.1137/S1064827598337373}
\showDOI{\tempurl}


\bibitem[\protect\citeauthoryear{Webber}{Webber}{2012}]%
        {10.1145/2384716.2384777}
\bibfield{author}{\bibinfo{person}{Jim Webber}.}
  \bibinfo{year}{2012}\natexlab{}.
\newblock \showarticletitle{A Programmatic Introduction to {N}eo4j}. In
  \bibinfo{booktitle}{\emph{Proceedings of the 3rd Annual Conference on
  Systems, Programming, and Applications: Software for Humanity}}
  \emph{(\bibinfo{series}{SPLASH ’12})}. \bibinfo{publisher}{ACM},
  \bibinfo{address}{New York, NY, USA}, \bibinfo{pages}{217–218}.
\newblock
\showISBNx{9781450315630}
\urldef\tempurl%
\url{https://doi.org/10.1145/2384716.2384777}
\showDOI{\tempurl}


\bibitem[\protect\citeauthoryear{Xie, Yan, Li, and Zhang}{Xie
  et~al\mbox{.}}{2014}]%
        {dbh}
\bibfield{author}{\bibinfo{person}{Cong Xie}, \bibinfo{person}{Ling Yan},
  \bibinfo{person}{Wu-Jun Li}, {and} \bibinfo{person}{Zhihua Zhang}.}
  \bibinfo{year}{2014}\natexlab{}.
\newblock \showarticletitle{Distributed Power-law Graph Computing: Theoretical
  and Empirical Analysis}.
\newblock In \bibinfo{booktitle}{\emph{Advances in Neural Information
  Processing Systems 27}}. \bibinfo{pages}{1673--1681}.
\newblock


\bibitem[\protect\citeauthoryear{Yang and Leskovec}{Yang and Leskovec}{2012}]%
        {6413740}
\bibfield{author}{\bibinfo{person}{Jaewon Yang} {and} \bibinfo{person}{Jure
  Leskovec}.} \bibinfo{year}{2012}\natexlab{}.
\newblock \showarticletitle{Defining and Evaluating Network Communities Based
  on Ground-Truth}. In \bibinfo{booktitle}{\emph{2012 IEEE 12th International
  Conference on Data Mining}}. \bibinfo{pages}{745--754}.
\newblock
\showISSN{1550-4786}
\urldef\tempurl%
\url{https://doi.org/10.1109/ICDM.2012.138}
\showDOI{\tempurl}


\bibitem[\protect\citeauthoryear{Zhang, Wei, Liu, Tang, and Li}{Zhang
  et~al\mbox{.}}{2017}]%
        {Zhang:2017:GEP:3097983.3098033}
\bibfield{author}{\bibinfo{person}{Chenzi Zhang}, \bibinfo{person}{Fan Wei},
  \bibinfo{person}{Qin Liu}, \bibinfo{person}{Zhihao~Gavin Tang}, {and}
  \bibinfo{person}{Zhenguo Li}.} \bibinfo{year}{2017}\natexlab{}.
\newblock \showarticletitle{Graph Edge Partitioning via Neighborhood
  Heuristic}. In \bibinfo{booktitle}{\emph{Proceedings of the 23rd ACM SIGKDD
  International Conference on Knowledge Discovery and Data Mining}}
  \emph{(\bibinfo{series}{KDD '17})}. \bibinfo{publisher}{ACM},
  \bibinfo{address}{New York, NY, USA}, \bibinfo{pages}{605--614}.
\newblock
\showISBNx{978-1-4503-4887-4}
\urldef\tempurl%
\url{https://doi.org/10.1145/3097983.3098033}
\showDOI{\tempurl}


\bibitem[\protect\citeauthoryear{Zhu, Han, and Chen}{Zhu et~al\mbox{.}}{2015}]%
        {gridgraph}
\bibfield{author}{\bibinfo{person}{Xiaowei Zhu}, \bibinfo{person}{Wentao Han},
  {and} \bibinfo{person}{Wenguang Chen}.} \bibinfo{year}{2015}\natexlab{}.
\newblock \showarticletitle{Grid{G}raph: Large-Scale Graph Processing on a
  Single Machine Using 2-Level Hierarchical Partitioning}. In
  \bibinfo{booktitle}{\emph{2015 {USENIX} Annual Technical Conference ({USENIX}
  {ATC} 15)}}. \bibinfo{publisher}{{USENIX}}, \bibinfo{pages}{375--386}.
\newblock
\showISBNx{978-1-931971-225}
\urldef\tempurl%
\url{https://www.usenix.org/conference/atc15/technical-session/presentation/zhu}
\showURL{%
\tempurl}


\end{thebibliography}

\end{document}